\documentclass[%
reprint,
superscriptaddress,
amsmath,amssymb,
aps,
pra,
]{revtex4-2}
\usepackage{graphicx}% Include figure files
\usepackage{dcolumn}% Align table columns on decimal point
\usepackage{bm}% bold math
\usepackage{bbm}
\graphicspath{{SupplFigures/}}
\usepackage{url}
\usepackage{xcolor}
\usepackage{amsthm}
\usepackage{slashed}
\usepackage{mathtools}
\usepackage{graphicx}
\usepackage{hyperref}
\usepackage{float}
\usepackage{comment}
\hypersetup{
  colorlinks   = true, 
  urlcolor     = blue, %Colour for external hyperlinks
  linkcolor    = blue, %Colour of internal links
  citecolor   = red %Colour of citations
}

\usepackage{minitoc}

\newcommand{\eqnref}[1]{Eq.~(\ref{#1})}
\newcommand{\figref}[1]{Fig.~\ref{#1}}
\newcommand{\sfigref}[2]{Fig.~\hyperref[#1]{\ref{#1}#2}}

\newcommand{\beq}{\begin{equation}}
	\newcommand{\eeq}{\end{equation}}
\newcommand{\beqd}{\begin{equation*}}
	\newcommand{\eeqd}{\end{equation*}}

\newcommand{\bpm}{\begin{pmatrix}}
	\newcommand{\epm}{\end{pmatrix}}

\DeclareSymbolFont{cmbrightop}{OT1}{cmbr}{m}{n}
\DeclareMathSymbol{\sfPsi}{\mathalpha}{cmbrightop}{9}
 \newcommand{\bcen}{\begin{center}}
 \newcommand{\ecen}{\end{center}}
 \newcommand{\btab}{\begin{tabular}}
 \newcommand{\etab}{\end{tabular}}
 \newcommand{\bdes}{\begin{description}}
 \newcommand{\edes}{\end{description}}

 \newcommand{\bea}{\begin{eqnarray}}
 \newcommand{\eea}{\end{eqnarray}}

 \newcommand{\half}{\frac{1}{2}}
 \newcommand{\bary}{\begin{array}}
 \newcommand{\eary}{\end{array}}
 \newcommand{\benum}{\begin{enumerate}}
 \newcommand{\eenum}{\end{enumerate}}
 \newcommand{\bitem}{\begin{itemize}}
 \newcommand{\eitem}{\end{itemize}}

 \newcommand{\bOne}{{\boldsymbol{1}}}
 \newcommand{\bG} { {\boldsymbol G}}
 \newcommand{\bH} { \mbox{\boldmath $H$}}

 \newcommand{\mean}[1]{\langle #1 \rangle}
 
 \newcommand{\bra}[1]{{{\langle #1 |}}}
 \newcommand{\ket}[1]{{| #1 \rangle}}

 \newcommand{\eqn}[1] {eqn.~(\ref{#1})}

 \newcommand{\GKSnote}[1]{} %Hides some of my comments/notes for now, can be unmasked by introducing an appropriate color
 \newcommand{\Reals}{{\mathbb{R}}}

 \newcommand{\ci}{\mathbbm{i}}

\newcommand{\oibook}[1]{}

\newcommand{\mytitle}{{Arboreal Obstructed Atomic Insulating and Metallic  Phases of Fermions}}
\newcommand{\myaffl}{{Centre for Condensed Matter Theory, Department of Physics, Indian Institute of Science, Bangalore 560012, India.}}

\definecolor{gkblue}{rgb}{0.294, 0.0, 0.506}

\newcommand{\Insulator}{{\bf I}}
\newcommand{\Metal}{{\bf M}}
\newcommand{\Critical}{{${\mathcal C}$}}
\newcommand{\SM}[1]{see \cite{SM}, sec.~\ref{#1}}

\begin{document}

\title{\mytitle}
\author{Gurkirat Singh}\email{gurkirats@iisc.ac.in}
\author{Surajit Bera}\email{surajit@iisc.ac.in}
\author{Vijay B.~Shenoy}\email{shenoy@iisc.ac.in}
\affiliation{\myaffl}

\date{\today{}}
\begin{abstract} 
We explore phases of free fermions on arenas that do not tessellate a manifold. Specializing to arboreal arenas described by tree graphs which possess a notion of translation symmetry, we study possible fermionic phases in the BDI symmetry class on the $p$-coordinated Bethe lattice. We find that there are $p$ distinct obstructed atomic insulating phases that are characterized by distinct edge states, pattern of entanglement, and a winding characteristic that we define here. These distinct insulting phases are always separated by a metallic region in the parameter space rather than isolated quantum critical points. The metallic region itself comprises several distinct metallic phases that are distinguished by the winding characteristic and correlation functions. The correlation functions of distinct subsystems display non-analytic behavior at distinct points in the metallic region, signaling a cascade of subsystem transitions. An intriguing feature of these arboreal metals is the presence of truncated subsystems with zero energy boundary modes despite being gapless. This work suggests new opportunities for synthetic quantum systems to realize these novel phases.
\end{abstract}


\maketitle 

%\tableofcontents{}

%\input{maintext_tenfold}

%\tableofcontents

%Example of Surajit comments

\section{Introduction}
\label{sec:Intro}

The discovery of the topological phases of matter\cite{vonKlitzing1980} in semiconductor heterostructures motivated activity\cite{Laughlin1981,TKNN1982,Haldane1988} that produced momentous advances\cite{KaneQSHE,Bernevig2006,KaneTI3D,KaneTIInv,MooreBalents2007,Roy2009,HasanKane2010,QiZhang2011} in the last decade leading to the classification\cite{KitaevPT,RyuTFW,ChiuRyu2016} of band insulating phases of free fermions based on their intrinsic symmetries.\cite{Altland1997,Zirnbauer2010} %Central to the classification scheme is intrinsic fermionic symmetries, including time reversal, charge conjugation, and sublattice symmetries. Fermions are viewed as projective representations of the symmetries leading to the ten symmetry classes\cite{Altland1997,Zirnbauer2010}, usually termed the ``tenfold way''. The topological classification proceeds by studying the homotopy classes of maps from the $d$-dimensional torus (the Brillouin zone of a $d$-dimensional crystal), which produces the classification in a ``periodic table''\cite{KitaevPT,RyuTFW}.
More recent work has advanced these findings into new directions. It is now realized that topological phases can be realized in quasicrystals\cite{Kraus2012,Goldman2015,Fulga2016,Rechtsman2016}, and even amorphous systems\cite{Agarwala2017,Mitchell2018}. Moreover, additional arena symmetries such as space group symmetries lead to more complex classifications\cite{Slager2013}. Along these lines, the notion of higher-order topological phases was  developed\cite{BB1,BB2,ZZC,Josias}, and a general class of obstructed atomic phases\cite{Schindler,Benalcazar2019} have come to prominence. Spurred by the promise of application\cite{Gilbert2021}, these developments have led to the search and study of materials that realize these novel phases in insulating crystals\cite{Bradlyn2017,Cano2021}.  

On the other hand, the contemporaneous advances in quantum technologies\cite{Cirac2012,Blais2020,Ozawa2019rmp,Lee2018,Roy2021,Souslov2022} offer unprecedented new opportunities to realize novel phases beyond those explored thus far. In particular, much of the advances in topological phases are based on systems whose long wavelength physics admits a quantum field theoretical description where the fields are associated with points on a manifold. The crystal (or the amorphous lattice) acts as an ultraviolet tessellation of this manifold. The new opportunities provided by synthetic quantum systems allow the exploration of systems beyond manifolds, i.~e., the creation of quantum many-body systems whose long wavelength description goes beyond the conventional quantum field theory. %with fields associated with points on a manifold. 
This motivates the study of many body systems defined on graphs (a collection of points and links) that are more general, not necessarily tessellating a manifold\cite{Nandagopal2023}. While there is a myriad of possibilities for such graphs, clearly defined and tractable systems emerge on focusing on those graphs that are trees (i.~e., graphs without closed loops) and, in addition, have a notion of ``translational symmetry'', i.e., that every point/site has a similar environment; we dub such tree graphs as ``arboreal arenas''. Infinite Bethe lattices are examples of arboreal arenae with the notion of ``translational symmetry''. %The aim of this paper is to explore phases of free fermions in these arboreal arenas. Previous works and even in disordered systems\cite{Tikhonov2016}.

\begin{figure*}
% \centerline{\includegraphics[width=0.3\textwidth]{NewNewFigures/fig1_a_InfiniteSystem.png}~\includegraphics[width=0.3\textwidth]{NewNewFigures/fig1_b_SSHchains.png}~\includegraphics[width=0.3\textwidth]{NewNewFigures/fig1_c_FiniteSubsystem.png}
% }
% \centerline{(a)~~~\hspace{0.25\textwidth}~~~~~~(b)~~~~~~~~\hspace{0.2\textwidth}~~~~~~~~(c)}

\includegraphics[width=\linewidth]{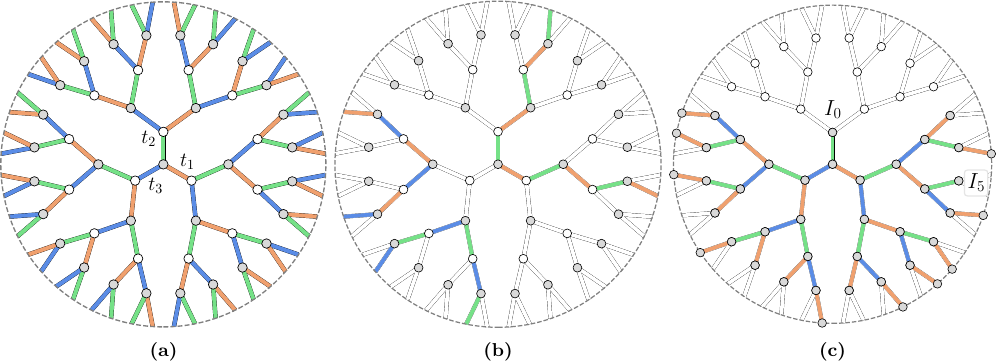}
    
\caption{{\bf Arboreal arena:} {\bf (a)} Infinite Bethe lattice with $p=3$. The links are color-coded according to the hopping value on them. {\bf (b)} Different types of chain subsystems used to characterize the entanglement structure of the ground state. For example, the orange-green chain shows a $[12 \ldots 21]$ SSH chain. {\bf (c)} $S_{5}$ subsystem (see end of section \ref{sec:Model:Methods} for the definition of $S_n$ subsystem). When $t_2 = 0$, sites $I_0$ and $I_5$ become isolated (no hopping in or out of the sites).
}

\label{fig:schematic}
\end{figure*}

This work aims to explore the many-body phases of free fermions in arboreal arenas. While a variety of arboreal arenas are available for study\cite{Nandagopal2023}, 
in this paper, we focus on topological phases realized %on those that are defined 
on a $p$-coordinated Bethe lattice. The system is described by $p$ real hopping parameters and at half filling falls in the class BDI \cite{Altland1997,Agarwala2017}. We uncover that this system hosts $p$-distinct arboreal obstructed atomic insulating (AOAI) phases, which are distinguished by their correlation functions, entanglement structure of subsystems, presence of protected boundary modes, and a new type of winding characteristic of certain Green's functions. These AOAI phases are separated from each other by a metallic region in the phase diagram. Remarkably, this metallic region itself consists of $p+1$ distinct gapless regimes. Of these, $p$ phases (gapless) are related to the $p$ insulating phases, carrying forth certain characteristics (such as the winding property). Finally, there is a metallic phase (which does not have the winding property), distinguished from the $p$ other metallic phases by the correlation functions. While all the insulating phases support zero energy boundary modes in a truncated system, the notable novelty of the arboreal arena is the presence of such zero modes even in the {\em gapless} metallic regimes. These findings uncoverd here are key new features of fermionic phases on arboreal arenas not discussed in previous works\cite{Mahan2001,Economou2006,Aryal2020,Weststrom2023,Tikhonov2016}. 

The bulk of the paper is organized in four sections, section.~\ref{sec:phase_diagram} - section.~\ref{sec:PD:Gapless}, which discuss the phase diagram, insulating phases, phase transitions and gapless phases in that sequence. In the main text, we present a comprehensive description of the main results, relegating, wherever possible, technical details to the Supplemental Material\cite{SM}. We conclude the paper in section~\ref{sec:Discussion} with a discussion and further perspective.

\section{Model and Methods}
\label{sec:Model}

\subsection{Model}
\label{sec:Model:Def}

%A general arboreal arena is realized on a graph with a set of sites and links that connect the sites, such that there are no closed loops. We will to restrict our attention to those arboreal arenas that realize some notion of ``translational symmetry'', which roughly means that every site has an identical environment. 

The arboreal arena chosen for study here that realizes the notion of ``translational symmetry'' is the $p$-coordinated Bethe lattice (see \figref{fig:schematic}), called $B(p)$. Each site of $B(p)$, labeled by $I$, is connected to $p$ (a positive integer, $p \ge 1$) other sites, by $p$ links. The sites can be organized starting from a root site $I_0$ (also called the generation 0 site), to which $p$ sites of generation $1$ are linked, and so on. Each site of a generation $g > 0$ is linked to $p-1$ sites of a generation $g+1$. Thus each site $I$ can be assigned a generation number $g(I)$. Links come in $p$ different colors labelled $\alpha=1,\dots,p$, with the additional constraint that every site has exactly one link of each color attached to it. A path from site $I$ to $J$ is specified by the sequence of sites and links that join $I$ and $J$. A closed loop at $I$ is a path such that no link is traversed twice and $J=I$. A central property of $B(p)$, already noted above, is that there are no closed loops. Further, $B(p)$ is connected in that there is a ``straight-path'' from any two distinct $I$ and $J$ in which no intervening sites or links are visited more than once. Finally, the notation $I+\alpha$ refers to the site reached by moving along the link of color $\alpha$ from $I$. 

We define a fermionic Hilbert space on this arena with the help for fermionic creation (annihilation) operators $c^\dagger_I (c_I)$ which creates (annihilates) a fermion at the site $I$ of $B(p)$.
The single particle state $\ket{I}  = c^\dagger_I\ket{\textup{vac}}$, where $\ket{\textup{vac}}$ is the vacuum state on $B(p)$.
Associating the hopping amplitude $t_\alpha$ with each link of color $\alpha$,  we introduce the hamiltonian
\beq\label{eqn:BetheHopping}
{\cal H} = -\sum_{I,\alpha} t_{\alpha} c^\dagger_{I+\alpha} c_{I} - \mu \sum_I c^\dagger_I c_I
\eeq
where $t_\alpha$ can be chosen to be non-negative numbers (\SM{SM:GaugeChoice}). The chemical potential $\mu$ determines the filling (number of fermions per site).

The model defined in \eqnref{eqn:BetheHopping} has time reversal symmetry ${\cal T}$, that is an antilinear operator that transforms ${\cal T} c^\dagger_I {\cal T}^{-1} = c^\dagger_I$. Further, when $\mu=0$ (which corresponds to half filling that is the focus here), the arboreal model \eqnref{eqn:BetheHopping} is endowed with a sublattice symmetry ${\cal S}$, which is an antilinear operator that transforms ${\cal S} c^\dagger_{I} {\cal S}^{-1} = (-1)^{g(I)} c_{I}$, where we remember that $g(I)$ is the generation of the site $I$. This puts the model in the symmetry class BDI \cite{Altland1997,Agarwala2017}. %\GKSaddition{The charge conjugation operator ${\cal C} = {\cal T} {\cal S}$ is also a symmetry with ${\cal C}^2 = 1$, which puts the model in the symmetry class BDI (\cite{Agarwala2017})}

\begin{figure*}

\includegraphics[width=\linewidth]{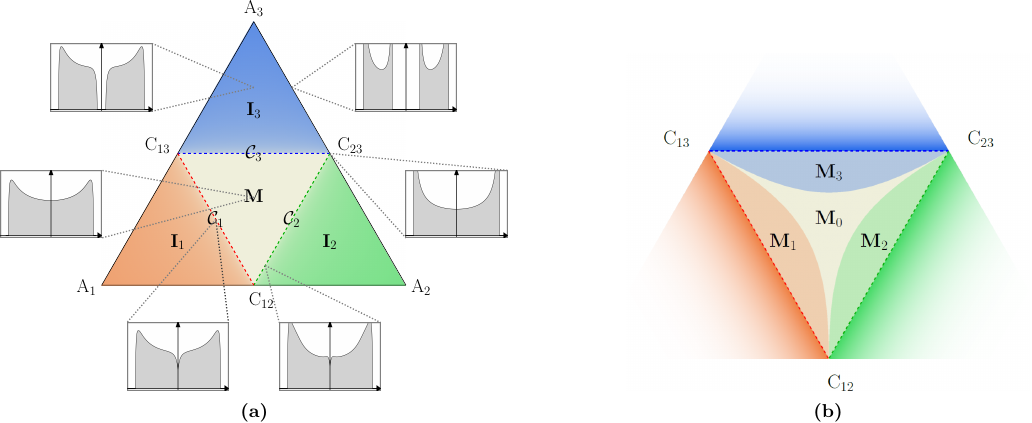}

\caption{
{\bf Phase Diagram:} {\bf (a)} Triangle shows the parameter simplex for $p=3$ Bethe lattice with $t_1^2 + t_2^2 + t_3^2 = 1$. Point $A_1$ corresponds to $(t_1, t_2, t_3) = (1,0,0)$. For a generic point in the simplex, $t^2_\alpha$ is obtained by the perpendicular distance to the side opposite to $A_\alpha$. Regions \Insulator$_\alpha$ host insulating phases, and region \Metal~is metallic. \Critical$_\alpha$ are critical surfaces where the energy gap closes, and are described by $t_\alpha^2 = \half$. The density of states ($\rho(\omega)$ vs.~$\omega$) for each region is shown. {\bf (b)} Blow-up of the \Metal~region. The metallic region \Metal~is made up of four distinct regions \Metal$_\alpha$ and \Metal$_0$, which are distinguished by the nature of their correlation functions and a winding characteristic (see \eqnref{eqn:WC}).
}

\label{fig:phase_diagram}

\end{figure*}

We determine the phase diagram of this model in the space of $t_\alpha \in \Reals^+ \cup \{0\}$. We work in a space $\sum_{\alpha} t^2_{\alpha} = 1$, which we find produces all the interesting features in an aesthetically appealing way. In other words, the parameters ${t_\alpha}$ can  parametrized by a point in a $p$-dimensional regular simplex which we dub as the ``parameter simplex'', defined by a set of apices A$_\alpha$ ($\alpha = 1, \ldots, p$). For example, for $p=3$, this parameter space is an equilateral triangle (as illustrated in \figref{fig:phase_diagram}) with vertices A$_1$, A$_2$, A$_3$,  while for $p=4$, this parameter space is a tetrahedron. For a generic point in the $p$ parameter simplex, the values of $t_\alpha^2$ is obtained by finding the shortest distance to the $(p-1)$ simplex defined by $\{A_{\beta \ne \alpha}\}$, such that $\sum_\alpha t^2_\alpha = 1$.
The apices A$_\alpha$ of the simplex, therefore, represent the points where $t^2_\alpha =1$ and all other $t_{\beta \ne \alpha} = 0$.
%\GKS{How is it parametrized? Not clear what the bijection is between $(t_1^2, t_2^2, t_3^2)$ and points on the triangle. Make comment regarding $t_\alpha^2$ being proportional to the distance to the opposite side}. 
We point out that for $p=2$, the model defined here the famous SSH model\cite{SSH1979,SSH1980} whose parameter space in our formulation is the line-segment  $t_1^2 = 1 - t_2^2 
\in [0,1]$, in which two gapped phases are separated by a critical point at $t_1^2 = t^2_2 = 1/2$. We determine and characterize all the phases and study the nature of transitions between them for arbitrary $p$, using $p=3$ to illustrate the key features. 

\subsection{Methods}
\label{sec:Model:Methods}
To achieve the ends discussed in the previous section we use the well known techniques of resolvent Green's functions\cite{Economou2006}. Working with the single-particle states, with the hamiltonian expressed as $\bH = -\sum_{I \alpha} t_\alpha \ket{I+\alpha}\bra{I}$ we define
\beq\label{eqn:ResG}
\bG(z) = (z \bOne - \bH)^{-1}
\eeq
where $\bOne = \sum_I \ket{I} \bra{I}$, and $z$ is a complex frequency.  Of particular interest are the components,
\beq\label{eqn:GJI}
G_{JI}(z) = \bra{J} \bG(z) \ket{I}
\eeq
with 
\beq\label{eqn:GII}
G(z) := G_{II}(z),
\eeq
the onsite Green's function playing a preeminent role (note that the quantity is independent of $I$, owing to the notion of translational symmetry introduced above). A second set of quantities pertains to the complex frequency-dependent amplitude $G_\alpha(z)$ of returning to site $I$, excluding all paths that traverse link $\alpha$, starting from site $I$. With these definitions, we have (\SM{SM:GreenFn_analytic})
\beq
G^{-1}(z) = z - \sum_{\alpha} t_\alpha^2 G_\alpha(z) \label{eqn:DysonG}
\eeq
and 
\beq
G_\alpha(z)^{-1} = z - \sum_{\beta \ne \alpha}  t_{\beta}^2 G_{\beta}(z). \label{eqn:DysonGalpha}
\eeq
%\GKS{Criticism of the way things were written: The models and methods section, as it currently stands, reads as if we have numerically computed quantities, and deduced the physics from there. The issue I feel with writing it like this, is that it will immediately cause mistrust in our statements such as "The physics is the same for arbitrary $p$", "This happens for ALL strong chains", "This happens for ALL weak chains", "Every point in $M_\alpha$ shows a transition". Although people will figure out that the results are analytically established while reading the paper, let us not give them an opportunity to raise their eyebrows}
This system of $p+1$ equations can be solved  analytically in certain limits, such as near phase boundaries and  for special frequencies $z$, which we exploit to establish the bulk of the results (valid for any $p$). We also obtain numerical results for arbitrary  $t_\alpha$, and corroborate the analytic findings, via solutions of $G$, $G_\alpha$ using the Newton-Raphson method (\SM{SM:NewtonRaphson}).

With the knowledge of $G(z)$ and $G_{\alpha}(z)$, we can obtain $G_{JI}(z)$ as
\beq\label{eqn:GJIsol}
    G_{JI}(z) = \left[\prod_{k=1}^n (-t_{\alpha_k} G_{\alpha_k}(z)) \right] G(z)
\eeq
where site $J$  is linked to site $I$ by the ``straight-path'' $[\alpha_1 \alpha_2 \ldots \alpha_n]$. These quantities allow us to obtain the information necessary to investigate the phases obtained. 

We obtain the density of single-particle states (dos)
\beq\label{eqn:dos}
\rho(\omega) = -\frac{1}{\pi} \Im{G(\omega^+)}
\eeq
where $\omega^+ = \omega + i \eta$, $\omega$ being a real frequency and $\eta$ a positive infinitesimal. %The dos provide crucial information such as the metallic or insulating nature of the phase; for example, $\rho(\omega=0) >0$ implies a gapless metallic state (recall we have set $\mu=0$).

A quantity that provides us critical information in understanding the many body states  in different regions of the parameter simplex is the pattern of entanglement of the subsystems of  the arboreal arena. Consider a sub-system consisting of  $N_s$ sites $I_l, l=1,\ldots,N_s$. Using ideas discussed in \cite{Peschel2003,Henley2004,Turner2010,Alexandradinata2011,HsiehFu2014}, we obtain information about the entanglement of the subsystem from the $N_s \times N_s$ correlation matrix 
\beq
C_{I_l I_{l'}} = \mean{c^\dagger_{I_l} c_{I_{l'}}}
\eeq
 whose eigenvalues $\lambda_n, n = 1, \ldots, N_s$  lie between $0$ and $1$. It is known \cite{Alexandradinata2011,HsiehFu2014} that the entanglement structure is revealed in a pattern of eigenvalues of the correlation matrix. For example, in a trivial insulator, the eigenvalues show a gapped spectrum (with an `entanglement gap' $\Delta_g$), with half of the eigengalues close to zero and the other half close to unity. On the other hand, in a metal, the spectrum of the correlation matrix is ``gapless'' and goes smoothly from zero to unity. Most interestingly, in a topologically insulating phase, one obtains ``mid-gap'' modes located near  $1/2$. These modes signal strong entanglement on some of the bonds that are cut to make up the subsystem and is a signature of nontrivial topology. Further, these midgap modes are stable, i.e., in a fully gapped phase, small changes of parameters (that maintain any required symmetry) cannot eliminate them; only a change of parameters that takes the system through a phase transition to a gapped trivial phase will eliminate them (\SM{SM:entanglement}). %\GKS{The previous statement is directly at odds with what we find in our work. In the gapped phases, mid-gap modes found along weak chains are \textbf{not} stable. This is clarified in a later section, but should be clear here too}. 
 Thus, the presence (or absence) of mid-gap modes in the correlation matrix can act as a diagnostic for nontrivial entanglement characterizing a topological phase. This goal of obtaining $\mean{c^\dagger_{I_l} c_{I_{l'}}}$ is achieved by noting that the resolvent Greens function $G_{I_l,I_{l'}}(z)$ evaluated at the fermionic Matsubara frequencies will obtain the {\em many-body} Greens function (as the problem is non-interacting). The correlation matrix  $\mean{c^\dagger_{I_l} c_{I_l'}}$ can now be determined by standard frequency summation techniques aided by techniques from complex analysis (\SM{SM:correlation_functions}). In the discussion below, we will show that the character of the entanglement spectrum is drastically different (from that discussed above for manifold systems) in arboreal arenas.

Another interesting quantity that we investigate is the filling anomaly introduced in \cite{Benalcazar2019} that is related to the presence of edge modes. The filling anomaly is obtained as a discontinuous jump in the number of occupied states as the chemical potential is tuned across $\mu = 0$. The filling anomaly is obtained by isolating a subsystem (i.~e., removing the links that connect it to $B(p)$), i.e., a truncated subsystem, and studying the number of occupied states as a function of the chemical potential. Presence of a filling anomaly indicates that the physics is controlled by an obstructed atomic limit\cite{Benalcazar2019}.

We will be interested in several types of subsystems to study entanglement properties and filling anomalies which are obtained by truncating the subsystem as discussed above. Chain subsystems like those displayed in \figref{fig:schematic}(b) play a preeminent role in the discussions that follow. Depending on the choice, these chain subsystems may be of the SSH type, denoted as $[\alpha \beta \alpha \ldots \beta \alpha]$ when the links forming the chain alternate between two colors $\alpha$ and $\beta$, and the terminal links of the subsystem are both of the color $\alpha$. We will find useful the notion of `strong', `intermediate` and `weak' chains in our discussion, which refer to the number of links associated with the largest hopping. If $t_1$ is the largest hopping, chains of the form $[1 \beta_2 1 \dots \beta_{n-1} 1$] are referred to as ``strong chains'', while chains having no links of the color $1$ are referred to as ``weak chains'', and all other chains are referred to as ``intermediate chains''. In $B(3), t_1 \geq t_2, t_3$, some examples of strong chains include $[12 \ldots 21]$ and $[13 \ldots 31]$, while there are only two kinds of weak chains, $[23 \ldots 32]$ and $[32 \ldots 23]$. Intermediate chains are exemplified by $[123 \ldots 123]$.  %\old{We will frequently use notions `strong', `intermediate' and `weak' chains. For example, in $B(3)$, with $t_1$ as the largest hopping, i.e, $t_1 > t_2 \gtrsim t_3$, a chain with with alternating 1 and 2 type links denoted $[12 \ldots 21]$, or with alternating 1 and 3 type links denoted [13..31] are strong chains. On the other hand a chain of the type [23..32] is a weak chain. A chain of the type [123...123] is termed moderate chain.} 
A second class of subsystems considered, dubbed $B_\alpha(p)$ (\SM{SM:subsystems}), is the subtree that is obtained by removing all links $\beta \ne \alpha$ at the generation 0 site $I_0$.
Finally, also consider another class of subsystems dubbed the  $S_n$ family ($n$ is odd), one of which is shown in \figref{fig:schematic}(c). These subsystems are designed to isolate the $t_2$ link (with $t_1$ being the largest hopping) connected to $I_0$, and study the edge modes that may be hosted on it when the subsystem is truncated. Furthermore, $S_n$ contains a $[21 \ldots 12]$ chain of length $n$ (from $I_0$ to $I_n$), which is the embedded SSH chain responsible for the filling anomaly of the truncated $S_n$ subsystem. In order to exclude interference from other potential edge modes and to have a sense of radial symmetry as $n \rightarrow \infty $, we construct $S_n$ by a recursive procedure (\SM{SM:subsystems}), taking note of possible zero energy modes in tree graphs\cite{IvanGutman2011}.

\begin{figure*}
% \centerline{
% \includegraphics[width = 0.4 \textwidth]{NewNewFigures/fig3_pd.png}~
% \includegraphics[width = 0.45 \textwidth]{NewNewFigures/fig3_EntSpecZoomIn.png}
% }
% \centerline{(a) \hspace{0.45 \textwidth} (b)}
% \centerline{
% \includegraphics[width = 0.45 \textwidth]{NewNewFigures/fig3_Winding.png}~
% \includegraphics[width = 0.45 \textwidth]{NewNewFigures/fig3_FillingAnomaly.png}
% }
% \centerline{(c) \hspace{0.45 \textwidth} (d)}
% \vspace{1 cm}
% \centerline{
% \includegraphics[width = 0.45 \textwidth]{NewNewFigures/fig3_EntBand.png} ~
% \includegraphics[width = 0.45 \textwidth]{NewNewFigures/fig3_EntSpecZoomOut.png}}
% \centerline{(e) \hspace{0.45 \textwidth} (f)}

\includegraphics[width=\linewidth]{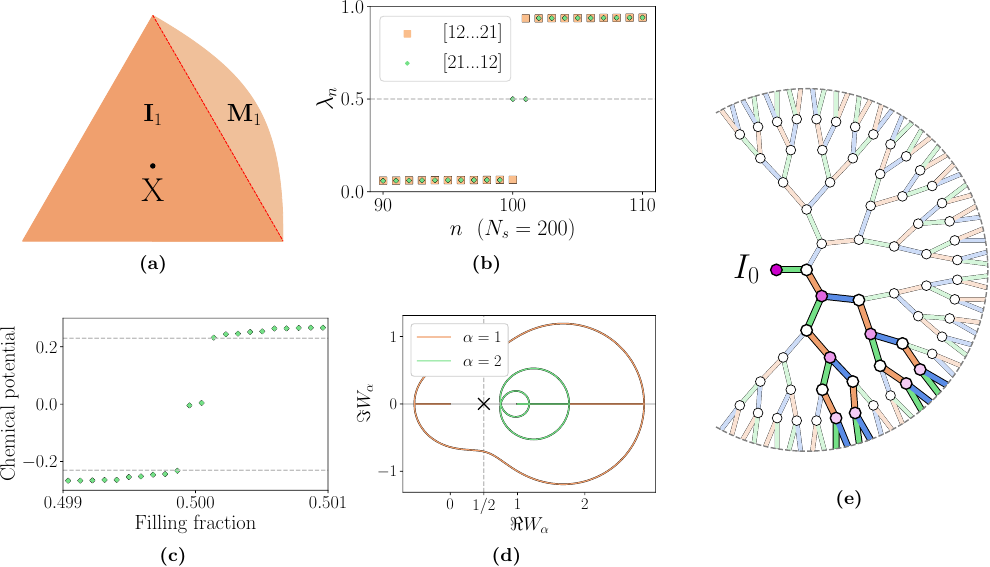}

\caption{{\bf Properties of Arboreal Obstructed Atomic Insulators:} {\bf (a)} AOAI realized at X: $(t_1 = 0.816, t_2= 0.408, t_3 =0.408)$. {\bf (b)} Entanglement spectrum of a finite $[12 \ldots 21]$ and $[21 \ldots 12]$ chains with $N_s=200$. Both show a gapped entanglement spectrum, but the latter shows mid-gap states. {\bf (c)} Filling anomaly of subsystem $S_{13}$ with boundary modes. {\bf (d)} Winding of the quantity $W_1 = G(z)/G_1(z)$ about the point $1/2$, to be contrasted with lack of similar winding for the quantity $W_2 = G(z)/G_2$. (e) Edge state wavefunction for $B_2(3)$ at X. Intensity of color of site indicates the weight of wavefunction, and is non-zero only at sites of the highlighted tree subsystem of $B_2(3)$.
}

\label{fig:gapped}

\end{figure*}

\section{Phase Diagram}
\label{sec:phase_diagram}

%We discuss the phase diagram at half filling $\mu=0$ in the $t^2_\alpha$ space, which for any is $p$ is characterized by a $p$-dimensional regular simplex (see \figref{fig:schematic}), such that the points in the simplex satisfy $\sum_\alpha t_\alpha^2 =1$. The apices of the simplex ($A_\alpha$ in \figref{fig:schematic}) represent the system where all but one of the $\alpha$ links have zero hoppings. For $p=3$, shown in \figref{fig:phase_diagram}, the $3$-simplex is an equilateral triangle. 
Fig.~\ref{fig:phase_diagram} shows the phase diagram for $p=3$. There are three insulating (gapped) regions in the parameter simplex, called \Insulator$_1$, \Insulator$_2$ and \Insulator$_3$, where the density of states vanishes around the chemical potential with a well-defined energy gap (see dos indicated in \Insulator$_3$ region in \figref{fig:phase_diagram}(a)). The phase diagram also contains a central metallic region denoted by \Metal, which possesses a finite density of states at the chemical potential. The regions are separated from each other via a set of critical surfaces \Critical$_\alpha$ which separate the insulator \Insulator$_{\alpha}$ and \Metal, such as those along the lines C$_{12}$-C$_{13}$ shown in \figref{fig:phase_diagram}. Interestingly, even though the critical points C$_{12}$ are metallic (finite density of states at the chemical potential), all other points on \Critical$_\alpha$ are semi-metallic with vanishing density states at zero energy. 
Remarkably, the picture is quite similar for larger values of $p$. There are $p$ insulating (gapped) regions in the parameter simplex called \Insulator$_\alpha$ characterized by $t^2_\alpha > 1/2$, while $t^2_\alpha = 1/2$ is the surface \Critical$_{\alpha}$ that separates insulator \Insulator$_\alpha$ from the metal \Metal. The mid-point C$_{\alpha \beta}$ $(\beta > \alpha)$ of the segment that joins the apex A$_\alpha$ to A$_\beta$ is metallic for generic $\alpha, \beta$, while all other critical points that lie on \Critical$_\alpha$ are semi-metallic. The central part of the simplex \Metal~characterized by $t_\alpha^2 < 1/2, ~\forall ~\alpha$, is metallic. 
Although each gapped phase \Insulator$_\alpha$ can be accessed from another gapped phase via the critical point C$_{\alpha \beta}$, a generic path out of the \Insulator$_\alpha$ region leads to the metallic phase \Metal~via a semi-metallic critical point through \Critical$_\alpha$. We will show later that the \Metal~region itself has a further structure as shown in \figref{fig:phase_diagram}(b).
%\GKS{Discuss the metallic regions \Metal$_\alpha$ and \Metal$_0$, and also revise the next paragraph accordingly}

These results pose intriguing questions. Are the gapped phases distinct? If yes, how? What is the nature of the critical points separating these phases, particularly the semi-metallic ones? The following sections address these questions. 

%\GKS{Should we state the kind of quantities we will be studying here itself?}

\section{Gapped Phases}
\label{sec:gapped_phases}

The gapped phases are characterized by a dos that has two energy scales. The first is the bottom of the conduction band $\varepsilon_g/2$ ($\varepsilon_g$ is the energy gap) and the other is the band top $\varepsilon_b$. Clearly, $\rho(\omega) = 0$ for $|\omega| < \varepsilon_g/2$ and $|\omega| > \varepsilon_b$. The dos at the bottom and the top of the conduction band is governed by two exponents (\SM{SM:band_properties}),
\begin{align}
    \rho(\omega) &= \Theta(|\omega| - \varepsilon_g/2) \left(|\omega| - \varepsilon_g/2 \right)^{\gamma_g}, & |\omega| \approx \varepsilon_g/2 \\
     &= \Theta(\varepsilon_b - |\omega|) (\varepsilon_b - |\omega|)^{\gamma_b}, & |\omega| \approx \varepsilon_b.
\end{align}
Along the lines A$_\alpha$-A$_\beta$ (\figref{fig:phase_diagram}(a)), we find $\gamma_g = \gamma_b = -\half$, while in the interior of the region \Insulator$_\alpha$, $\gamma_g = \gamma_b = + \half$.

%\textbf{\GKS{ENTANGLEMENT:}}
Turning to the entanglement properties, for a point in the \Insulator$_1$ region, the chain subsystem of the type $[12 \ldots 21]$ has a gapped, but otherwise featureless, entanglement spectrum as is evident from \figref{fig:gapped}(b).  On the other hand, chains subsystems of the type $[21 \ldots 12]$ show clear mid-gap states that signal a nontrivial entanglement on the links that are cut to make up this subsystem. At a similar point in the region \Insulator$_2$, the $[12 \ldots 21]$ chains and $[21 \ldots 12]$ have exactly the opposite type of entanglement spectrum.   These two entanglement structures cannot be smoothly deformed to each other, i.e., moving from a point in \Insulator$_1$ to \Insulator$_2$ necessary encounters gapless phases where the entanglement pattern is restructured. Thus, the insulating phases \Insulator$_1$ and \Insulator$_2$ are distinguished by the entanglement structure of the $[12 \ldots 21]$, $[21 \ldots 12]$ subsystems. In fact, the phase \textbf{I}$_\alpha$ is characterized by the entanglement spectrum of the strong chain subsystem $[\alpha \beta \alpha \ldots \beta \alpha]$ being gapped and featureless, while that of the subsystem $[\beta \alpha \beta \alpha \ldots \alpha \beta]$ will be gapped with a pair of mid-gap modes (\SM{SM:entanglement}).

We now show that the three gapped phases in \figref{fig:phase_diagram}(a) are three distinct obstructed atomic insulating phases (OAI)\cite{Benalcazar2019}, possessing distinct types of edge modes in truncated subsystems. To this end, we consider a semi-infinite subsystem $B_2(3)$, which consists of a single site $I_0$ connected to the rest of the semi-infinite tree solely via a link of type $2$. The on-site Green's function $\Gamma_2(z)$  (or $\Gamma_\beta(z)$, for analogously defined $B_\beta(z)$) for the $I_0$ site in $B_2(3)$ satisfies
\beq\label{eqn:Gambeta}
\Gamma^{-1}_\beta(z) = z - t^2_\beta G_\beta(z).
\eeq
We find that in the gapped phase \Insulator$_1$, $\Gamma_2(z)$ possesses a pole at $z = 0$, which indicates the presence of a ``zero energy edge state'' localized at the $I_0$ site in $B_2(3)$. A similar pole is shown by analogously defined $\Gamma_3 (\omega^+)$, while $\Gamma_{1}(\omega^+)$ possesses no such poles. By analytically solving for the associated wavefunction (\SM{SM:edge_mode}), depicted in \figref{fig:gapped}(e), we find that this edge state survives weak disorder in the nearest neighbour hoppings $t_\alpha$ (hence preserving the BDI symmetry). This corroborates the expectation of topologically protected edge modes in phases \Insulator$_\alpha$, which is further confirmed by the vanishing of the pole at the critical surface \Critical$_\alpha$ (as will be discussed in the next section).  Such boundary states also appear in truncated subsystems, which are finite and give rise to a filling anomaly as shown for $S_{13}$ subsystem in \figref{fig:gapped}(c)  in \Insulator$_1$ (\SM{SM:filling_anomaly}). The key point is that distinct subsystems show filling anomalies in different insulators \Insulator$_\alpha$ owing to their distinct obstructed atomic insulating character.

Remarkably, these results and ideas readily generalize to any $p$. Each \Insulator$_\alpha,\, \alpha=1,\ldots,p$ is an insulating phase representing a distinct atomic limit, none of which can be deformed smoothly into each other. Qualitatively, these \Insulator$_\alpha$ phases are those whose Wannier centers \cite{BB2} are located in the $\alpha$-links, leading to the $p$ distinct arboreal obstructed atomic insulators. It is, therefore, natural to explore the possibility of characterizing the topology via a quantity such as a winding number. We show (\SM{SM:winding_characteristic}) that the quantity 
\begin{equation}\label{eqn:WC}
W_\alpha(z)=\frac{G(z)}{G_\alpha(z)}
\end{equation}
 possesses characteristics promising to fill this role. Our main observation is that in each gapped phase \Insulator$_\alpha$, the quantity $W_\alpha(z)$ winds around the point $W_\alpha = \half$ in the complex $W_\alpha$ plane as $z$ traverses the real frequency axis $z = \omega + \ci \eta$ as $\omega$ goes form $-\infty$ to $\infty$. Interestingly, in the insulating phase \Insulator$_\alpha$, all other $W_\beta(z)$ with $\beta \ne \alpha$ {\em do not } wind the point $1/2$ in the complex plane.  \figref{fig:gapped}(d) shows a plot for $W_1(z)$ and $W_2(z)$ in the \Insulator$_1$ phase. A physical motivation to explore this quantity is that $W_\alpha(z)$ is a dimensionless measure of the correlation function (\eqn{eqn:GJIsol}) between two sites separated by an $\alpha$ link; indeed (\SM{SM:winding_characteristic}) 
\beq
\frac{(1 - W_\alpha)}{t_\alpha} = G_{I,I+\alpha}(z)
\eeq
which is a qualitative proxy for the Wannier localization on the link $\alpha$.

%\end{document}
We conclude the discussion of the insulating phases by further exploring the nature of the correlation functions $C_{IJ} = \langle c^\dagger_I c_J \rangle$. In a usual manifold arena, such a correlation function will be determined by the energy gap $\epsilon_g$, which provides the key length scale in the insulating phase. We point out that the situation is different and interesting in an arboreal OAI phase, where the correlations functions are determined by {\em multiple length scales} depending on the path separating the two sites in question. We illustrate this explicitly through the closed-form expression of the correlation function obtained for weak and strong chains (see \SM{SM:correlation_lengths}). Considering first the weak chains (in the \Insulator$_1$ phase, these will be $[23 \ldots 32]$ type chains, for example), the correlation function between two sites separated by $n_{\beta}$ links of the $\beta$ type  $\beta \ne \alpha$ in the \Insulator$_\alpha$ phase, we obtain for large separation
\beq\label{eqn:ins_cor_weak}
\lvert C_{\textup{weak}}(\{n_\beta\}) \rvert \sim \prod_{\beta  \neq \alpha}  \left( \frac{t_\beta/t_\alpha}{1 + \sqrt{1 - (t_\beta/t_\alpha)^2}} \right)^{n_\beta}.
\eeq
We see that the correlations fall off exponentially in weak chain determined by length scales $\left[ \ln\left((1+\sqrt{1- (t_\beta/t_\alpha)^2})/(t_\beta/t_\alpha) \right) \right]^{-1}$ (there are $p-1$ such scales). On the other hand, considering two sites separated by a strong chain having $n_\beta$ links of the type $\beta \neq \alpha$, we get
\beq\label{eqn:ins_cor_strong}
\lvert C_{\textup{strong}}(\{ n_\beta \}) \rvert \sim \prod_{\beta \ne \alpha} \left( \frac{t_\beta}{t_\alpha}\right)^{n_\beta}
\eeq
which is again exponentially decaying, but with a different set of length scales $2 \left[\ln(t_\alpha/t_\beta) \right]^{-1}$. This reveals a key feature of the arboreal AOI; the {\em presence of multiple length scales (distinct from the scale given by the energy gap)} that characterize the correlations. This is to be contrasted with what is found on manifold systems where the scale obtained from the gap determines key decay rates of all correlation functions.
%\SB{Doubt: If we consider the line along $A_1-A_2$, we encounter the SSH model critical point \(C_{12}\), where I think the correlation length diverges as \((t_{\alpha}-t_{\beta})^{-1}\). If I now consider the strong chain along the $A_1-A_2$ line, the expression for the correlation length suggests that it will diverge logarithmically. Is there anything else we need to take into consideration? }

%\GKSnote{Insert this line somewhere: intermediate chains do not admit a closed form formula, but can be analysed through similar arguments}\VBS{Not needed.}

\begin{figure*}
% \centerline{
% \includegraphics[width = 0.3 \textwidth]{NewNewFigures/fig4_pd.png} \hspace{0.2 \textwidth}
% \includegraphics[width = 0.5 \textwidth]{NewNewFigures/fig4_CorrLen.png}
% }
% \centerline{
% (a) \hspace{0.5 \textwidth} (b)
% }
% \centerline{
% \includegraphics[width = 0.2 \textwidth]{NewNewFigures/fig4_EdgeDosA.png}
% \includegraphics[width = 0.2 \textwidth]{NewNewFigures/fig4_EdgeDosB.png}
% \includegraphics[width = 0.2 \textwidth]{NewNewFigures/fig4_EdgeDosC.png}
% \includegraphics[width = 0.5 \textwidth]{NewNewFigures/fig4_EntGapDerivative.png}
% }
% \centerline{(c) \hspace{0.5 \textwidth} (d)}
% \vspace{1 cm}
% \centerline{\includegraphics[width = 0.5 \textwidth]{NewNewFigures/fig4_EntGap.png}}
% \centerline{(e)}

\includegraphics[width=\linewidth]{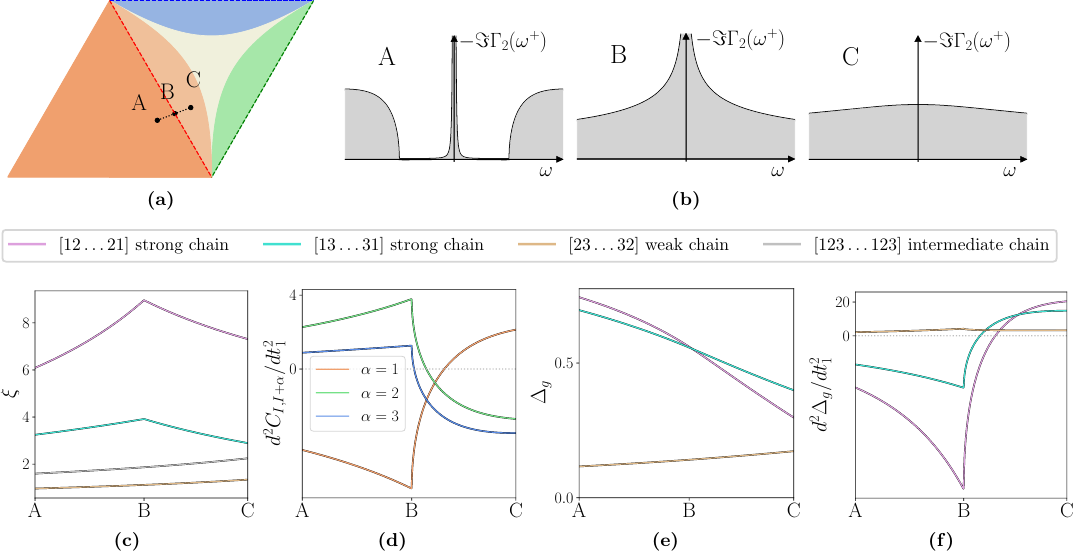}

\caption{ {\bf Critical Properties:} {\bf (a)} Critical properties studied along segment ABC, where the parameters $(t_1, t_2, t_3)$ are A: (0.743, 0.535, 0.401), B: (0.707, 0.565, 0.424) and C: (0.673, 0.592, 0.444). {\bf (b)} Evolution of pole of $\Gamma_1$ indicating loss of boundary state across \Critical$_1$.  {\bf (c)} Evolution of the correlation length along different chain subsystems from A to C. {\bf (d)} Second derivative of the nearest neighbour correlation function $\langle c^\dagger_I c_{I+\alpha} \rangle$ showing a non-analyticity. {\bf (e)} Evolution of the entanglement gap $\Delta_g$. {\bf (f)} Second derivative of the entanglement gap showing non-analyticity.
}
\label{fig:critical}

\end{figure*}

\section{Phase Transitions}
\label{sec:PT}

There are two types of critical points where the energy gap closes in the phase diagram \figref{fig:phase_diagram}. The first ones are  C$_{\alpha\beta}$ which are midpoints of A$_\alpha - $ A$_\beta$ segment such as C$_{12}$ in \figref{fig:phase_diagram}, and the second type are all other points that lie on the critical surfaces \Critical$_\alpha$. 

The first type of critical point (such as C$_{12}$) is the well-known critical point of the one-dimensional SSH model with a finite density of states at zero-energy, which has a dynamical exponent $z=1$, i.e, $\epsilon_g \sim |t_1 - t_2|$. On the other hand, the critical point that transitions from the \Insulator$_1$ gapped phase to the metal \Metal, such as a point along the surface \Critical$_1$ (excluding points C$_{12}$, C$_{13}$), corresponds to a semi-metal with a vanishing density of states at the chemical potential, and a power-law dos at low energies 
\beq
\rho(\omega) \sim (t_1^4 - t_2^4 - t_3^4)^{-1/3} |\omega|^{1/3}, ~~~ |\omega| \ll (t_1^4 - t_2^4 - t_3^4)^{1/4}.
\eeq
%\GKSaddition{Where \# is a constant numerical coefficient independent of $p$}.
Most interestingly, the exponent $1/3$ is unchanged along the critical surface $\mathcal{C}_1$, although the coefficient is determined by the specific point on the critical surface. The coefficient can be seen to diverge upon approaching the points $C_{12}$, $C_{13}$.
Further, the dynamical exponent for the critical surface is $z = 3/2$, i.e., 
\begin{equation}
    \varepsilon_g \sim \Theta \left( t_1^2 - \frac{1}{2} \right) ~ \left\lvert t_1^2 - \frac{1}{2}\right\rvert^{3/2} 
\end{equation}
and the density of states at zero energy also vanishes upon approaching $C_\alpha$ as
\begin{equation}
    \rho(0) \sim \Theta \left(\frac{1}{2} - t_1^2 \right) ~ \left\lvert t_1^2 - \frac{1}{2}\right\rvert^{1/2}.
\end{equation}
Together, these three exponents characterize the critical properties of the transition and demonstrate that they lie in a distinct universality class from the SSH critical points $C_{\alpha \beta}$. 
A notable aspect we uncover is that these findings are independent of $p$, i.e., there are two classes of critical points for any $p\ge3$, and all of them have the same critical properties! This can be understood using analytical considerations (\SM{SM:critical_properties}).

%\GKSnote{Do we need to take the approach of discussing $p = 3$ before arbitrary $p$ in all contexts? I would think that some straightforward cases, like this one, we can discuss the arbitrary $p$ case in full generality }

Studying a phase transition from \Insulator$_\alpha$ to \Metal~across \Critical$_\alpha$, we find that the pole structure of the quantity $\Gamma_\beta(z), \beta \ne \alpha$ introduced in \eqnref{eqn:Gambeta} undergoes a marked change across \Critical$_\alpha$. For example, as shown in \figref{fig:critical}(b), the spectral function $\Im \Gamma_2(\omega^+)$ has a delta function structure at $\omega = 0$ at point A in \Insulator$_1$, which evolves to a power law behaviour at the point B on \Critical$_1$ described by $ \Im \Gamma_2(\omega^+) \sim |\omega|^{-1/3} $, and further evolves to a broad feature at point C in \Metal. This is a generic feature of $\Gamma_\beta, ~ \beta \neq \alpha$ for transitions across $\mathcal{C}_\alpha$, and indicates dissolution of the protected edge mode of $B_\alpha(p)$.

Turning to the correlation function introduced in the last section, we can extract a correlation length $\xi$ associated with each type of chain subsystem. This is achieved using eqns.~\ref{eqn:ins_cor_weak} and \ref{eqn:ins_cor_strong}, respectively for weak and strong chains on the insulating side of \Critical$_\alpha$. Analogous formulae are derived on the metal side of \Critical$_\alpha$ (\SM{SM:correlation_lengths}). \figref{fig:critical}(c) shows the evolution of $\xi$ across \Critical$_\alpha$ for various chains. Several important features are to be noted. First, $\xi$ {\em does not diverge} for any chain across \Critical$_\alpha$. Second, the correlation lengths of the strong chains show a {\em non-analyticity} (more specifically, a kink) across \Critical$_\alpha$, while the correlation length associated with the weak and intermediate chains evolves smoothly oblivious of the gap closing and phase transition!

Another interesting quantity to study is the nature of the correlation function across a single link $\alpha$, for which we find 
\begin{equation}
(C_{I, I + \alpha} - C^{\textrm{smooth}}_{I, I + \alpha}) \sim \Theta{\left( \frac{1}{2} - t_\alpha^2 \right)} ~ \left\lvert t_\alpha^2 - \frac{1}{2} \right\rvert^{5/2}
\end{equation}
where the analytic dependence on parameters $C^{\textrm{smooth}}_{I, I + \alpha}$ has been subtracted to isolate the non-analyticity. Remarkably, the same critical exponent of $5/2$ is seen for $C_{I, I + \beta}$, with arbitrary $\beta$, across the $\mathcal{C}_\alpha$ phase transition, see \figref{fig:critical}(d) (also,  \SM{SM:critical_properties}).

%\GKSaddition{(Crudely written at the moment, to serve as a skeleton). Studying the two-point correlation function $C_{IJ}$ across the transition, one finds that for any choice of $I$ and $J$, $C_{IJ}$ shows a non-analyticity at points on $\mathcal{C}_\alpha$, where the exponent characterizing the singularity is always a half-integer and depends on the choice of $I$ and $J$. This is substantiated and discussed upon more in \SM{SM:critical_properties}). As a particular case, we find that for the correlation function between nearest neighbour sites, we have:}
%\begin{equation}
%C_{I, I + \beta} = C^{\textrm{smooth}}_{I, I + \beta} + (\cdots) \Theta{[D^{(2)}_\alpha]} ~ [D^{(2)}_\alpha]^{5/2}  ~~~~~~~ \forall ~ \beta  
%\end{equation} \GKSaddition{
%where $D^{(2)}_\alpha = -t^2_\alpha + \sum_{\beta \neq \alpha} t^2_\beta$ is the distance to the transition $\mathcal{C}_\alpha$. Interestingly, this non-analyticity vanishes in the limit where $I$ and $J$ are far apart, as the (suppressed) term $(\cdots)$ goes to $0$
%}

The evolution of entanglement across SSH chains also sheds light on the nature of the transition. Remarkably, the entanglement gap of all the chains evolves continuously across \Critical$_\alpha$ (see \figref{fig:critical}(e)), with a notable feature of a ``level crossing'' between strong chains. %\GKSaddition{Re`gardless of the chain we look at, we find an associated non-analyticity in the entanglement gap of the form:} 
Interestingly, the entanglement gap for each SSH chain subsystem shows a non-analyticity
\begin{equation}
    (\Delta_g - \Delta^{\textrm{smooth}}_{g}) \sim  \Theta{\left( \frac{1}{2} - t_\alpha^2 \right)} ~ \left\lvert t_\alpha^2 - \frac{1}{2} \right\rvert^{5/2} 
\end{equation}
%\GKSaddition{where $D^{(2)}_\alpha = -t^2_\alpha + \sum_{\beta \neq \alpha} t^2_\beta$ is the distance to the transition $\mathcal{C}_\alpha$. This is further substantiated in \SM{SM:critical_properties})}
which is uncovered numerically by examining the second derivative of $\Delta_g$ along the path ABC indicated in \figref{fig:critical}(a) which reveals the expected cusp for all chains at \Critical$_\alpha$ as shown in \figref{fig:critical}(e). When taken together with the fact that the same exponent was found in $C_{I, I + \beta}$, these results reveal an intriguing picture, non-analyticities in $\Delta_g$ are simply a reflection of those shown by the short-range correlation functions $C_{I, I + \alpha}$, while the true long-range behaviour is instead captured by the non-analyticites in $\xi$. %\old{Noting that $\xi$ captures the long-distance physics, while $\Delta_g$ depends on both short and long distance physics, we conclude that the short distance physics changes for all chains across the transition, while long-distance physics of only the strong chains is changed at \Critical$_\alpha$} 

Finally, we note that the winding of $W_\beta(z)$ (for all $\beta$), as well as the presence of mid-gap modes in the entanglement spectrum of chains of form $[\beta \alpha \ldots \alpha \beta]$, is unchanged across the surface \Critical$_\alpha$. This is an indication that some of the topological character of the \Insulator$_1$ phase survives into the metal.

\begin{figure*}
    % \centerline{
    % \includegraphics[width = 0.4\textwidth]{NewNewFigures/fig5_pd.png} ~~~~~~~~
    % \includegraphics[width = 0.5\textwidth]{NewNewFigures/fig5_CorrLen.png}
    % }
    % \centerline{(a) \hspace{0.5 \textwidth} (b)}
    % \centerline{
    % \includegraphics[width = 0.2\textwidth]{NewNewFigures/fig5_WindingB.png}~~
    % \includegraphics[width = 0.2\textwidth]{NewNewFigures/fig5_WindingC.png}~~~
    % \includegraphics[width = 0.2\textwidth]{NewNewFigures/fig5_WindingD.png}~~~
    % \includegraphics[width = 0.2\textwidth]{NewNewFigures/fig5_WindingE.png}}
    % \centerline{(c)}

    \includegraphics[width=\linewidth]{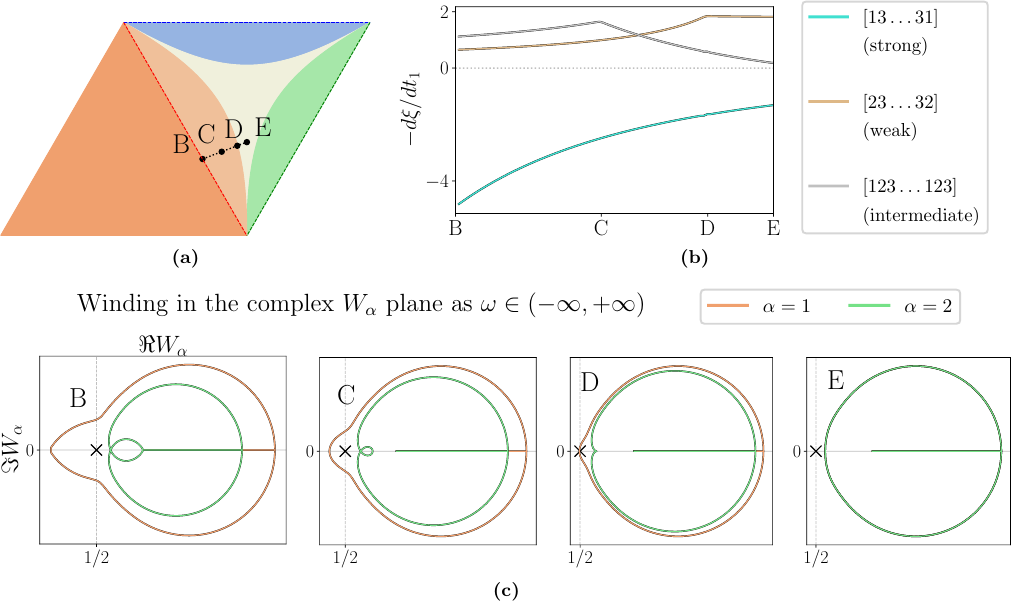}
    \caption{{\bf Properties of metal \Metal$_1$:} {\bf (a)} Properties studied along the segment BCDE, where B and C are same as in \figref{fig:critical}(a), D: (0.642, 0.614, 0.460) and E: (0.625, 0.625, 0.468). (b) Derivative of correlation length showing non-analytic behavior at different points for different chains, indicating a cascade of subsystem transitions along this path. (c) The winding characteristic of $W_1(z)$ changes across the \Metal$_1$-\Metal$_0$ boundary, while that of $W_2(z)$ is unchanged.}
    \label{fig:M1}
\end{figure*}

\section{Gapless Phases}
\label{sec:PD:Gapless}

% \DOTL{
% \begin{enumerate}
%     \item Discuss entanglement spectrum of different chains and show that there are subsystems that appear topological (see \figref{fig:ESGapless}).
%     \item Show that there are subsystems which host edge states, and demonstrate filling anomaly \figref{fig:FAGapless} and \figref{fig:FiniteEdgeState}.

%     %\item Discuss correlation functions and their properties.
% \end{enumerate}

% }

The metallic phase in region \Metal~is characterized by a non-vanishing density of states at zero energy. The gapless region \Metal~is 
 a (simply) connected region of the parameter space in that any two points in \Metal~can be connected by a path that is gapless everywhere along it. The density of states near the band-top $\varepsilon_b$ is similar to that in the gapped case, $\rho(\omega) \sim (\varepsilon_b - \omega)^{1/2}$. This observation might tempt one to concludethat all points in the region \Metal~are identical and correspond to the same phase. A more detailed study below shows that this is far from the case.

 Examining the long-range correlation functions across weak chains reveals $p + 1$ distinct functional forms that can be obtained within \Metal, and hence further divide it into regions denoted \Metal$_\alpha$ and \Metal$_0$ (see \figref{fig:phase_diagram}(b)). The \Metal$_\alpha$ phase is closely associated with the \Insulator$_\alpha$ and is characterized by correlation functions of the form
\begin{align}
	\lvert C_{\textrm{weak}}(\{ n_\beta \}) \rvert &\sim \prod_{\beta \neq \alpha}  \left( \frac{t_\beta/t_\alpha}{1 + \sqrt{1 - (t_\beta/t_\alpha)^2}} \right)^{n_\beta}, \\
	\lvert C_{\textrm{strong}}(\{ n_\beta \}) \rvert &\sim \prod_{\beta \neq \alpha} \left( \frac{t_\beta}{t_\alpha} ~ \frac{1 + \sqrt{1 - (2 \pi \rho_0 t_\alpha)^2}}{1 + \sqrt{1 - (2 \pi \rho_0 t_\beta)^2}} \right)^{n_\beta},
\end{align}
where the formula for strong chain has picked up an explicit dependence on the density of states at the chemical potential $\rho(0) = \rho_0$, while weak chains in \Metal$_\alpha$ have the same form as in \Insulator$_\alpha$. The locus of points corresponding to non-analyticity of $C_{\textrm{weak}}$ in the parameter simplex defines the boundary between the \Metal$_\alpha$ and \Metal$_0$ region.

For any type of intermediate chain, a generic path in parameter space from a point in \Insulator$_\alpha$ to a point in \Metal$_0$ must display a non-analyticity in the correlation function $C_{\textrm{intermediate}}$ at some point in \Metal$_\alpha$ (see \SM{SM:correlation_lengths}). This is a remarkable feature of the arboreal arena, in that along a generic path from \Insulator$_\alpha$ to \Metal$_0$, the system undergoes a {\em cascade of subsystem transitions} where the correlation lengths across distinct chains show non-analyticity at different points. For the path BCDE in $B(3)$ (see \figref{fig:M1}(a)), the strong chains show non-analyticity in their correlations at B (on \Critical$_\alpha$) (\figref{fig:critical}(c)). Subsequently, at point C the intermediate chain $[123 \ldots 123]$ chain suffers a non-analyticity, followed by a non-analyticity in the weak chain $[23 \ldots 32]$ at the point D (\figref{fig:M1}(b)). 
% In fact, it can be shown that for any choice of intermediate chain, the associated correlation length  shows a non-analyticity in going from B to D \SM{SM:correlation_lengths}.
%Based on numerical evidence, we surmise that for any choice of a point along BD, we can find an associated chain, whose correlation function displays a non analyticity at that point. 

\begin{figure*}
    \includegraphics[width=\linewidth]{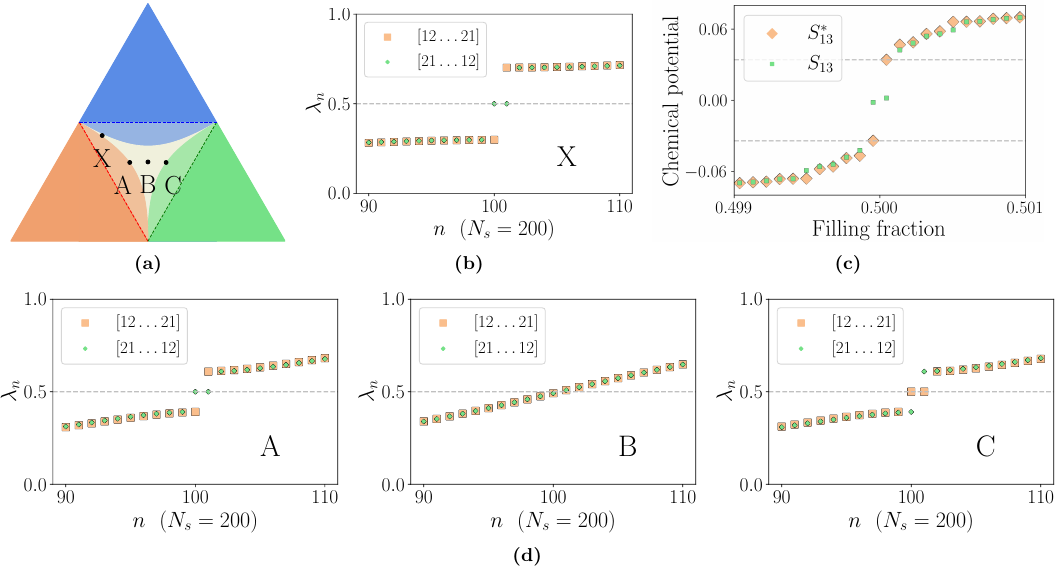}

    \caption{{\bf Properties of metal \Metal$_0$:} {\bf (a)} Points A: ($t_1=0.518$, $t_2=0.633$, $t_3=0.575$), B: (0.577, 0.577, 0.577), C: (0.633, 0.518, 0.575) and X: (0.667, 0.333, 0.667) .  {\bf (b)} Chain subsystem $[21 \ldots 12]$ shows midgap states in the entanglement spectrum at X. {\bf (c)} Subsystem $S_{13}$  shows a filling anomaly at X. $S^*_{13}$ subsystem is the $S_{13}$ subsystem with the two boundary sites $I_0$ $I_n$ removed. $S^*_{13}$ does not show filling anomaly. {\bf (d)} Evaluation of the entanglement spectrum of $[12 \ldots 21]$ and $[21 \ldots 12]$ chains from A to C. } 
    \label{fig:M0}  
\end{figure*}

Notably, the entanglement spectrum for the aforementioned chains varies {\em smoothly} across the cascade of subsystem transitions. In fact, analysis and numerics both show that the entanglement spectrum for any SSH chain is a smooth function of the parameters in the interior of \Metal~phase, oblivious to distinctions between \Metal$_\alpha$, \Metal$_0$. Closing of the entanglement gap for the chain $[12 \ldots 21]$ occurs at points with $t_1 = t_2$ in a smooth fashion, with the subsystem possessing topological characteristics for $t_2 > t_1$ and trivial otherwise ($t_1 < t_2$) (see \figref{fig:M0} (d)). 

Turning to the winding characteristic \eqn{eqn:WC}, remarkably, $W_\alpha(z)$ winds about the point $1/2$ even in the \Metal$_\alpha$ phase, while the $W_\beta(z)$ for $\beta \neq \alpha$ do not show such characteristics, similar to what we found in the \Insulator$_\alpha$ phase. A change in this winding characteristic occurs only across the \Metal$_\alpha$ - \Metal$_0$ boundary, where none of the $W_\alpha(z)$ wind about $1/2$ in \Metal$_0$  (see \figref{fig:M1}(c)). Thus, each \Metal$_\alpha$ phase is characterised by a non-trivial winding of $W_\alpha$, which vanishes at the \Metal$_\alpha$ - \Metal$_0$ boundary given by (\SM{SM:winding_characteristic})
\begin{equation}
    p - 2 = \sum_{\beta \neq \alpha} \sqrt{1 - (t_\beta / t_\alpha)^2}~.
\end{equation}
Interestingly, this surface in the parameter-simplex are also where the correlation of weak changes undergo non-analyticity.

Turning to the \Metal$_0$ phase, the central feature here is that its correlation functions do not explicitly distinguish between strong, intermediate, and weak chains. Instead of the largest hopping $t_\alpha$, the inverse of the density of states at zero energy $\rho^{-1}_0=\rho^{-1}(0)$ explicitly enters the correlation functions for {\em all} chains, and evolves smoothly as a function of parameters in the \Metal$_0$ phase. Between any two sites separated by a path consisting of $n_\alpha$ links of the $\alpha$-type (see \SM{SM:correlation_lengths}), we have
\begin{equation}
    \lvert C(\{n_\alpha\}) \rvert \sim \prod_\alpha \left(\frac{ 2 \pi \rho_0 t_\alpha }{1 + \sqrt{1 - (2 \pi \rho_0 t_\alpha)^2}} \right)^{n_\alpha}.
\end{equation}
This is clear evidence that \Metal$_\alpha$ and \Metal$_0$ are distinguished by the nature of correlations in the chain subsystems. More explicitly, in \Metal$_\alpha$ the correlations along strong, moderate and weak chains are governed by distinct scales, while in \Metal$_0$ the key scale that determines the correlations is $\rho_0$. 

Despite the absence of distinct scales, the apparently innocuous metallic region \Metal$_0$ (with no winding of $W_\alpha(z)$) hosts intriguing physics! At the point X shown in the parameter simplex (see \figref{fig:M0}(a)), we see that the $[12 \ldots 21]$ and  $[21 \ldots 12]$ chains have an entanglement gap (see \figref{fig:M0}(b)), in sharp contrast to a metal on a manifold arena (\SM{SM:manifold_entanglement}). Interestingly, the $[21 \ldots 12]$ chain also has mid-gap states in the entanglement spectrum, suggesting that the $[21 \ldots 12]$ system still hosts nontrivial topology. It is natural to ask if there are other subsystems that host ``boundary states''; indeed, we have studied truncated $S_{13}$ subsystem, which shows a clear signature (see \figref{fig:M0}(c)) of a filling anomaly (which has to be carefully defined for a metallic phase taking into account the ``Kubo gap'', i.e, the spacing of energy levels of a finite system, \SM{SM:filling_anomaly}). It must be emphasized that an arbitrary truncated subsystem will not show a filling anomaly, but the key point is that there exists a special class of truncated subsystems which host protected edge modes, and have filling anomaly {\em even in the \Metal$_0$} phase, a truly remarkable feature of the arboreal arena. The origin of these can be traced to the topological nature of the strong chains, which is still preserved in the \Metal$_0$ phase.

\section{Discussion and Perspectives}
\label{sec:Discussion}
% \DOTL{
% \begin{enumerate}
% \item Unique/Interesting/Puzzling features of arboreal arenas
% \begin{enumerate}
% \item distinct oai phases
% \item no direct transition
% \item Existnace of many metallic phases
% \item Nature of correlation functions and transitions
% \item \GKS{Arboreal universality! Properties are generic for all P}
% \end{enumerate}
% \item New opportunities for experimental work
% \begin{enumerate}
% \item Possibility of boundary states even in a gapless system. It will be interesting to realize these in variety of experimental systems (light, topotronics etc.)
% \end{enumerate}
% \end{enumerate}
% }

As is evident, the arboreal arena hosts a variety of new features that are different and distinct from those found in manifold systems, as discussed below.

\begin{enumerate}
\item An interesting finding here is number of distinct insulating phases -- $p$ distinct AOAI phases in the arboreal arena $B(p)$. While our result is obtained based on the BDI symmetry class, we believe that this is likely a generic feature. This is suggested by the fact that  the pattern of entanglement realized in these interesting insulating phases can be realized in {\em many distinct} yet ``not smoothly connected'' ways, leading to larger possibilities for the number of insulating phases.  While this issue has to be settled by further work, what is clear is that the arboreal arena offers more in the way of possibilities. Clearly, the classification of gapped phases a la \cite{KitaevPT,RyuTFW} would be an interesting direction of research. 

\item Another intriguing feature realized in arboreal arenas is that no two AOAI phases are directly connected by a quantum critical point -- they are always separated by a gapless metallic region in the phase diagram. We surmise that this is again a generic feature of the arboreal arena (distinct from what is found in lattices that possess wavelength description as a field theory on a manifold). On a manifold system, the entanglement pattern restructuring can be done at a quantum critical point.
On the other hand, in the arboreal arena, the pattern of entanglement has to ``unsewn'' and ``resewn'' subsystem by subsystem, leading to the cascade of transitions (characterized by non-analyticities of the correlations) of distinct subsystems. 
This also corroborates the presence of many distinct metallic phases which host the cascade of subsystem transitions where the correlation functions in different subsystems change in non-analytic fashion at different points in the phase diagram. These metallic regions are ``related'' to their adjacent AOAI phase, for example, the winding characteristic in \Insulator$_1$ is preserved even in \Metal$_1$. We are not aware of such phenomena found in manifold systems.

\item Yet another remarkable aspect is that the phase diagram (\figref{fig:phase_diagram}) does not have any points (other that C$_{\alpha \beta}$) where the correlation length diverges -- even on the critical surfaces \Critical$_\alpha$ the correlation length is finite. Similar results are known for other models on the Bethe lattice\cite{Nagaj2008,Laumann2009,Pollmann2013,Kimchi2014}. The correlation functions in the arboreal arena also hold other interesting features, particularly in the metallic phases. In a manifold arena, correlation functions in metal are usually power laws owing to their gaplessness. Remarkably, even in \Metal$_0$, the correlation functions decay exponentially. The physics behind this may be traced to a sum rule (\SM{SM:correlation_lengths}).

%Finally, we comment on the nature of the critical points on the arboreal arena (where the energy gap of single fermion excitation closes). An interesting finding here is that there is no diverging correlation length scale, i.e., all correlation lengths are finite (see also \cite{IsingModelOnTrees}). Further, even in the metallic phase, the creation functions in the arboreal arena decay exponentially with distance. We believe that many of these features are generic characteristics of the physics in arboreal arenae and warrant further investigation.

\item One of the most intriguing results found in this work is the property of arboreal metals to support protected edge states in a class of truncated subsystems. Another remarkable feature is the presence of subsystems in an arboreal metal that mimic the properties of topological phases on a manifold. For example, SSH chains in the metallic phase \Metal$_0$ continue to show mid-gap states in the entanglement spectrum (see X in \figref{fig:M0}(a)). We are not aware of this kind of physics being realized in a manifold setting; for example, the SSH chain subsystems in models on the square lattice (\SM{SM:manifold_entanglement}) do not possess mid-gap modes in a metallic phase.

\item An interesting qualitative observation from our calculations is that different diagnostics of topology/entanglement take on an ``independent character'' in the arboreal arena, in contrast to manifold systems. In a manifold system, quantities such as density of states, entanglement gap, and topological indices all change simultaneously at a critical point. In contact, on the arboreal arena, one can have a gap-closing transition, without the change of a winding characteristic/entanglement. This is indeed the origin of the new gapless phases such as \Metal$_\alpha$, and even the presence of midgap modes in the \Metal$_0$ phase. 

\item Finally, we wish to point out that the physics in different arboreal arenas have the ``same universaility''. In other words, the nature of the non-analyticities found on the critical surfaces \Critical$_\alpha$, does not depend on $p$ as long as $p \ge 3$ (i.e., the arena is arboreal). On the other hand $p=2$, which also does not have closed loops, has a manifold description at long wavelengths and hence belongs to a different universality class (displayed by points $C_{\alpha \beta}$). Similar results have been obtained for topological order on other arboreal arenas\cite{Nandagopal2023}. This phenomenon, therefore, seems to be a general feature of arboreal arenas. We surmise that this arises from the non-abelian translational symmetry group associated with Bethe lattices\cite{Armstrong1988}, and are investigating this further.

\end{enumerate}

Clearly, these intriguing points offer new directions for further theoretical investigations. 
As already noted, a full classification of insulating arboreal phases based on intrinsic symmetries is a well-defined research direction. More specific directions include uncovering the deeper underpinnings of the winding characteristic, in particular, addressing the question of indices/measures that will provide a unified description of the insulating phases on either the arboreal or manifold arena.

We conclude the paper by noting some of the new experimental opportunities provided by our investigation. As noted in the introduction, synthetic quantum systems offer a new era of realizing designer arenas, and in fact novel manifold arenas such as hyperbolic lattices have already been realized\cite{Kollar2019}. Systems based on topological fibre optics\cite{Souslov2022}, topoelectric circuits\cite{circuit,Weststrom2023} etc., are the likely the best platforms to experimental realization of the phenomena discovered here. An interesting direction would be the realization of protected boundary states in an arboreal metal.

Acknowledgement: VBS thanks SERB, DST for support. GKS is supported by KVPY program of DST, India. VBS thanks Prateek Mukati for the verification of some numerical results.

\bibliographystyle{apsrev4-2}
\bibliography{ref}

\begin{thebibliography}{61}%
\makeatletter
\providecommand \@ifxundefined [1]{%
 \@ifx{#1\undefined}
}%
\providecommand \@ifnum [1]{%
 \ifnum #1\expandafter \@firstoftwo
 \else \expandafter \@secondoftwo
 \fi
}%
\providecommand \@ifx [1]{%
 \ifx #1\expandafter \@firstoftwo
 \else \expandafter \@secondoftwo
 \fi
}%
\providecommand \natexlab [1]{#1}%
\providecommand \enquote  [1]{``#1''}%
\providecommand \bibnamefont  [1]{#1}%
\providecommand \bibfnamefont [1]{#1}%
\providecommand \citenamefont [1]{#1}%
\providecommand \href@noop [0]{\@secondoftwo}%
\providecommand \href [0]{\begingroup \@sanitize@url \@href}%
\providecommand \@href[1]{\@@startlink{#1}\@@href}%
\providecommand \@@href[1]{\endgroup#1\@@endlink}%
\providecommand \@sanitize@url [0]{\catcode `\\12\catcode `\$12\catcode
  `\&12\catcode `\#12\catcode `\^12\catcode `\_12\catcode `\%12\relax}%
\providecommand \@@startlink[1]{}%
\providecommand \@@endlink[0]{}%
\providecommand \url  [0]{\begingroup\@sanitize@url \@url }%
\providecommand \@url [1]{\endgroup\@href {#1}{\urlprefix }}%
\providecommand \urlprefix  [0]{URL }%
\providecommand \Eprint [0]{\href }%
\providecommand \doibase [0]{https://doi.org/}%
\providecommand \selectlanguage [0]{\@gobble}%
\providecommand \bibinfo  [0]{\@secondoftwo}%
\providecommand \bibfield  [0]{\@secondoftwo}%
\providecommand \translation [1]{[#1]}%
\providecommand \BibitemOpen [0]{}%
\providecommand \bibitemStop [0]{}%
\providecommand \bibitemNoStop [0]{.\EOS\space}%
\providecommand \EOS [0]{\spacefactor3000\relax}%
\providecommand \BibitemShut  [1]{\csname bibitem#1\endcsname}%
\let\auto@bib@innerbib\@empty
%</preamble>
\bibitem [{\citenamefont {Klitzing}\ \emph {et~al.}(1980)\citenamefont
  {Klitzing}, \citenamefont {Dorda},\ and\ \citenamefont
  {Pepper}}]{vonKlitzing1980}%
  \BibitemOpen
  \bibfield  {author} {\bibinfo {author} {\bibfnamefont {K.~v.}\ \bibnamefont
  {Klitzing}}, \bibinfo {author} {\bibfnamefont {G.}~\bibnamefont {Dorda}},\
  and\ \bibinfo {author} {\bibfnamefont {M.}~\bibnamefont {Pepper}},\ }\href
  {https://doi.org/10.1103/PhysRevLett.45.494} {\bibfield  {journal} {\bibinfo
  {journal} {Phys. Rev. Lett.}\ }\textbf {\bibinfo {volume} {45}},\ \bibinfo
  {pages} {494} (\bibinfo {year} {1980})}\BibitemShut {NoStop}%
\bibitem [{\citenamefont {Laughlin}(1981)}]{Laughlin1981}%
  \BibitemOpen
  \bibfield  {author} {\bibinfo {author} {\bibfnamefont {R.~B.}\ \bibnamefont
  {Laughlin}},\ }\href {https://doi.org/10.1103/PhysRevB.23.5632} {\bibfield
  {journal} {\bibinfo  {journal} {Phys. Rev. B}\ }\textbf {\bibinfo {volume}
  {23}},\ \bibinfo {pages} {5632} (\bibinfo {year} {1981})}\BibitemShut
  {NoStop}%
\bibitem [{\citenamefont {Thouless}\ \emph {et~al.}(1982)\citenamefont
  {Thouless}, \citenamefont {Kohmoto}, \citenamefont {Nightingale},\ and\
  \citenamefont {den Nijs}}]{TKNN1982}%
  \BibitemOpen
  \bibfield  {author} {\bibinfo {author} {\bibfnamefont {D.~J.}\ \bibnamefont
  {Thouless}}, \bibinfo {author} {\bibfnamefont {M.}~\bibnamefont {Kohmoto}},
  \bibinfo {author} {\bibfnamefont {M.~P.}\ \bibnamefont {Nightingale}},\ and\
  \bibinfo {author} {\bibfnamefont {M.}~\bibnamefont {den Nijs}},\ }\href
  {https://doi.org/10.1103/PhysRevLett.49.405} {\bibfield  {journal} {\bibinfo
  {journal} {Phys. Rev. Lett.}\ }\textbf {\bibinfo {volume} {49}},\ \bibinfo
  {pages} {405} (\bibinfo {year} {1982})}\BibitemShut {NoStop}%
\bibitem [{\citenamefont {Haldane}(1988)}]{Haldane1988}%
  \BibitemOpen
  \bibfield  {author} {\bibinfo {author} {\bibfnamefont {F.~D.~M.}\
  \bibnamefont {Haldane}},\ }\href
  {https://doi.org/10.1103/PhysRevLett.61.2015} {\bibfield  {journal} {\bibinfo
   {journal} {Phys. Rev. Lett.}\ }\textbf {\bibinfo {volume} {61}},\ \bibinfo
  {pages} {2015} (\bibinfo {year} {1988})}\BibitemShut {NoStop}%
\bibitem [{\citenamefont {Kane}\ and\ \citenamefont {Mele}(2005)}]{KaneQSHE}%
  \BibitemOpen
  \bibfield  {author} {\bibinfo {author} {\bibfnamefont {C.~L.}\ \bibnamefont
  {Kane}}\ and\ \bibinfo {author} {\bibfnamefont {E.~J.}\ \bibnamefont
  {Mele}},\ }\href {https://doi.org/10.1103/PhysRevLett.95.146802} {\bibfield
  {journal} {\bibinfo  {journal} {Phys. Rev. Lett.}\ }\textbf {\bibinfo
  {volume} {95}},\ \bibinfo {pages} {146802} (\bibinfo {year}
  {2005})}\BibitemShut {NoStop}%
\bibitem [{\citenamefont {Bernevig}\ \emph {et~al.}(2006)\citenamefont
  {Bernevig}, \citenamefont {Hughes},\ and\ \citenamefont
  {Zhang}}]{Bernevig2006}%
  \BibitemOpen
  \bibfield  {author} {\bibinfo {author} {\bibfnamefont {B.~A.}\ \bibnamefont
  {Bernevig}}, \bibinfo {author} {\bibfnamefont {T.~L.}\ \bibnamefont
  {Hughes}},\ and\ \bibinfo {author} {\bibfnamefont {S.-C.}\ \bibnamefont
  {Zhang}},\ }\href {https://doi.org/10.1126/science.1133734} {\bibfield
  {journal} {\bibinfo  {journal} {Science}\ }\textbf {\bibinfo {volume}
  {314}},\ \bibinfo {pages} {1757} (\bibinfo {year} {2006})},\ \Eprint
  {https://arxiv.org/abs/https://www.science.org/doi/pdf/10.1126/science.1133734}
  {https://www.science.org/doi/pdf/10.1126/science.1133734} \BibitemShut
  {NoStop}%
\bibitem [{\citenamefont {Fu}\ \emph {et~al.}(2007)\citenamefont {Fu},
  \citenamefont {Kane},\ and\ \citenamefont {Mele}}]{KaneTI3D}%
  \BibitemOpen
  \bibfield  {author} {\bibinfo {author} {\bibfnamefont {L.}~\bibnamefont
  {Fu}}, \bibinfo {author} {\bibfnamefont {C.~L.}\ \bibnamefont {Kane}},\ and\
  \bibinfo {author} {\bibfnamefont {E.~J.}\ \bibnamefont {Mele}},\ }\href
  {https://doi.org/10.1103/PhysRevLett.98.106803} {\bibfield  {journal}
  {\bibinfo  {journal} {Phys. Rev. Lett.}\ }\textbf {\bibinfo {volume} {98}},\
  \bibinfo {pages} {106803} (\bibinfo {year} {2007})}\BibitemShut {NoStop}%
\bibitem [{\citenamefont {Fu}\ and\ \citenamefont {Kane}(2007)}]{KaneTIInv}%
  \BibitemOpen
  \bibfield  {author} {\bibinfo {author} {\bibfnamefont {L.}~\bibnamefont
  {Fu}}\ and\ \bibinfo {author} {\bibfnamefont {C.~L.}\ \bibnamefont {Kane}},\
  }\href {https://doi.org/10.1103/PhysRevB.76.045302} {\bibfield  {journal}
  {\bibinfo  {journal} {Phys. Rev. B}\ }\textbf {\bibinfo {volume} {76}},\
  \bibinfo {pages} {045302} (\bibinfo {year} {2007})}\BibitemShut {NoStop}%
\bibitem [{\citenamefont {Moore}\ and\ \citenamefont
  {Balents}(2007)}]{MooreBalents2007}%
  \BibitemOpen
  \bibfield  {author} {\bibinfo {author} {\bibfnamefont {J.~E.}\ \bibnamefont
  {Moore}}\ and\ \bibinfo {author} {\bibfnamefont {L.}~\bibnamefont
  {Balents}},\ }\href {https://doi.org/10.1103/PhysRevB.75.121306} {\bibfield
  {journal} {\bibinfo  {journal} {Phys. Rev. B}\ }\textbf {\bibinfo {volume}
  {75}},\ \bibinfo {pages} {121306} (\bibinfo {year} {2007})}\BibitemShut
  {NoStop}%
\bibitem [{\citenamefont {Roy}(2009)}]{Roy2009}%
  \BibitemOpen
  \bibfield  {author} {\bibinfo {author} {\bibfnamefont {R.}~\bibnamefont
  {Roy}},\ }\href {https://doi.org/10.1103/PhysRevB.79.195322} {\bibfield
  {journal} {\bibinfo  {journal} {Phys. Rev. B}\ }\textbf {\bibinfo {volume}
  {79}},\ \bibinfo {pages} {195322} (\bibinfo {year} {2009})}\BibitemShut
  {NoStop}%
\bibitem [{\citenamefont {Hasan}\ and\ \citenamefont
  {Kane}(2010)}]{HasanKane2010}%
  \BibitemOpen
  \bibfield  {author} {\bibinfo {author} {\bibfnamefont {M.~Z.}\ \bibnamefont
  {Hasan}}\ and\ \bibinfo {author} {\bibfnamefont {C.~L.}\ \bibnamefont
  {Kane}},\ }\href {https://doi.org/10.1103/RevModPhys.82.3045} {\bibfield
  {journal} {\bibinfo  {journal} {Rev. Mod. Phys.}\ }\textbf {\bibinfo {volume}
  {82}},\ \bibinfo {pages} {3045} (\bibinfo {year} {2010})}\BibitemShut
  {NoStop}%
\bibitem [{\citenamefont {Qi}\ and\ \citenamefont {Zhang}(2011)}]{QiZhang2011}%
  \BibitemOpen
  \bibfield  {author} {\bibinfo {author} {\bibfnamefont {X.-L.}\ \bibnamefont
  {Qi}}\ and\ \bibinfo {author} {\bibfnamefont {S.-C.}\ \bibnamefont {Zhang}},\
  }\href {https://doi.org/10.1103/RevModPhys.83.1057} {\bibfield  {journal}
  {\bibinfo  {journal} {Rev. Mod. Phys.}\ }\textbf {\bibinfo {volume} {83}},\
  \bibinfo {pages} {1057} (\bibinfo {year} {2011})}\BibitemShut {NoStop}%
\bibitem [{\citenamefont {Kitaev}(2009)}]{KitaevPT}%
  \BibitemOpen
  \bibfield  {author} {\bibinfo {author} {\bibfnamefont {A.}~\bibnamefont
  {Kitaev}},\ }\href {https://doi.org/10.1063/1.3149495} {\bibfield  {journal}
  {\bibinfo  {journal} {AIP Conference Proceedings}\ }\textbf {\bibinfo
  {volume} {1134}},\ \bibinfo {pages} {22} (\bibinfo {year}
  {2009})}\BibitemShut {NoStop}%
\bibitem [{\citenamefont {Ryu}\ \emph {et~al.}(2010)\citenamefont {Ryu},
  \citenamefont {Schnyder}, \citenamefont {Furusaki},\ and\ \citenamefont
  {Ludwig}}]{RyuTFW}%
  \BibitemOpen
  \bibfield  {author} {\bibinfo {author} {\bibfnamefont {S.}~\bibnamefont
  {Ryu}}, \bibinfo {author} {\bibfnamefont {A.~P.}\ \bibnamefont {Schnyder}},
  \bibinfo {author} {\bibfnamefont {A.}~\bibnamefont {Furusaki}},\ and\
  \bibinfo {author} {\bibfnamefont {A.~W.~W.}\ \bibnamefont {Ludwig}},\ }\href
  {https://doi.org/10.1088/1367-2630/12/6/065010} {\bibfield  {journal}
  {\bibinfo  {journal} {New Journal of Physics}\ }\textbf {\bibinfo {volume}
  {12}},\ \bibinfo {pages} {065010} (\bibinfo {year} {2010})}\BibitemShut
  {NoStop}%
\bibitem [{\citenamefont {Chiu}\ \emph {et~al.}(2016)\citenamefont {Chiu},
  \citenamefont {Teo}, \citenamefont {Schnyder},\ and\ \citenamefont
  {Ryu}}]{ChiuRyu2016}%
  \BibitemOpen
  \bibfield  {author} {\bibinfo {author} {\bibfnamefont {C.-K.}\ \bibnamefont
  {Chiu}}, \bibinfo {author} {\bibfnamefont {J.~C.~Y.}\ \bibnamefont {Teo}},
  \bibinfo {author} {\bibfnamefont {A.~P.}\ \bibnamefont {Schnyder}},\ and\
  \bibinfo {author} {\bibfnamefont {S.}~\bibnamefont {Ryu}},\ }\href
  {https://doi.org/10.1103/RevModPhys.88.035005} {\bibfield  {journal}
  {\bibinfo  {journal} {Rev. Mod. Phys.}\ }\textbf {\bibinfo {volume} {88}},\
  \bibinfo {pages} {035005} (\bibinfo {year} {2016})}\BibitemShut {NoStop}%
\bibitem [{\citenamefont {Altland}\ and\ \citenamefont
  {Zirnbauer}(1997)}]{Altland1997}%
  \BibitemOpen
  \bibfield  {author} {\bibinfo {author} {\bibfnamefont {A.}~\bibnamefont
  {Altland}}\ and\ \bibinfo {author} {\bibfnamefont {M.~R.}\ \bibnamefont
  {Zirnbauer}},\ }\href {https://doi.org/10.1103/PhysRevB.55.1142} {\bibfield
  {journal} {\bibinfo  {journal} {Phys. Rev. B}\ }\textbf {\bibinfo {volume}
  {55}},\ \bibinfo {pages} {1142} (\bibinfo {year} {1997})}\BibitemShut
  {NoStop}%
\bibitem [{\citenamefont {{Zirnbauer}}(2010)}]{Zirnbauer2010}%
  \BibitemOpen
  \bibfield  {author} {\bibinfo {author} {\bibfnamefont {M.~R.}\ \bibnamefont
  {{Zirnbauer}}},\ }\href@noop {} {\bibfield  {journal} {\bibinfo  {journal}
  {ArXiv e-prints}\ } (\bibinfo {year} {2010})},\ \Eprint
  {https://arxiv.org/abs/1001.0722} {arXiv:1001.0722 [math-ph]} \BibitemShut
  {NoStop}%
\bibitem [{\citenamefont {Kraus}\ \emph {et~al.}(2012)\citenamefont {Kraus},
  \citenamefont {Lahini}, \citenamefont {Ringel}, \citenamefont {Verbin},\ and\
  \citenamefont {Zilberberg}}]{Kraus2012}%
  \BibitemOpen
  \bibfield  {author} {\bibinfo {author} {\bibfnamefont {Y.~E.}\ \bibnamefont
  {Kraus}}, \bibinfo {author} {\bibfnamefont {Y.}~\bibnamefont {Lahini}},
  \bibinfo {author} {\bibfnamefont {Z.}~\bibnamefont {Ringel}}, \bibinfo
  {author} {\bibfnamefont {M.}~\bibnamefont {Verbin}},\ and\ \bibinfo {author}
  {\bibfnamefont {O.}~\bibnamefont {Zilberberg}},\ }\href
  {https://doi.org/10.1103/PhysRevLett.109.106402} {\bibfield  {journal}
  {\bibinfo  {journal} {Phys. Rev. Lett.}\ }\textbf {\bibinfo {volume} {109}},\
  \bibinfo {pages} {106402} (\bibinfo {year} {2012})}\BibitemShut {NoStop}%
\bibitem [{\citenamefont {Tran}\ \emph {et~al.}(2015)\citenamefont {Tran},
  \citenamefont {Dauphin}, \citenamefont {Goldman},\ and\ \citenamefont
  {Gaspard}}]{Goldman2015}%
  \BibitemOpen
  \bibfield  {author} {\bibinfo {author} {\bibfnamefont {D.-T.}\ \bibnamefont
  {Tran}}, \bibinfo {author} {\bibfnamefont {A.}~\bibnamefont {Dauphin}},
  \bibinfo {author} {\bibfnamefont {N.}~\bibnamefont {Goldman}},\ and\ \bibinfo
  {author} {\bibfnamefont {P.}~\bibnamefont {Gaspard}},\ }\href
  {https://doi.org/10.1103/PhysRevB.91.085125} {\bibfield  {journal} {\bibinfo
  {journal} {Phys. Rev. B}\ }\textbf {\bibinfo {volume} {91}},\ \bibinfo
  {pages} {085125} (\bibinfo {year} {2015})}\BibitemShut {NoStop}%
\bibitem [{\citenamefont {Fulga}\ \emph {et~al.}(2016)\citenamefont {Fulga},
  \citenamefont {Pikulin},\ and\ \citenamefont {Loring}}]{Fulga2016}%
  \BibitemOpen
  \bibfield  {author} {\bibinfo {author} {\bibfnamefont {I.~C.}\ \bibnamefont
  {Fulga}}, \bibinfo {author} {\bibfnamefont {D.~I.}\ \bibnamefont {Pikulin}},\
  and\ \bibinfo {author} {\bibfnamefont {T.~A.}\ \bibnamefont {Loring}},\
  }\href {https://doi.org/10.1103/PhysRevLett.116.257002} {\bibfield  {journal}
  {\bibinfo  {journal} {Phys. Rev. Lett.}\ }\textbf {\bibinfo {volume} {116}},\
  \bibinfo {pages} {257002} (\bibinfo {year} {2016})}\BibitemShut {NoStop}%
\bibitem [{\citenamefont {Bandres}\ \emph {et~al.}(2016)\citenamefont
  {Bandres}, \citenamefont {Rechtsman},\ and\ \citenamefont
  {Segev}}]{Rechtsman2016}%
  \BibitemOpen
  \bibfield  {author} {\bibinfo {author} {\bibfnamefont {M.~A.}\ \bibnamefont
  {Bandres}}, \bibinfo {author} {\bibfnamefont {M.~C.}\ \bibnamefont
  {Rechtsman}},\ and\ \bibinfo {author} {\bibfnamefont {M.}~\bibnamefont
  {Segev}},\ }\href {https://doi.org/10.1103/PhysRevX.6.011016} {\bibfield
  {journal} {\bibinfo  {journal} {Phys. Rev. X}\ }\textbf {\bibinfo {volume}
  {6}},\ \bibinfo {pages} {011016} (\bibinfo {year} {2016})}\BibitemShut
  {NoStop}%
\bibitem [{\citenamefont {Agarwala}\ and\ \citenamefont
  {Shenoy}(2017)}]{Agarwala2017}%
  \BibitemOpen
  \bibfield  {author} {\bibinfo {author} {\bibfnamefont {A.}~\bibnamefont
  {Agarwala}}\ and\ \bibinfo {author} {\bibfnamefont {V.~B.}\ \bibnamefont
  {Shenoy}},\ }\href {https://doi.org/10.1103/PhysRevLett.118.236402}
  {\bibfield  {journal} {\bibinfo  {journal} {Phys. Rev. Lett.}\ }\textbf
  {\bibinfo {volume} {118}},\ \bibinfo {pages} {236402} (\bibinfo {year}
  {2017})}\BibitemShut {NoStop}%
\bibitem [{\citenamefont {Mitchell}\ \emph {et~al.}(2018)\citenamefont
  {Mitchell}, \citenamefont {Nash}, \citenamefont {Hexner}, \citenamefont
  {Turner},\ and\ \citenamefont {Irvine}}]{Mitchell2018}%
  \BibitemOpen
  \bibfield  {author} {\bibinfo {author} {\bibfnamefont {N.~P.}\ \bibnamefont
  {Mitchell}}, \bibinfo {author} {\bibfnamefont {L.~M.}\ \bibnamefont {Nash}},
  \bibinfo {author} {\bibfnamefont {D.}~\bibnamefont {Hexner}}, \bibinfo
  {author} {\bibfnamefont {A.~M.}\ \bibnamefont {Turner}},\ and\ \bibinfo
  {author} {\bibfnamefont {W.~T.~M.}\ \bibnamefont {Irvine}},\ }\href
  {https://doi.org/10.1038/s41567-017-0024-5} {\bibfield  {journal} {\bibinfo
  {journal} {Nature Physics}\ }\textbf {\bibinfo {volume} {14}},\ \bibinfo
  {pages} {380} (\bibinfo {year} {2018})}\BibitemShut {NoStop}%
\bibitem [{\citenamefont {Slager}\ \emph {et~al.}(2013)\citenamefont {Slager},
  \citenamefont {Mesaros}, \citenamefont {Juri{\v{c}}i{\'{c}}},\ and\
  \citenamefont {Zaanen}}]{Slager2013}%
  \BibitemOpen
  \bibfield  {author} {\bibinfo {author} {\bibfnamefont {R.-J.}\ \bibnamefont
  {Slager}}, \bibinfo {author} {\bibfnamefont {A.}~\bibnamefont {Mesaros}},
  \bibinfo {author} {\bibfnamefont {V.}~\bibnamefont {Juri{\v{c}}i{\'{c}}}},\
  and\ \bibinfo {author} {\bibfnamefont {J.}~\bibnamefont {Zaanen}},\ }\href
  {https://doi.org/10.1038/nphys2513} {\bibfield  {journal} {\bibinfo
  {journal} {Nature Physics}\ }\textbf {\bibinfo {volume} {9}},\ \bibinfo
  {pages} {98} (\bibinfo {year} {2013})}\BibitemShut {NoStop}%
\bibitem [{\citenamefont {Benalcazar}\ \emph
  {et~al.}(2017{\natexlab{a}})\citenamefont {Benalcazar}, \citenamefont
  {Bernevig},\ and\ \citenamefont {Hughes}}]{BB1}%
  \BibitemOpen
  \bibfield  {author} {\bibinfo {author} {\bibfnamefont {W.~A.}\ \bibnamefont
  {Benalcazar}}, \bibinfo {author} {\bibfnamefont {B.~A.}\ \bibnamefont
  {Bernevig}},\ and\ \bibinfo {author} {\bibfnamefont {T.~L.}\ \bibnamefont
  {Hughes}},\ }\href {https://doi.org/10.1126/science.aah6442} {\bibfield
  {journal} {\bibinfo  {journal} {Science}\ }\textbf {\bibinfo {volume}
  {357}},\ \bibinfo {pages} {61} (\bibinfo {year}
  {2017}{\natexlab{a}})}\BibitemShut {NoStop}%
\bibitem [{\citenamefont {Benalcazar}\ \emph
  {et~al.}(2017{\natexlab{b}})\citenamefont {Benalcazar}, \citenamefont
  {Bernevig},\ and\ \citenamefont {Hughes}}]{BB2}%
  \BibitemOpen
  \bibfield  {author} {\bibinfo {author} {\bibfnamefont {W.~A.}\ \bibnamefont
  {Benalcazar}}, \bibinfo {author} {\bibfnamefont {B.~A.}\ \bibnamefont
  {Bernevig}},\ and\ \bibinfo {author} {\bibfnamefont {T.~L.}\ \bibnamefont
  {Hughes}},\ }\href {https://doi.org/10.1103/PhysRevB.96.245115} {\bibfield
  {journal} {\bibinfo  {journal} {Phys. Rev. B}\ }\textbf {\bibinfo {volume}
  {96}},\ \bibinfo {pages} {245115} (\bibinfo {year}
  {2017}{\natexlab{b}})}\BibitemShut {NoStop}%
\bibitem [{\citenamefont {Song}\ \emph {et~al.}(2017)\citenamefont {Song},
  \citenamefont {Fang},\ and\ \citenamefont {Fang}}]{ZZC}%
  \BibitemOpen
  \bibfield  {author} {\bibinfo {author} {\bibfnamefont {Z.}~\bibnamefont
  {Song}}, \bibinfo {author} {\bibfnamefont {Z.}~\bibnamefont {Fang}},\ and\
  \bibinfo {author} {\bibfnamefont {C.}~\bibnamefont {Fang}},\ }\href
  {https://doi.org/10.1103/PhysRevLett.119.246402} {\bibfield  {journal}
  {\bibinfo  {journal} {Phys. Rev. Lett.}\ }\textbf {\bibinfo {volume} {119}},\
  \bibinfo {pages} {246402} (\bibinfo {year} {2017})}\BibitemShut {NoStop}%
\bibitem [{\citenamefont {Langbehn}\ \emph {et~al.}(2017)\citenamefont
  {Langbehn}, \citenamefont {Peng}, \citenamefont {Trifunovic}, \citenamefont
  {von Oppen},\ and\ \citenamefont {Brouwer}}]{Josias}%
  \BibitemOpen
  \bibfield  {author} {\bibinfo {author} {\bibfnamefont {J.}~\bibnamefont
  {Langbehn}}, \bibinfo {author} {\bibfnamefont {Y.}~\bibnamefont {Peng}},
  \bibinfo {author} {\bibfnamefont {L.}~\bibnamefont {Trifunovic}}, \bibinfo
  {author} {\bibfnamefont {F.}~\bibnamefont {von Oppen}},\ and\ \bibinfo
  {author} {\bibfnamefont {P.~W.}\ \bibnamefont {Brouwer}},\ }\href
  {https://doi.org/10.1103/PhysRevLett.119.246401} {\bibfield  {journal}
  {\bibinfo  {journal} {Phys. Rev. Lett.}\ }\textbf {\bibinfo {volume} {119}},\
  \bibinfo {pages} {246401} (\bibinfo {year} {2017})}\BibitemShut {NoStop}%
\bibitem [{\citenamefont {Schindler}\ \emph {et~al.}(2018)\citenamefont
  {Schindler}, \citenamefont {Cook}, \citenamefont {Vergniory}, \citenamefont
  {Wang}, \citenamefont {Parkin}, \citenamefont {Bernevig},\ and\ \citenamefont
  {Neupert}}]{Schindler}%
  \BibitemOpen
  \bibfield  {author} {\bibinfo {author} {\bibfnamefont {F.}~\bibnamefont
  {Schindler}}, \bibinfo {author} {\bibfnamefont {A.~M.}\ \bibnamefont {Cook}},
  \bibinfo {author} {\bibfnamefont {M.~G.}\ \bibnamefont {Vergniory}}, \bibinfo
  {author} {\bibfnamefont {Z.}~\bibnamefont {Wang}}, \bibinfo {author}
  {\bibfnamefont {S.~S.~P.}\ \bibnamefont {Parkin}}, \bibinfo {author}
  {\bibfnamefont {B.~A.}\ \bibnamefont {Bernevig}},\ and\ \bibinfo {author}
  {\bibfnamefont {T.}~\bibnamefont {Neupert}},\ }\href
  {https://doi.org/10.1126/sciadv.aat0346} {\bibfield  {journal} {\bibinfo
  {journal} {Science Advances}\ }\textbf {\bibinfo {volume} {4}},\ \bibinfo
  {pages} {eaat0346} (\bibinfo {year} {2018})}\BibitemShut {NoStop}%
\bibitem [{\citenamefont {Benalcazar}\ \emph {et~al.}(2019)\citenamefont
  {Benalcazar}, \citenamefont {Li},\ and\ \citenamefont
  {Hughes}}]{Benalcazar2019}%
  \BibitemOpen
  \bibfield  {author} {\bibinfo {author} {\bibfnamefont {W.~A.}\ \bibnamefont
  {Benalcazar}}, \bibinfo {author} {\bibfnamefont {T.}~\bibnamefont {Li}},\
  and\ \bibinfo {author} {\bibfnamefont {T.~L.}\ \bibnamefont {Hughes}},\
  }\href {https://doi.org/10.1103/PhysRevB.99.245151} {\bibfield  {journal}
  {\bibinfo  {journal} {Phys. Rev. B}\ }\textbf {\bibinfo {volume} {99}},\
  \bibinfo {pages} {245151} (\bibinfo {year} {2019})}\BibitemShut {NoStop}%
\bibitem [{\citenamefont {Gilbert}(2021)}]{Gilbert2021}%
  \BibitemOpen
  \bibfield  {author} {\bibinfo {author} {\bibfnamefont {M.~J.}\ \bibnamefont
  {Gilbert}},\ }\href {https://doi.org/10.1038/s42005-021-00569-5} {\bibfield
  {journal} {\bibinfo  {journal} {Communications Physics}\ }\textbf {\bibinfo
  {volume} {4}},\ \bibinfo {pages} {70} (\bibinfo {year} {2021})}\BibitemShut
  {NoStop}%
\bibitem [{\citenamefont {Bradlyn}\ \emph {et~al.}(2017)\citenamefont
  {Bradlyn}, \citenamefont {Elcoro}, \citenamefont {Cano}, \citenamefont
  {Vergniory}, \citenamefont {Wang}, \citenamefont {Felser}, \citenamefont
  {Aroyo},\ and\ \citenamefont {Bernevig}}]{Bradlyn2017}%
  \BibitemOpen
  \bibfield  {author} {\bibinfo {author} {\bibfnamefont {B.}~\bibnamefont
  {Bradlyn}}, \bibinfo {author} {\bibfnamefont {L.}~\bibnamefont {Elcoro}},
  \bibinfo {author} {\bibfnamefont {J.}~\bibnamefont {Cano}}, \bibinfo {author}
  {\bibfnamefont {M.~G.}\ \bibnamefont {Vergniory}}, \bibinfo {author}
  {\bibfnamefont {Z.}~\bibnamefont {Wang}}, \bibinfo {author} {\bibfnamefont
  {C.}~\bibnamefont {Felser}}, \bibinfo {author} {\bibfnamefont {M.~I.}\
  \bibnamefont {Aroyo}},\ and\ \bibinfo {author} {\bibfnamefont {B.~A.}\
  \bibnamefont {Bernevig}},\ }\href {https://doi.org/10.1038/nature23268}
  {\bibfield  {journal} {\bibinfo  {journal} {Nature}\ }\textbf {\bibinfo
  {volume} {547}},\ \bibinfo {pages} {298} (\bibinfo {year}
  {2017})}\BibitemShut {NoStop}%
\bibitem [{\citenamefont {Cano}\ and\ \citenamefont
  {Bradlyn}(2021)}]{Cano2021}%
  \BibitemOpen
  \bibfield  {author} {\bibinfo {author} {\bibfnamefont {J.}~\bibnamefont
  {Cano}}\ and\ \bibinfo {author} {\bibfnamefont {B.}~\bibnamefont {Bradlyn}},\
  }\href {https://doi.org/10.1146/annurev-conmatphys-041720-124134} {\bibfield
  {journal} {\bibinfo  {journal} {Annual Review of Condensed Matter Physics}\
  }\textbf {\bibinfo {volume} {12}},\ \bibinfo {pages} {225} (\bibinfo {year}
  {2021})}\BibitemShut {NoStop}%
\bibitem [{\citenamefont {Cirac}\ and\ \citenamefont
  {Zoller}(2012)}]{Cirac2012}%
  \BibitemOpen
  \bibfield  {author} {\bibinfo {author} {\bibfnamefont {J.~I.}\ \bibnamefont
  {Cirac}}\ and\ \bibinfo {author} {\bibfnamefont {P.}~\bibnamefont {Zoller}},\
  }\href {https://doi.org/10.1038/nphys2275} {\bibfield  {journal} {\bibinfo
  {journal} {Nature Physics}\ }\textbf {\bibinfo {volume} {8}},\ \bibinfo
  {pages} {264} (\bibinfo {year} {2012})}\BibitemShut {NoStop}%
\bibitem [{\citenamefont {Blais}\ \emph {et~al.}(2020)\citenamefont {Blais},
  \citenamefont {Girvin},\ and\ \citenamefont {Oliver}}]{Blais2020}%
  \BibitemOpen
  \bibfield  {author} {\bibinfo {author} {\bibfnamefont {A.}~\bibnamefont
  {Blais}}, \bibinfo {author} {\bibfnamefont {S.~M.}\ \bibnamefont {Girvin}},\
  and\ \bibinfo {author} {\bibfnamefont {W.~D.}\ \bibnamefont {Oliver}},\
  }\href {https://doi.org/10.1038/s41567-020-0806-z} {\bibfield  {journal}
  {\bibinfo  {journal} {Nature Physics}\ }\textbf {\bibinfo {volume} {16}},\
  \bibinfo {pages} {247} (\bibinfo {year} {2020})}\BibitemShut {NoStop}%
\bibitem [{\citenamefont {Ozawa}\ \emph {et~al.}(2019)\citenamefont {Ozawa},
  \citenamefont {Price}, \citenamefont {Amo}, \citenamefont {Goldman},
  \citenamefont {Hafezi}, \citenamefont {Lu}, \citenamefont {Rechtsman},
  \citenamefont {Schuster}, \citenamefont {Simon}, \citenamefont {Zilberberg},\
  and\ \citenamefont {Carusotto}}]{Ozawa2019rmp}%
  \BibitemOpen
  \bibfield  {author} {\bibinfo {author} {\bibfnamefont {T.}~\bibnamefont
  {Ozawa}}, \bibinfo {author} {\bibfnamefont {H.~M.}\ \bibnamefont {Price}},
  \bibinfo {author} {\bibfnamefont {A.}~\bibnamefont {Amo}}, \bibinfo {author}
  {\bibfnamefont {N.}~\bibnamefont {Goldman}}, \bibinfo {author} {\bibfnamefont
  {M.}~\bibnamefont {Hafezi}}, \bibinfo {author} {\bibfnamefont
  {L.}~\bibnamefont {Lu}}, \bibinfo {author} {\bibfnamefont {M.~C.}\
  \bibnamefont {Rechtsman}}, \bibinfo {author} {\bibfnamefont {D.}~\bibnamefont
  {Schuster}}, \bibinfo {author} {\bibfnamefont {J.}~\bibnamefont {Simon}},
  \bibinfo {author} {\bibfnamefont {O.}~\bibnamefont {Zilberberg}},\ and\
  \bibinfo {author} {\bibfnamefont {I.}~\bibnamefont {Carusotto}},\ }\href
  {https://doi.org/10.1103/RevModPhys.91.015006} {\bibfield  {journal}
  {\bibinfo  {journal} {Rev. Mod. Phys.}\ }\textbf {\bibinfo {volume} {91}},\
  \bibinfo {pages} {015006} (\bibinfo {year} {2019})}\BibitemShut {NoStop}%
\bibitem [{\citenamefont {Lee}\ \emph {et~al.}(2018)\citenamefont {Lee},
  \citenamefont {Imhof}, \citenamefont {Berger}, \citenamefont {Bayer},
  \citenamefont {Brehm}, \citenamefont {Molenkamp}, \citenamefont {Kiessling},\
  and\ \citenamefont {Thomale}}]{Lee2018}%
  \BibitemOpen
  \bibfield  {author} {\bibinfo {author} {\bibfnamefont {C.~H.}\ \bibnamefont
  {Lee}}, \bibinfo {author} {\bibfnamefont {S.}~\bibnamefont {Imhof}}, \bibinfo
  {author} {\bibfnamefont {C.}~\bibnamefont {Berger}}, \bibinfo {author}
  {\bibfnamefont {F.}~\bibnamefont {Bayer}}, \bibinfo {author} {\bibfnamefont
  {J.}~\bibnamefont {Brehm}}, \bibinfo {author} {\bibfnamefont {L.~W.}\
  \bibnamefont {Molenkamp}}, \bibinfo {author} {\bibfnamefont {T.}~\bibnamefont
  {Kiessling}},\ and\ \bibinfo {author} {\bibfnamefont {R.}~\bibnamefont
  {Thomale}},\ }\href {https://doi.org/10.1038/s42005-018-0035-2} {\bibfield
  {journal} {\bibinfo  {journal} {Communications Physics}\ }\textbf {\bibinfo
  {volume} {1}},\ \bibinfo {pages} {39} (\bibinfo {year} {2018})}\BibitemShut
  {NoStop}%
\bibitem [{\citenamefont {Dong}\ \emph {et~al.}(2021)\citenamefont {Dong},
  \citenamefont {Juri\ifmmode \check{c}\else \v{c}\fi{}i\ifmmode~\acute{c}\else
  \'{c}\fi{}},\ and\ \citenamefont {Roy}}]{Roy2021}%
  \BibitemOpen
  \bibfield  {author} {\bibinfo {author} {\bibfnamefont {J.}~\bibnamefont
  {Dong}}, \bibinfo {author} {\bibfnamefont {V.}~\bibnamefont {Juri\ifmmode
  \check{c}\else \v{c}\fi{}i\ifmmode~\acute{c}\else \'{c}\fi{}}},\ and\
  \bibinfo {author} {\bibfnamefont {B.}~\bibnamefont {Roy}},\ }\href
  {https://doi.org/10.1103/PhysRevResearch.3.023056} {\bibfield  {journal}
  {\bibinfo  {journal} {Phys. Rev. Res.}\ }\textbf {\bibinfo {volume} {3}},\
  \bibinfo {pages} {023056} (\bibinfo {year} {2021})}\BibitemShut {NoStop}%
\bibitem [{\citenamefont {Roberts}\ \emph {et~al.}(2022)\citenamefont
  {Roberts}, \citenamefont {Baardink}, \citenamefont {Nunn}, \citenamefont
  {Mosley},\ and\ \citenamefont {Souslov}}]{Souslov2022}%
  \BibitemOpen
  \bibfield  {author} {\bibinfo {author} {\bibfnamefont {N.}~\bibnamefont
  {Roberts}}, \bibinfo {author} {\bibfnamefont {G.}~\bibnamefont {Baardink}},
  \bibinfo {author} {\bibfnamefont {J.}~\bibnamefont {Nunn}}, \bibinfo {author}
  {\bibfnamefont {P.~J.}\ \bibnamefont {Mosley}},\ and\ \bibinfo {author}
  {\bibfnamefont {A.}~\bibnamefont {Souslov}},\ }\href
  {https://doi.org/10.1126/sciadv.add3522} {\bibfield  {journal} {\bibinfo
  {journal} {Science Advances}\ }\textbf {\bibinfo {volume} {8}},\ \bibinfo
  {pages} {eadd3522} (\bibinfo {year} {2022})},\ \Eprint
  {https://arxiv.org/abs/https://www.science.org/doi/pdf/10.1126/sciadv.add3522}
  {https://www.science.org/doi/pdf/10.1126/sciadv.add3522} \BibitemShut
  {NoStop}%
\bibitem [{\citenamefont {Manoj}\ and\ \citenamefont
  {Shenoy}(2023)}]{Nandagopal2023}%
  \BibitemOpen
  \bibfield  {author} {\bibinfo {author} {\bibfnamefont {N.}~\bibnamefont
  {Manoj}}\ and\ \bibinfo {author} {\bibfnamefont {V.~B.}\ \bibnamefont
  {Shenoy}},\ }\href {https://doi.org/10.1103/PhysRevB.107.165136} {\bibfield
  {journal} {\bibinfo  {journal} {Phys. Rev. B}\ }\textbf {\bibinfo {volume}
  {107}},\ \bibinfo {pages} {165136} (\bibinfo {year} {2023})}\BibitemShut
  {NoStop}%
\bibitem [{\citenamefont {Mahan}(2001)}]{Mahan2001}%
  \BibitemOpen
  \bibfield  {author} {\bibinfo {author} {\bibfnamefont {G.~D.}\ \bibnamefont
  {Mahan}},\ }\href {https://doi.org/10.1103/PhysRevB.63.155110} {\bibfield
  {journal} {\bibinfo  {journal} {Phys. Rev. B}\ }\textbf {\bibinfo {volume}
  {63}},\ \bibinfo {pages} {155110} (\bibinfo {year} {2001})}\BibitemShut
  {NoStop}%
\bibitem [{\citenamefont {Economou}(2006)}]{Economou2006}%
  \BibitemOpen
  \bibfield  {author} {\bibinfo {author} {\bibfnamefont {E.~N.}\ \bibnamefont
  {Economou}},\ }\href@noop {} {\emph {\bibinfo {title} {Green's Functions in
  Quantum Physics}}},\ \bibinfo {edition} {3rd}\ ed.,\ Springer Series in
  Solid-State Sciences\ (\bibinfo  {publisher} {Springer, New York},\ \bibinfo
  {year} {2006})\BibitemShut {NoStop}%
\bibitem [{\citenamefont {Aryal}\ and\ \citenamefont
  {Kettemann}(2020)}]{Aryal2020}%
  \BibitemOpen
  \bibfield  {author} {\bibinfo {author} {\bibfnamefont {D.}~\bibnamefont
  {Aryal}}\ and\ \bibinfo {author} {\bibfnamefont {S.}~\bibnamefont
  {Kettemann}},\ }\href {https://doi.org/10.1088/2399-6528/abc1c3} {\bibfield
  {journal} {\bibinfo  {journal} {Journal of Physics Communications}\ }\textbf
  {\bibinfo {volume} {4}},\ \bibinfo {pages} {105010} (\bibinfo {year}
  {2020})}\BibitemShut {NoStop}%
\bibitem [{\citenamefont {{Weststr{\"o}m}}\ \emph {et~al.}(2023)\citenamefont
  {{Weststr{\"o}m}}, \citenamefont {{Duan}}, \citenamefont {{Yao}},
  \citenamefont {{Wang}}, \citenamefont {{Liu}},\ and\ \citenamefont
  {{Li}}}]{Weststrom2023}%
  \BibitemOpen
  \bibfield  {author} {\bibinfo {author} {\bibfnamefont {A.}~\bibnamefont
  {{Weststr{\"o}m}}}, \bibinfo {author} {\bibfnamefont {W.}~\bibnamefont
  {{Duan}}}, \bibinfo {author} {\bibfnamefont {K.}~\bibnamefont {{Yao}}},
  \bibinfo {author} {\bibfnamefont {X.}~\bibnamefont {{Wang}}}, \bibinfo
  {author} {\bibfnamefont {J.}~\bibnamefont {{Liu}}},\ and\ \bibinfo {author}
  {\bibfnamefont {J.}~\bibnamefont {{Li}}},\ }\href
  {https://doi.org/10.48550/arXiv.2302.03166} {\bibfield  {journal} {\bibinfo
  {journal} {arXiv e-prints}\ ,\ \bibinfo {eid} {arXiv:2302.03166}} (\bibinfo
  {year} {2023})},\ \Eprint {https://arxiv.org/abs/2302.03166}
  {arXiv:2302.03166 [cond-mat.mes-hall]} \BibitemShut {NoStop}%
\bibitem [{\citenamefont {Tikhonov}\ and\ \citenamefont
  {Mirlin}(2016)}]{Tikhonov2016}%
  \BibitemOpen
  \bibfield  {author} {\bibinfo {author} {\bibfnamefont {K.~S.}\ \bibnamefont
  {Tikhonov}}\ and\ \bibinfo {author} {\bibfnamefont {A.~D.}\ \bibnamefont
  {Mirlin}},\ }\href {https://doi.org/10.1103/PhysRevB.94.184203} {\bibfield
  {journal} {\bibinfo  {journal} {Phys. Rev. B}\ }\textbf {\bibinfo {volume}
  {94}},\ \bibinfo {pages} {184203} (\bibinfo {year} {2016})}\BibitemShut
  {NoStop}%
\bibitem [{SM()}]{SM}%
  \BibitemOpen
  \href@noop {} {}\bibinfo {note} {Supplemental Material.}\BibitemShut {Stop}%
\bibitem [{\citenamefont {Su}\ \emph {et~al.}(1979)\citenamefont {Su},
  \citenamefont {Schrieffer},\ and\ \citenamefont {Heeger}}]{SSH1979}%
  \BibitemOpen
  \bibfield  {author} {\bibinfo {author} {\bibfnamefont {W.~P.}\ \bibnamefont
  {Su}}, \bibinfo {author} {\bibfnamefont {J.~R.}\ \bibnamefont {Schrieffer}},\
  and\ \bibinfo {author} {\bibfnamefont {A.~J.}\ \bibnamefont {Heeger}},\
  }\href {https://doi.org/10.1103/PhysRevLett.42.1698} {\bibfield  {journal}
  {\bibinfo  {journal} {Phys. Rev. Lett.}\ }\textbf {\bibinfo {volume} {42}},\
  \bibinfo {pages} {1698} (\bibinfo {year} {1979})}\BibitemShut {NoStop}%
\bibitem [{\citenamefont {Su}\ \emph {et~al.}(1980)\citenamefont {Su},
  \citenamefont {Schrieffer},\ and\ \citenamefont {Heeger}}]{SSH1980}%
  \BibitemOpen
  \bibfield  {author} {\bibinfo {author} {\bibfnamefont {W.~P.}\ \bibnamefont
  {Su}}, \bibinfo {author} {\bibfnamefont {J.~R.}\ \bibnamefont {Schrieffer}},\
  and\ \bibinfo {author} {\bibfnamefont {A.~J.}\ \bibnamefont {Heeger}},\
  }\href {https://doi.org/10.1103/PhysRevB.22.2099} {\bibfield  {journal}
  {\bibinfo  {journal} {Phys. Rev. B}\ }\textbf {\bibinfo {volume} {22}},\
  \bibinfo {pages} {2099} (\bibinfo {year} {1980})}\BibitemShut {NoStop}%
\bibitem [{\citenamefont {Peschel}(2003)}]{Peschel2003}%
  \BibitemOpen
  \bibfield  {author} {\bibinfo {author} {\bibfnamefont {I.}~\bibnamefont
  {Peschel}},\ }\href {https://doi.org/10.1088/0305-4470/36/14/101} {\bibfield
  {journal} {\bibinfo  {journal} {Journal of Physics A: Mathematical and
  General}\ }\textbf {\bibinfo {volume} {36}},\ \bibinfo {pages} {L205}
  (\bibinfo {year} {2003})}\BibitemShut {NoStop}%
\bibitem [{\citenamefont {Cheong}\ and\ \citenamefont
  {Henley}(2004)}]{Henley2004}%
  \BibitemOpen
  \bibfield  {author} {\bibinfo {author} {\bibfnamefont {S.-A.}\ \bibnamefont
  {Cheong}}\ and\ \bibinfo {author} {\bibfnamefont {C.~L.}\ \bibnamefont
  {Henley}},\ }\href {https://doi.org/10.1103/PhysRevB.69.075111} {\bibfield
  {journal} {\bibinfo  {journal} {Phys. Rev. B}\ }\textbf {\bibinfo {volume}
  {69}},\ \bibinfo {pages} {075111} (\bibinfo {year} {2004})}\BibitemShut
  {NoStop}%
\bibitem [{\citenamefont {Turner}\ \emph {et~al.}(2010)\citenamefont {Turner},
  \citenamefont {Zhang},\ and\ \citenamefont {Vishwanath}}]{Turner2010}%
  \BibitemOpen
  \bibfield  {author} {\bibinfo {author} {\bibfnamefont {A.~M.}\ \bibnamefont
  {Turner}}, \bibinfo {author} {\bibfnamefont {Y.}~\bibnamefont {Zhang}},\ and\
  \bibinfo {author} {\bibfnamefont {A.}~\bibnamefont {Vishwanath}},\ }\href
  {https://doi.org/10.1103/PhysRevB.82.241102} {\bibfield  {journal} {\bibinfo
  {journal} {Phys. Rev. B}\ }\textbf {\bibinfo {volume} {82}},\ \bibinfo
  {pages} {241102} (\bibinfo {year} {2010})}\BibitemShut {NoStop}%
\bibitem [{\citenamefont {Alexandradinata}\ \emph {et~al.}(2011)\citenamefont
  {Alexandradinata}, \citenamefont {Hughes},\ and\ \citenamefont
  {Bernevig}}]{Alexandradinata2011}%
  \BibitemOpen
  \bibfield  {author} {\bibinfo {author} {\bibfnamefont {A.}~\bibnamefont
  {Alexandradinata}}, \bibinfo {author} {\bibfnamefont {T.~L.}\ \bibnamefont
  {Hughes}},\ and\ \bibinfo {author} {\bibfnamefont {B.~A.}\ \bibnamefont
  {Bernevig}},\ }\href {https://doi.org/10.1103/PhysRevB.84.195103} {\bibfield
  {journal} {\bibinfo  {journal} {Phys. Rev. B}\ }\textbf {\bibinfo {volume}
  {84}},\ \bibinfo {pages} {195103} (\bibinfo {year} {2011})}\BibitemShut
  {NoStop}%
\bibitem [{\citenamefont {Hsieh}\ and\ \citenamefont {Fu}(2014)}]{HsiehFu2014}%
  \BibitemOpen
  \bibfield  {author} {\bibinfo {author} {\bibfnamefont {T.~H.}\ \bibnamefont
  {Hsieh}}\ and\ \bibinfo {author} {\bibfnamefont {L.}~\bibnamefont {Fu}},\
  }\href {https://doi.org/10.1103/PhysRevLett.113.106801} {\bibfield  {journal}
  {\bibinfo  {journal} {Phys. Rev. Lett.}\ }\textbf {\bibinfo {volume} {113}},\
  \bibinfo {pages} {106801} (\bibinfo {year} {2014})}\BibitemShut {NoStop}%
\bibitem [{\citenamefont {Ivan~Gutman}(2011)}]{IvanGutman2011}%
  \BibitemOpen
  \bibfield  {author} {\bibinfo {author} {\bibfnamefont {B.~B.}\ \bibnamefont
  {Ivan~Gutman}},\ }\href {http://eudml.org/doc/256788} {\bibfield  {journal}
  {\bibinfo  {journal} {Zbornik Radova}\ }\textbf {\bibinfo {volume} {22}},\
  \bibinfo {pages} {137} (\bibinfo {year} {2011})}\BibitemShut {NoStop}%
\bibitem [{\citenamefont {Nagaj}\ \emph {et~al.}(2008)\citenamefont {Nagaj},
  \citenamefont {Farhi}, \citenamefont {Goldstone}, \citenamefont {Shor},\ and\
  \citenamefont {Sylvester}}]{Nagaj2008}%
  \BibitemOpen
  \bibfield  {author} {\bibinfo {author} {\bibfnamefont {D.}~\bibnamefont
  {Nagaj}}, \bibinfo {author} {\bibfnamefont {E.}~\bibnamefont {Farhi}},
  \bibinfo {author} {\bibfnamefont {J.}~\bibnamefont {Goldstone}}, \bibinfo
  {author} {\bibfnamefont {P.}~\bibnamefont {Shor}},\ and\ \bibinfo {author}
  {\bibfnamefont {I.}~\bibnamefont {Sylvester}},\ }\href
  {https://doi.org/10.1103/PhysRevB.77.214431} {\bibfield  {journal} {\bibinfo
  {journal} {Phys. Rev. B}\ }\textbf {\bibinfo {volume} {77}},\ \bibinfo
  {pages} {214431} (\bibinfo {year} {2008})}\BibitemShut {NoStop}%
\bibitem [{\citenamefont {Laumann}\ \emph {et~al.}(2009)\citenamefont
  {Laumann}, \citenamefont {Parameswaran},\ and\ \citenamefont
  {Sondhi}}]{Laumann2009}%
  \BibitemOpen
  \bibfield  {author} {\bibinfo {author} {\bibfnamefont {C.~R.}\ \bibnamefont
  {Laumann}}, \bibinfo {author} {\bibfnamefont {S.~A.}\ \bibnamefont
  {Parameswaran}},\ and\ \bibinfo {author} {\bibfnamefont {S.~L.}\ \bibnamefont
  {Sondhi}},\ }\href {https://doi.org/10.1103/PhysRevB.80.144415} {\bibfield
  {journal} {\bibinfo  {journal} {Phys. Rev. B}\ }\textbf {\bibinfo {volume}
  {80}},\ \bibinfo {pages} {144415} (\bibinfo {year} {2009})}\BibitemShut
  {NoStop}%
\bibitem [{\citenamefont {Depenbrock}\ and\ \citenamefont
  {Pollmann}(2013)}]{Pollmann2013}%
  \BibitemOpen
  \bibfield  {author} {\bibinfo {author} {\bibfnamefont {S.}~\bibnamefont
  {Depenbrock}}\ and\ \bibinfo {author} {\bibfnamefont {F.}~\bibnamefont
  {Pollmann}},\ }\href {https://doi.org/10.1103/PhysRevB.88.035138} {\bibfield
  {journal} {\bibinfo  {journal} {Phys. Rev. B}\ }\textbf {\bibinfo {volume}
  {88}},\ \bibinfo {pages} {035138} (\bibinfo {year} {2013})}\BibitemShut
  {NoStop}%
\bibitem [{\citenamefont {Kimchi}\ \emph {et~al.}(2014)\citenamefont {Kimchi},
  \citenamefont {Analytis},\ and\ \citenamefont {Vishwanath}}]{Kimchi2014}%
  \BibitemOpen
  \bibfield  {author} {\bibinfo {author} {\bibfnamefont {I.}~\bibnamefont
  {Kimchi}}, \bibinfo {author} {\bibfnamefont {J.~G.}\ \bibnamefont
  {Analytis}},\ and\ \bibinfo {author} {\bibfnamefont {A.}~\bibnamefont
  {Vishwanath}},\ }\href {https://doi.org/10.1103/PhysRevB.90.205126}
  {\bibfield  {journal} {\bibinfo  {journal} {Phys. Rev. B}\ }\textbf {\bibinfo
  {volume} {90}},\ \bibinfo {pages} {205126} (\bibinfo {year}
  {2014})}\BibitemShut {NoStop}%
\bibitem [{\citenamefont {Armstrong}(1988)}]{Armstrong1988}%
  \BibitemOpen
  \bibfield  {author} {\bibinfo {author} {\bibfnamefont {M.~A.}\ \bibnamefont
  {Armstrong}},\ }\href@noop {} {\emph {\bibinfo {title} {Groups and
  symmetry}}},\ Undergraduate Texts in Mathematics\ (\bibinfo  {publisher}
  {Springer},\ \bibinfo {year} {1988})\BibitemShut {NoStop}%
\bibitem [{\citenamefont {Koll{\'a}r}\ \emph {et~al.}(2019)\citenamefont
  {Koll{\'a}r}, \citenamefont {Fitzpatrick},\ and\ \citenamefont
  {Houck}}]{Kollar2019}%
  \BibitemOpen
  \bibfield  {author} {\bibinfo {author} {\bibfnamefont {A.~J.}\ \bibnamefont
  {Koll{\'a}r}}, \bibinfo {author} {\bibfnamefont {M.}~\bibnamefont
  {Fitzpatrick}},\ and\ \bibinfo {author} {\bibfnamefont {A.~A.}\ \bibnamefont
  {Houck}},\ }\href {https://doi.org/10.1038/s41586-019-1348-3} {\bibfield
  {journal} {\bibinfo  {journal} {Nature}\ }\textbf {\bibinfo {volume} {571}},\
  \bibinfo {pages} {45} (\bibinfo {year} {2019})}\BibitemShut {NoStop}%
\bibitem [{\citenamefont {Imhof}\ \emph {et~al.}(2018)\citenamefont {Imhof},
  \citenamefont {Berger}, \citenamefont {Bayer}, \citenamefont {Brehm},
  \citenamefont {Molenkamp}, \citenamefont {Kiessling}, \citenamefont
  {Schindler}, \citenamefont {Lee}, \citenamefont {Greiter}, \citenamefont
  {Neupert},\ and\ \citenamefont {Thomale}}]{circuit}%
  \BibitemOpen
  \bibfield  {author} {\bibinfo {author} {\bibfnamefont {S.}~\bibnamefont
  {Imhof}}, \bibinfo {author} {\bibfnamefont {C.}~\bibnamefont {Berger}},
  \bibinfo {author} {\bibfnamefont {F.}~\bibnamefont {Bayer}}, \bibinfo
  {author} {\bibfnamefont {J.}~\bibnamefont {Brehm}}, \bibinfo {author}
  {\bibfnamefont {L.~W.}\ \bibnamefont {Molenkamp}}, \bibinfo {author}
  {\bibfnamefont {T.}~\bibnamefont {Kiessling}}, \bibinfo {author}
  {\bibfnamefont {F.}~\bibnamefont {Schindler}}, \bibinfo {author}
  {\bibfnamefont {C.~H.}\ \bibnamefont {Lee}}, \bibinfo {author} {\bibfnamefont
  {M.}~\bibnamefont {Greiter}}, \bibinfo {author} {\bibfnamefont
  {T.}~\bibnamefont {Neupert}},\ and\ \bibinfo {author} {\bibfnamefont
  {R.}~\bibnamefont {Thomale}},\ }\href
  {https://doi.org/10.1038/s41567-018-0246-1} {\bibfield  {journal} {\bibinfo
  {journal} {Nature Physics}\ }\textbf {\bibinfo {volume} {14}},\ \bibinfo
  {pages} {925} (\bibinfo {year} {2018})}\BibitemShut {NoStop}%
\end{thebibliography}%
%\end{document}

\def\makeSM{1}
\ifdefined\makeSM

\clearpage
\newpage
\appendix

\renewcommand{\appendixname}{}
\renewcommand{\thesection}{{S\arabic{section}}}
\renewcommand{\thefigure}{S\arabic{figure}}
\renewcommand{\theequation}{\thesection.\arabic{equation}}
 
\setcounter{page}{1}
\setcounter{figure}{0}

\begin{widetext}

\maketitle

\centerline{\bf Supplemental Material}
\centerline{\bf for}
\centerline{\bf \mytitle}
\medskip
\centerline{by Gurkirat Singh, Surajit Bera, Vijay B. Shenoy}
\bigskip
\end{widetext}

\newcommand{\GreenFn}{Green's function}
\newcommand{\citeeq}[1]{(Eqn. \eqref{#1})}

\section{Choice of hopping amplitudes}
\label{SM:GaugeChoice}

Starting with a more general hamiltonian (compared to \ref{eqn:BetheHopping}) with complex hoppings $\tilde{t}_\alpha$
\beq\label{eqn:BetheHoppingGeneral}
{\cal H} = -\sum_{I,\alpha} 
\tilde{t}_{\alpha} c^\dagger_{I+\alpha} c_{I} - \mu \sum_I c^\dagger_I c_I
\eeq
where
\beq
\tilde{t}_\alpha = t_\alpha e^{\ci A_\alpha}
\eeq
and $t_\alpha$ are non-negative real quantities.
Here the phase of $\tilde{t}_\alpha$ depends on $I$, and is fixed by a suitably chosen convention. For example, suppose $I+\alpha=J$, then $\tilde{t}_\alpha c^\dagger_{I+\alpha} c_I$, will have a hermitian conjugate counterpart  $\tilde{t}_\alpha c^\dagger_{J+\alpha} c_J$ (note, $J+\alpha = I$), but in the latter term $\tilde{t}_\alpha$ is taken as the complex conjugate of that in the former term.
We now show that \eqnref{eqn:BetheHoppingGeneral} is equivalent to \eqnref{eqn:BetheHopping}. To this end, we choose a reference site $I_0$, which is the root site of $B(p)$. Any other site $J$ in generation $g(J)$ can be reached by a path specified by the sequence of links  $[\alpha_1,\ldots,\alpha_{l(J)}]$. The hoppings along these links are  $\tilde{t}_\alpha = t_\alpha e^{\ci A_\alpha}$ ($\ci=\sqrt{-1}, A_\alpha \in \Reals$). Now, we define
\beq
\tilde{c}^\dagger_J = e^{\ci \sum_{k=1}^{g(J)} A_{\alpha_k}} c^\dagger_J, \;\;\; \tilde{c}^\dagger_{I_0} = c^\dagger_{I_0}
\eeq
in terms of which the hamiltonian becomes
\beq
H = - \sum_{I \alpha} t_\alpha \tilde{c}^\dagger_{I+\alpha} \tilde{c}_{I} - \mu \sum_{I} \tilde{c}^\dagger_I \tilde{c}_I
\eeq
We thus see that the phases of $\tilde{t}_\alpha$ can be gauged away, and the choice $t_\alpha \ge 0$ in \eqnref{eqn:BetheHopping} is thus justified, after redefining $\tilde{c}_\alpha$ as $c_\alpha$.

\section{ \GreenFn, analytic structure}
\label{SM:GreenFn_analytic}
\subsection{Formulation}

For the on-site Green's functions $G(z)$ and $G_\alpha(z)$ defined in the main text (Eq. \eqref{eqn:GII}), the respective Dyson equations read
\begin{align}
	G^{-1}(z) &= z - \sum_{\beta} t_\beta^2 G_\beta(z) \label{eqn:SM_DysonG} \\
	G_\alpha^{-1}(z) &= z - \sum_{\beta \neq \alpha} t_\beta^2 G_\beta(z). \label{eqn:SM_DysonGalpha}
\end{align}
Eliminating $z$ from the above, one obtains
\begin{equation}
	G_\alpha^{-1} - G^{-1} = t_\alpha^2 G_\alpha, \label{eqn:GalphaG}
\end{equation}
yielding a solution for $G_\alpha(z)$ in terms of $G(z)$
\begin{equation}
	G_\alpha(z) = \frac{ -1 \pm \sqrt{1 + 4 t_\alpha^2 G(z)^2} }{2 t_\alpha^2 G(z)}.
\end{equation}

For the remainder of the supplementary material, we adopt the convention  $\sqrt{r e^{\ci \theta}} = \sqrt{r} e^{\ci \theta /2} ~~ \forall \theta \in (-\pi, \pi]$. The multivaluedness of the complex square-root function is instead absorbed into an auxiliary function $\sigma_\alpha(z) \in \{+1, -1\} ~~  \forall z$, which keeps track of which Riemann sheet the point $z$ corresponds to. In terms of $\sigma_\alpha$ the solution can then be written
\begin{equation}
	G_\alpha = \frac{ -1 + \sigma_\alpha \sqrt{1 + 4 t_\alpha^2 G^2} }{ 2 t_\alpha^2 G}, \label{eqn:SM_Galpha}
\end{equation}
where the dependence on $z$ has been suppressed. From \eqref{eqn:SM_DysonG} and \eqref{eqn:SM_Galpha}, one obtains an equation for $z$ in terms of $G(z)$
\begin{equation}
	z = \frac{2 - p}{2G} +  \sum_{\alpha} \sigma_\alpha \frac{\sqrt{1 + 4 t_\alpha^2 G^2}}{2G}. \label{eqn:G}
\end{equation}
However to utilize the same, one needs knowledge of $\sigma_\alpha(z)$, which we develop.

\begin{figure*}
\includegraphics[width = \textwidth]{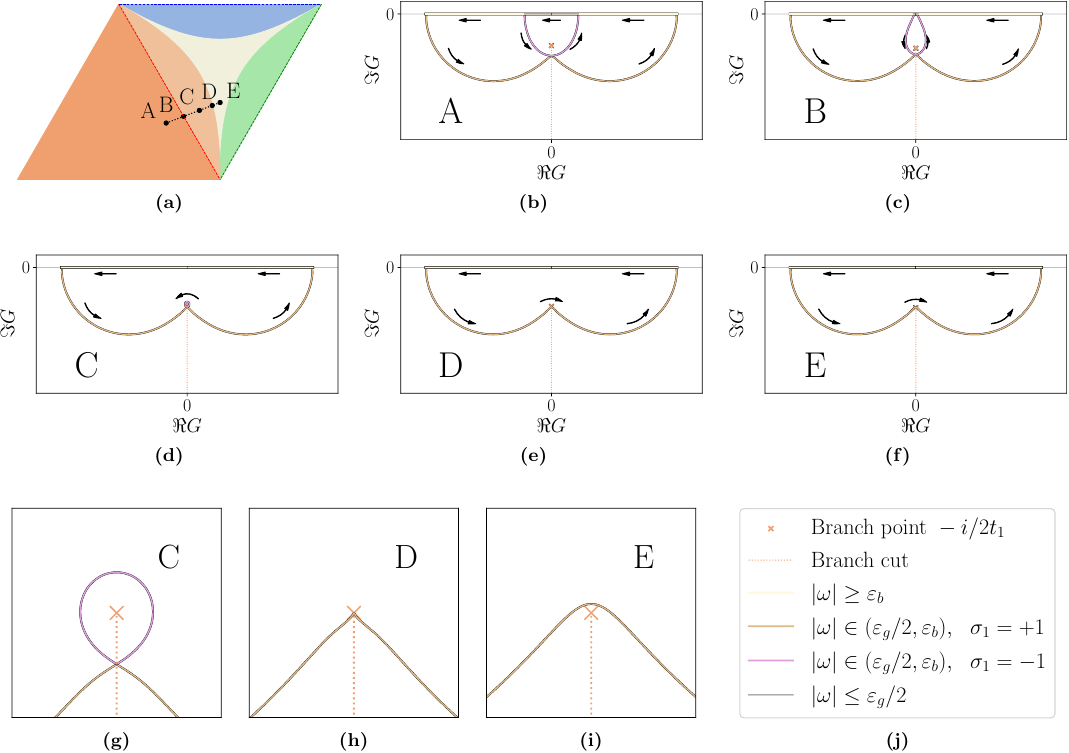}

    \caption{\textbf{(a)} Points A: (0.743, 0.535, 0.401), B: (0.707, 0.565, 0.424), C: (0.673, 0.592, 0.444), D: (0.642, 0.614, 0.460) and E: (0.625, 0.625, 0.468) in phase diagram. \textbf{(b-f)} The complex contours traced out by $G(\omega^+)$ for $\omega \in (-\infty, \infty)$ in the complex $G$ plane. Shown separately for points ABCDE. \textbf{(g-i)} Behaviour of $G(\omega^+)$ near branch point at $-\ci/2t_1$. Zoomed in versions of (d), (e) and (f) respectively. \textbf{(j)} Collective legend for figures (b) to (i), indicating different segments of the contour as well as the branch structure.}

    \label{fig:G_contours}

\end{figure*}

\subsection{Determining $\sigma_\alpha(z)$}
Discontinuities in $\sigma_\alpha(z)$ are associated with the chosen branch cut of the function $\sqrt{1 + 4 t_\alpha^2 G^2}$, which is along the imaginary axis and has branch points at $G = \pm \ci /2 t_\alpha$. Thus, along the curve $G(\omega^+)$ in the complex $G$ plane, $\sigma_\alpha$ changes sign only when the curve crosses the branch cut. First, the behaviour of $\sigma_\alpha(z)$ for large $z$ is discussed, and subsequently the conditions for the function $\sigma_\alpha(\omega^+)$ to change sign is discussed. \\

For any system with a bounded spectrum, in the limit of $z \rightarrow \infty$, the respective resolvent Green's function approaches the form $\mathbf{G}(z) = (z \mathbf{1} - \mathbf{H})^{-1} \sim \mathbf{1} z^{-1}$. Accordingly, we have $G_\alpha(z) \sim G(z) \sim 1/z$ as $z \rightarrow \infty$. This boundary condition along with \eqref{eqn:G} leads to the conclusion that $\sigma(z) = +1$ for $|z| \gg \epsilon_b$.\\

For the sign $\sigma_\alpha$ to change from $+1$, the curve $G(\omega^+)$ must intercept on the imaginary $G$ axis, and pass through the branch cut discussed earlier. To calculate this intercept, we substitute $G = - \ci g$ for $g \in \mathbb{R}$ and obtain the equation
\begin{equation}
	\omega^+ = +\ci \frac{2 - p}{2g} + \ci \sum_{\alpha} \frac{\sqrt{1 - 4 t_\alpha^2 g^2}}{2 g}. \label{eqn:SM_intercept}
\end{equation} 
Without loss of generality, assume $t_1 \geq t_2 \geq  \cdots \geq t_p$. One can then assert that $g < \frac{1}{2 t_2}$, as assuming the negation leads to a contradiction, which can be seen by examining the imaginary part of \eqnref{eqn:SM_intercept}
\begin{equation}
	(p-2) + 2g \eta = \sum_{\alpha} \Re{\sqrt{1 - 4 t_\alpha^2 g^2}} < (p-2).
\end{equation}
This indicates that the curve $G(\omega^+)$ can only pass through the branch cut corresponding to a change in sign of $\sigma_1$. A similar analysis can be performed to establish that starting from a point with $\sigma_1 = -1 = -\sigma_{\beta \neq 1}$, the curve $G(\omega^+)$ can only pass through the branch cut corresponding to a change in sign of $\sigma_1$. Thus, in the regime when $t_1 \geq t_\beta ~~ \forall \beta$, we have $\sigma_\beta(\omega^+) = +1 ~~ \forall \omega, ~ \forall \beta \neq 1$. \\

In general, depending on which of the $p$ hoppings is the largest, we can have either of:
\begin{enumerate}
    \item $\sigma_\alpha(\omega^+) = +1 ~~ \forall \alpha ~~ \forall \omega \in (-\infty, \infty)$
    \item $\sigma_\alpha(\omega^+) = -1$ for some $\omega \in \mathbb{R}$, \\ $\sigma_\beta(\omega^+) = +1 ~~ \forall \beta \neq \alpha ~~ \forall \omega \in (-\infty, \infty)$ \\ where $t_\alpha$ refers to the largest hopping
\end{enumerate}
The first case is a characteristic of \Metal$_0$ phase, while the second case characterizes \Insulator$_\alpha$, \Critical$_\alpha$ and also \Metal$_\alpha$.

\section{Numerical solution through Newton-Raphson method}
\label{SM:NewtonRaphson}
The set of $p$ independent equations in \citeeq{eqn:DysonGalpha} is iteratively solved for the $p$-element column vector $\{G^{-1}_\alpha(\omega^+)\}_{\alpha = 1}^{p}$ denoted as $\overrightarrow{X}$. The problem can be stated in the form $\overrightarrow{X} = F(\overrightarrow{X})$, and subsequently solved using Newton-Raphson methods
\begin{equation}
	\left( \mathbf{1} - \mathbf{DF}(\overrightarrow{X}_{n}) \right) \left(\overrightarrow{X}_{n+1} - \overrightarrow{X}_{n} \right) = \alpha \left(\overrightarrow{X}_{n} - F(\overrightarrow{X}_{n})\right),  
\end{equation}
where $\mathbf{1}$ is the $p \times p$ identity matrix and $\mathbf{DF}$ is the total derivative of the vector valued function $F$. We numerically investigate a wide range of points for the models $p = 3,4,5$ and find the parameters $\eta = 10^{-10}$, an absolute tolerance of $| \overrightarrow{X} - F(\overrightarrow{X}) | < 10^{-13}$ and Newton-Raphson parameter of $\alpha = 0.05$ to be sufficient for most points of interest, with less than $10^4$ iterations required. Due to slower convergence near the critical surfaces $\mathcal{C}_\alpha$, especially at low frequencies $\omega \ll 1$, an accelerated version of the algorithm is used with $\alpha = 0.2$ and $10^6$ iterations. \\

Once the vector $\overrightarrow{X}$ is obtained, $G(z)$ and subsequently $\rho(\omega) = -\frac{1}{\pi} \Im{G(\omega^+}) $ is computed using \eqref{eqn:DysonG}. Various points in the $p = 3, 4, 5$ parameter spaces have been investigated to verify the analytic claims put forward.

\section{Calculation of entanglement properties}
\label{SM:entanglement}

\subsection{Brief review of previous work}
It is established in \cite{Peschel2003,Henley2004} that for a non-interacting fermionic system, the reduced density matrix of subsystem can be related to its correlation matrix. For a subsystem consisting of $N_s$ sites $I_l$, where $l = 1, \ldots, N_s$ with the correlation matrix defined $C_{I_l,I_{l'}} = \langle c^\dagger_{I_l} c_{I_{l'}} \rangle$, the reduced density matrix reads
\begin{equation}
    \rho = \det{(1-C)} \exp \left( \sum_{l, l'} \left [ \ln C (1-C)^{-1} \right]_{I_l, I_{l'}} c^\dagger_{I_l} c_{I_{l'}}  \right).
\end{equation}
Thus, properties of the reduced density matrix can be read off from the spectrum of $C$, referred to as the entanglement spectrum. Since the many body density matrix can be obtained from the $N_s \times N_s$ matrix $C$, we will often refer to $C$ as the {\em``entanglement hamiltonian"} of the subsystem. \\

As an example, we demonstrate the entanglement spectrum for a segment of an infinite SSH chain in \figref{fig:SM_EntSpecSSH}

\begin{figure*}
    \centering
    \includegraphics[width = \textwidth]{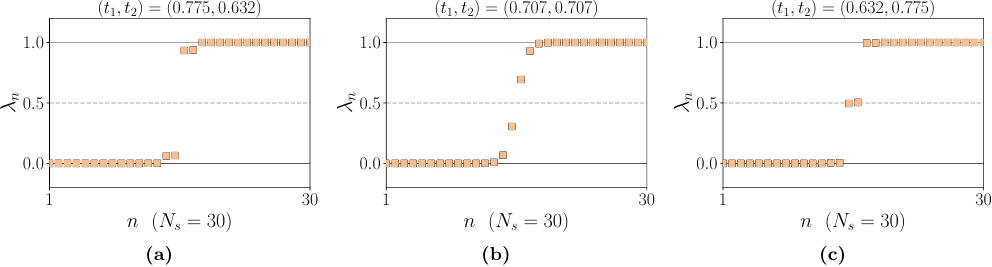}
    \caption{Entanglement spectrum for the $[12 \ldots 21]$ subsystem of the infinite $1d$ SSH chain at various parameters. \textbf{(a)} Trivial gapped phase, with no mid-gap modes in the entanglement spectrum. \textbf{(b)} Gapless phase, where the entanglement spectrum is also gapless. \textbf{(c)} Topological gapped phase, with mid-gap modes in the entanglement spectrum. }
    \label{fig:SM_EntSpecSSH}
\end{figure*}

\subsection{Numerical evaluation for finite chains in $B(p)$}
For a finite subsystem of the infinite tree (${I_1}, {I_2}, \cdots, {I_{N_s}}$) , the entanglement spectrum is obtained by constructing the correlation matrix $C_{I, J} = \langle c_{I}^\dagger c_{J} \rangle$ and subsequently diagonalizing it for its eigenvalues $\lambda_n$ for $n \in \{1, 2, \cdots N_s \}$. For numerical computation, we explicitly construct this matrix for chain-like subsystems upto $1000$ sites, and diagonalize it using standard routines from the Python package \href{https://numpy.org/}{NumPy}.

\subsection{SSH type infinite chains in $B(p)$}

\subsubsection{Entanglement hamiltonian}
To make analytic progress, we consider subsystems with infinite number of sites and a notion of translational symmetry. In particular, we look at SSH chain subsystems of $B(p)$, such as the infinite $[12\ldots21]$ chain. The correlation matrix of this subsystem is interpreted as a single particle hamiltonian $\tilde{H}$
\begin{figure}
    \centering
    \includegraphics[width = \columnwidth]{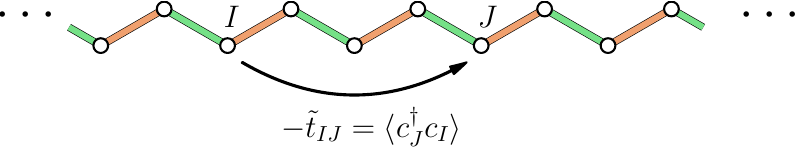}
    \caption{Graphical depiction of the entanglement hamiltonian $\tilde{H}$ of the $[12 \ldots 21]$ SSH chain.}
    \label{fig:Entanglement_hamiltonian}
\end{figure}
\begin{align}
	\tilde{H}_{I, {I'}} &= C_{I, {I'}}  & \\
	\tilde{\mu} &= C_{I,I'} = 1/2 & \textrm{Chemical potential} \\
	-\tilde{t}_{I, {I'}} &= C_{I, {I'}}  &  \textrm{Hopping element for $I \neq I'$}
\end{align}

We re-iterate that $\tilde{H}$ is referred to as the entanglement hamiltonian, as its spectrum corresponds exactly to the entanglement spectrum. The advantage of this interpretation is that $\tilde{H}$ describes a 1d chain with translation symmetry as shown in \figref{fig:Entanglement_hamiltonian}, which implies applicability of techniques from band theory. We define a unit cell consisting of sites labelled A and B, which are connected internally through a link of type $\alpha = 1$, and connected to the other unit cells through links of type $\alpha = 2$. Unit cells are labelled with a unique integer $n \in \mathbb{Z}$, which allows us to write the entanglement hamiltonian as
\begin{align*}
	\tilde{H} &= \sum_n \sum_\delta \tilde{t}_{AB}(\delta) \tilde{c}^\dagger_{n+\delta, B} \tilde{c}_{n, A} + \tilde{t}_{BA}(\delta) \tilde{c}^\dagger_{n+\delta, A} \tilde{c}_{n, B}\\  
	&+ \frac{1}{2} \sum_n  \tilde{c}^\dagger_{n,A} \tilde{c}_{n,A} + \tilde{c}^\dagger_{n,B} \tilde{c}_{n,B}.
\end{align*}
Fourier transforming to the basis $\tilde{c}_{k,s} = \frac{1}{\sqrt{N_s}} \sum_{n} e^{-\ci k n} \tilde{c}_{n,s}$, we obtain the band hamiltonian
\begin{align}
	\tilde{H} &= \sum_k \begin{pmatrix} \tilde{c}^\dagger_{k, A} &	 \tilde{c}^\dagger_{k, B} \end{pmatrix} \tilde{H}(k) \begin{pmatrix} \tilde{c}_{k, A} \\  \tilde{c}_{k, B} \end{pmatrix}, \\
	\tilde{H}(k) &= \frac{\mathbf{I}}{2} + h_x(k) ~ \mathbf{\sigma_x} + h_y(k) ~ \mathbf{\sigma_y},
\end{align}
where
\begin{align}
\tilde{h}(k) &= h_x(k) + \ci h_y(k) \\
&= \sum_\delta t_{AB}(\delta) e^{\ci k \delta}.
\end{align}
Then entanglement bands $\lambda(k)$ can be subsequently obtained as the eigenvalues of the band hamiltonian $\tilde{H}(k)$,
\begin{equation}
    \lambda_{\pm}(k) = \frac{1}{2} \pm |\tilde{h}(k)|.
\end{equation}

Using the integral form for correlation functions developed in \ref{SM:correlation_functions} \citeeq{eqn:corrfn_intform}, we can write $t_{AB}(\delta)$ as
\begin{align}
t_{AB}(\delta) &= (-1)^\delta \int dg ~ f(g) ~ (g_1 g_2)^\delta g_1 & (\delta \geq 0) \\
&= (-1)^{-\delta - 1} \int dg ~ f(g) ~ (g_1 g_2)^{-\delta - 1} g_2  & (\delta < 0)
\end{align}
where the the domain of integration for $\int dg$ depends on the region on the phase diagram one is in, as discussed in \ref{SM:correlation_functions}. \\

Fourier transforming the integral form, and subsequently interchanging the order of summation and integration, we write
\begin{align}
\tilde{h}(k) &= \int dg f(g) ~~ \left[ \frac{g_1}{1 + g_1 g_2 e^{\ci k}} + \frac{g_2 e^{-\ci k}}{1 + g_1 g_2 e^{-\ci k}} \right] \\
 &= e^{-\ci k/2} \left( h_+(k) \cos(k/2) + \ci h_-(k) \sin(k/2) \right), \label{eqn:htilde}
\end{align}
where $h_\pm(k)$ are integrals of strictly positive or strictly negative quantities. 
\begin{equation}
h_\pm (k) =  \int dg f(g) ~~ \frac{(g_1 \pm g_2)(1 \pm g_1 g_2)}{1 + 2 g_1 g_2 \cos(k) + g^2_1 g^2_2} \label{eqn:SM_hpmdef}
\end{equation}
Absence of sign changes makes them easy to compute numerically through adaptive numerical integration methods. \\
\begin{figure*}

    \includegraphics[width = 0.72\textwidth]{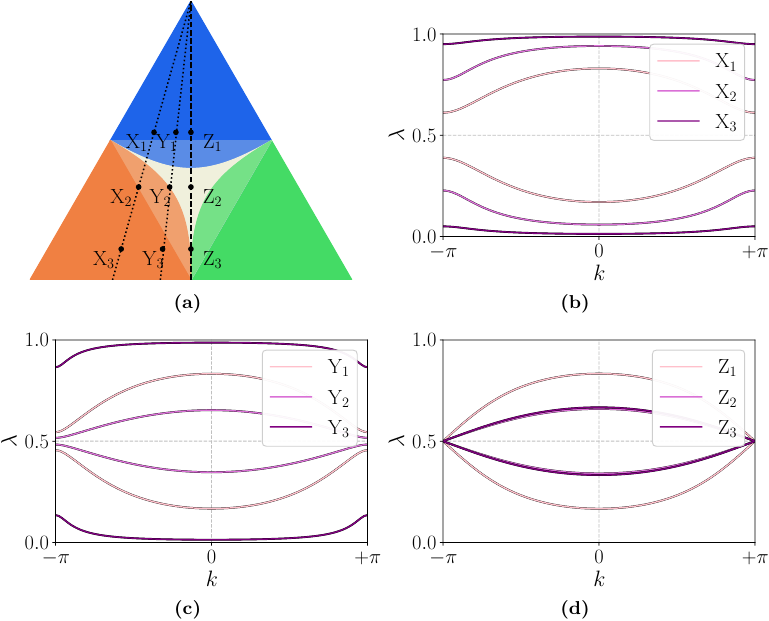}
    
    \label{fig: Entanglement_Bands}

    \caption{ \textbf{(a)} Entanglement band structures for the infinite $[12 \ldots 21]$ chains in $B(3)$ evaluated at points $X_1$ : $ ( 0.592 , 0.347 , 0.728 )$, $X_2$ : $ ( 0.704 , 0.413 , 0.577 )$, $X_3$ : $ ( 0.813 , 0.477 , 0.333 )$, $Y_1$ : $ ( 0.531 , 0.434 , 0.728 )$, $Y_2$ : $ ( 0.632 , 0.517 , 0.577 )$, $Y_3$ : $ ( 0.730 , 0.597 , 0.333 )$, $Z_1$ : $ ( 0.485 , 0.485 , 0.728 )$, $Z_2$ : $ ( 0.577 , 0.577 , 0.577 )$, $Z_3 $ : $ ( 0.667 , 0.667 , 0.333 )$. \textbf{(b)} Gapped spectra at $X_1, X_2$ and $X_3$, despite their being in distinct phases. \textbf{(c)} Gapped spectra at $Y_1, Y_2$ and $Y_3$. \textbf{(d)} Gap closing transition along the line $t_1 = t_2$, gaplessness at the point $k = \pi$ at all such points.  }
\end{figure*}
\subsubsection{Condition for gaplessness of spectrum}
For the entanglement spectrum to be gapless, we require $| \tilde{h}(k)| = 0$ for some $k$ in the Brillouin zone $(-\pi, \pi]$, which occurs when
\begin{equation}
	h_+(k)\cos(k/2) = h_-(k)\sin(k/2) = 0.
\end{equation}
For all points in parameter space, $h_+(k)$ is an integral of positive quantities, and thus the spectrum can only be gapless at the point $k = \pi$, with a sufficient and necessary condition for gaplessness being $h_-(\pi) = 0$. \\
Through algebraic manipulations it can be shown that for all values of $g$ in the domain of integration, we have the inequalities:
\begin{align}
	\label{eqn: g1g2signs}
	g_1(g) > g_2(g) & ~~~~~~ \textrm{if } t_1 > t_2, \\
	g_1(g) < g_2(g) & ~~~~~~ \textrm{if } t_1 < t_2, \\
	g_1(g) g_2(g) \leq 1 & ~~~~~~ \forall t_\alpha.
\end{align}
These  inequalities, along with \citeeq{eqn:SM_hpmdef} imply $h_-(\pi)$ is a strictly positive quantity when $t_1 > t_2$ and is strictly negative when $t_1 < t_2$. Thus, a gapless spectrum for the $[12 \ldots 21]$ chain is obtained if and only if $t_1 = t_2$. This suggests that for generic points in the \Metal ~ region of the phase diagram, there exist chain subsystems with gapped entanglement spectra, a departure from manifolds (see \ref{SM:manifold_entanglement}).

\subsubsection{Entanglement gap}
Numerical investigation suggests that the gap in the entanglement spectrum corresponds to the point $k = \pi$ of the Brillouin zone, and allows a convenient expression under this assumption
\begin{equation}
	\Delta_g = 2 \int dg f(g) ~ \frac{g_1 - g_2}{1 - g_1 g_2}.
\end{equation}
Similarly, the eigenvalues associated with band bottom and band top $(= \pm \Delta_b)$ correspond to $k = 0$,
\begin{equation}
	\Delta_b = \int dg f(g) ~ \frac{g_1 + g_2}{1 + g_1 g_2}.
\end{equation}

\subsubsection{Topology of the chain subsystem}
Since $\tilde{h}(k)$ determines the entanglement hamiltonian's band structure, one can define an `entanglement winding number' for the chosen chain subsystem to be $w_{\textrm{ent}} =  \int_{0}^{2 \pi} dk ~ \frac{\tilde{h}'(k)}{\tilde{h}(k)}$. From \eqref{eqn:htilde}, \eqref{eqn:SM_hpmdef} and \eqref{eqn: g1g2signs}, one can reason that
\begin{align}
	w_{\textrm{ent}} = 0 ~~~~~~~~~\textrm{for } t_1 > t_2, \\
	w_{\textrm{ent}} = 1 ~~~~~~~~~\textrm{for } t_1 < t_2.
\end{align}
This is highly reminiscent of the phase diagram of an individual SSH chain, and suggests that the topology of the truncated subsystem $[1 2 \ldots 2 1]$ manifests itself in the entanglement spectrum. It also asserts the presence of a single pair of midgap modes for $t_1 < t_2$ and the lack thereof when $t_1 > t_2$. \\
An equivalent way of distinguishing the cases $t_1 < t_2$ and $t_1 > t_2$ includes examining the symmetry representation carried by the ground state at $k = \pi$ point.

\section{Calculation of correlation functions}
\label{SM:correlation_functions}

Consider two-sites $I$ and $J$ on the tree, and let the shortest path between the two be described through the sequence of links traversed as $[\alpha_1 \alpha_2 \ldots \alpha_k \ldots \alpha_n]$. Then the corresponding fermionic two-point function in frequency space is
\begin{equation}
G_{JI}(z) = G(z) \left[ \prod_{k = 1}^n (- t_{\alpha_k} G_{\alpha_k}(z) ) \right]. \label{eqn:freq_2pt}
\end{equation}
Since $G(z)$ is a scalar function of $z$, the order in which the links $\alpha_k$ are traversed does not matter for the two point function. Rather, it only depends on the number of links $n_\alpha$ of each type $\alpha$ traversed from $I$ to $J$
\begin{equation}
G_{\{n_\alpha\}}(z) = G(z) \left[ \prod_{\alpha} (- t_{\alpha} G_{\alpha}(z) )^{n_\alpha} \right].
\end{equation}

To obtain the equal-time correlation function, we perform Matsubara summation and obtain

\begin{widetext}
\begin{equation}
\langle c_I^\dagger c_J \rangle = - \frac{1}{\pi} \Im \int_{-\infty}^{\infty} d \omega ~~ G(\omega^+) \left[ \prod_{\alpha} (- t_{\alpha} G_{\alpha}(\omega^+) )^{n_\alpha} \right] n_F(\omega - \mu)
\end{equation}
\end{widetext}
where $n_F$ is the Fermi function. Restricting our attention to half filling ($\mu = 0$) and zero temperature ($n_F(\omega) = \theta(-\omega)$), the correlation function can thus be written

\begin{equation}
\label{eqn:Matsubara}
C_{IJ} = - \frac{1}{\pi} \Im \int_{-\infty}^{0} d \omega ~~ G(\omega^+) \left[ \prod_{\alpha} (- t_{\alpha} G_{\alpha}(\omega^+) )^{n_\alpha} \right]. 
\end{equation}
It is difficult to analyse the above due to the lack of closed form expressions for $G(z)$ and $G_\alpha(z)$. Furthermore, numerical issues due to the highly oscillatory nature of the integrand arise when $n_\alpha$ are large. This suggests a change of complex variables from $z$ to $G$, exploiting the known analytic structure of $z(G)$ and $G_\alpha(G)$, which leads to
\begin{equation}
C_{IJ} = - \frac{1}{\pi} \Im \int_{G(-\infty + \ci \eta)}^{G(0 + \ci \eta)} dG ~ G\frac{dz}{dG} \left[ \prod_{\alpha} (- t_{\alpha} G_{\alpha}(G) )^{n_\alpha} \right].
\end{equation}
This is a complex line integral from the point $G = 0$ (at $\omega = -\infty$) to the point $G = -\ci \pi \rho_0$ (at $\omega = 0$) along the curve traced by $G(\omega)$ as shown in \figref{fig:G_domains}(b - f). One can smoothly deform this curve to a more convenient one in the $G$ - plane without changing the value of the integral, as long as the end-points are unchanged and the curve does not cross branch points or pole singularities of the integrand. A careful analysis of the analytic structure of the integrand reveals that (with the exception of the trivial case where $n_\alpha = 0 ~~ \forall \alpha$) one only has to deal with the branch point singularities at $G = \pm \ci /2t_\alpha$, and thus the integral can be deformed to one along the imaginary $G$ axis as in \figref{fig:G_domains}(h - j). \\

Without any loss of generality, we assume that $t_1 \geq t_2 \geq \cdots \geq t_p$. The function $\sigma_\alpha(\omega^+)$ can exhibit four distinct possibilities for our model:

\begin{itemize}
\item Metal  with $\sigma_1(0^+) = +1$, corresponds to \textbf{M}$_0$ region
\item Metal with $\sigma_1(0^+) = -1$, corresponds to \textbf{M}$_1$ region
\item Semimetal with $\sigma_1(0^+) = -1$, corresponds to $\mathcal{C}_1$ critical surface
\item Insulator with $\sigma_1(0^+) = -1$, corresponds to \textbf{I}$_1$ phase
\end{itemize}

\begin{figure*}

    \includegraphics[width = \textwidth]{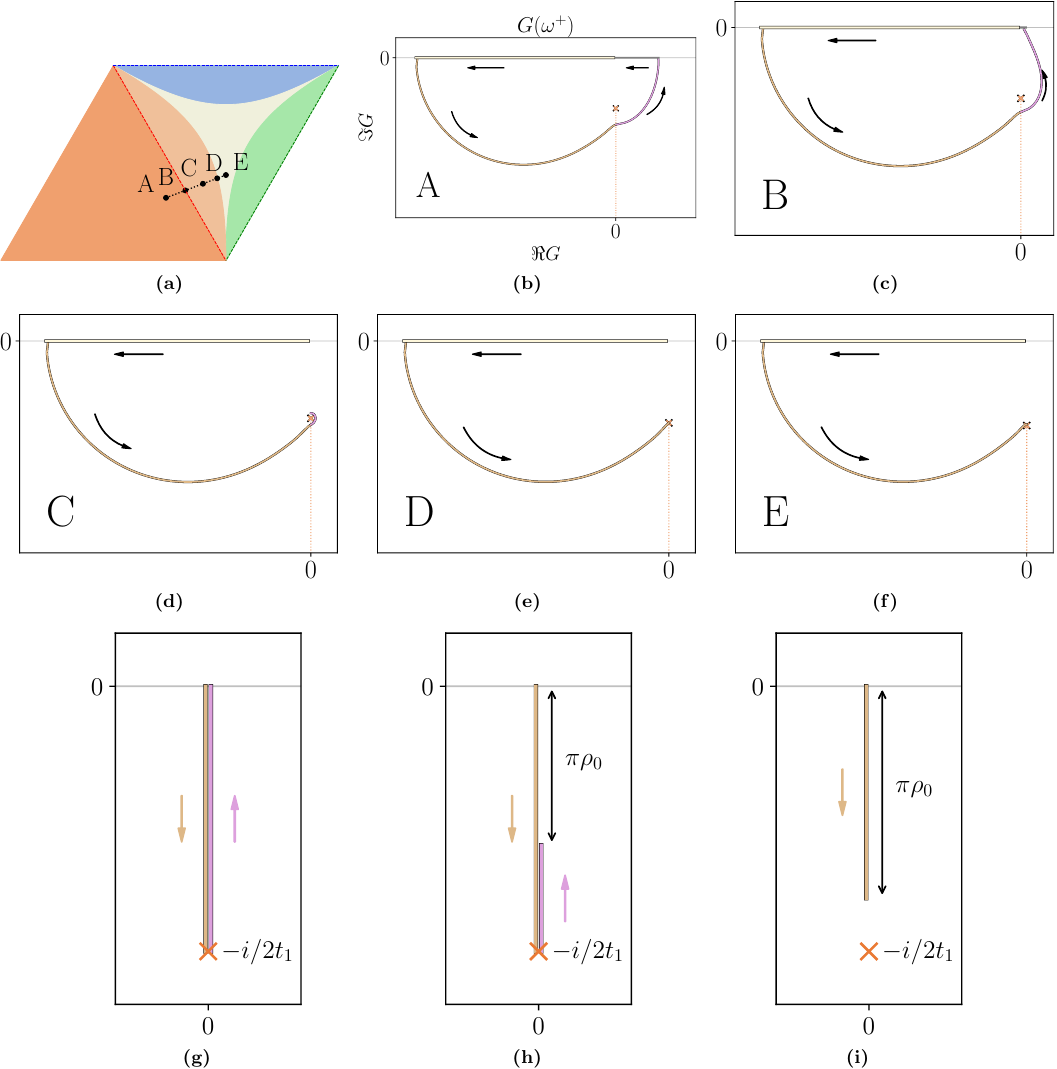}

    \caption{\textbf{(a)} Points A: (0.743, 0.535, 0.401), B: (0.707, 0.565, 0.424), C: (0.673, 0.592, 0.444), D: (0.642, 0.614, 0.460) and E: (0.625, 0.625, 0.468) in phase diagram. \textbf{(b-f)} The curves traced out by $G(\omega^+)$ for $\omega \in (-\infty, 0]$ for points ABCDE. \textbf{(g-i)} Sketch of integration domain for \Insulator$_1$, \Metal$_1$ and \Metal$_0$ phases respectively.}
    \label{fig:G_domains}

\end{figure*}

In each case, the integrand may be parametrized using $G = - \ci g$ where $g \in [0,1/2 t_1]$ is a real parameter. We define $g_\alpha(g) = \ci t_\alpha G_\alpha = \frac{1 - \sigma_\alpha \sqrt{1 - 4 g^2 t_\alpha^2}}{2 t_\alpha g}$, where $\sigma_\alpha$ is no longer an explicit function of $z$, and takes value $+1$ or $-1$ depending on the segment of the line integral. Under this parametrization, we also find it useful to define

\begin{equation}
f(g) = \frac{\ci g}{\pi} \frac{dz}{dg} = \frac{1}{2 \pi g} \left[ \sum_\alpha \frac{ \sigma_\alpha } {\sqrt{1 - 4 t^2_\alpha g^2}} - (p - 2) \right]. 
\end{equation}
In terms of the new variables, we can write $C_{IJ}$ as
\begin{equation}
	C_{IJ} = \Im \int f(g) dg  ~~ \prod_{\alpha} (\ci g_\alpha)^{n_\alpha}, \label{eqn:corrfn_intform}   
\end{equation}
where the limits of the line integral depend on the value of $\sigma_1(0^+)$.

\begin{widetext}

\begin{align}
\sigma_1(0^+) = +1 : ~~~~~~~~ \int f(g) dg ~~ (\cdots) & = \int_0^{\pi \rho_0} f(g) dg ~~ (\cdots)_{\sigma_1 = +1} \\
\sigma_1(0^+) = -1 : ~~~~~~~~ \int f(g) dg ~~ (\cdots) & = \int_0^\frac{1}{2 t} f(g) dg ~~ (\cdots)_{\sigma_1 = +1} - \int_{\pi \rho_0}^\frac{1}{2 t} f(g) dg ~~ (\cdots)_{\sigma_1 = -1}
\end{align}
\end{widetext}

The overall pre-factor $\Im \left[ ~ \ci^{~\sum n_\alpha} \right]$ is a manifestation of the sublattice symmetry of the problem, and ensures that the two point function between two distinct sites belonging to the same sub-lattice will always be $0$. Aside from this sign factor, the integrand in this new formalism can be chosen to be positive, i.e.

\begin{align}
\sigma_1(0^+) = +1 :  & + f(g) \prod_\alpha (g_\alpha)^{n_\alpha} > 0 &\forall ~ g \in [0,1/2t_1] \label{eqn:SM_PosinPos} \\ 
\sigma_1(0^+) = -1 : & -f(g) \prod_\alpha (g_\alpha)^{n_\alpha} > 0 & \forall ~ g \in [\rho_0,1/2t_1] \label{eqn:SM_PosinNeg}
\end{align}
The result \eqref{eqn:SM_PosinPos} follows from the relevant definitions, while the result \eqref{eqn:SM_PosinNeg} requires more careful analysis which we defer to the next subsection. Since $f(g)$ always occurs in the integrand regardless of the choice of points $I$ and $J$, it is helpful to interpret $f(g)dg$ as a positive integration measure, once the factor of $-1$ associated with the $\sigma_\alpha = -1$ segment has been absorbed. \\

Thus, a highly oscillatory complex integral \eqref{eqn:Matsubara} has been reduced to a well - behaved real integral. We use this advantage to numerically compute correlation functions for points up to a thousand sites apart using the integrate function from the Python package \href{https://scipy.org}{Scipy}, while specifying the (integrable) square root divergence of $f(g)$ at $g = 1/2 t_1$. This formulation leads to analytic progress too, as discussed in sections \ref{SM:entanglement} and \ref{SM:correlation_lengths}.
\subsubsection{Showing positivity of integrand in branch $\sigma_1 = -1$}
Extensive numerical investigation suggests that $f(g)$ has a unique zero in the interval $(0,1/2t_1)$, regardless of all the other parameters $t_2, t_3, \ldots t_p \leq t_1$. Assuming this statement to be true in general, we can derive the positivity of integrand. From the definition of $f(g) = g \frac{dz}{dg}$, the identity
\begin{equation}
	\int_{0}^{\pi \rho_0} dg ~~ f(g)/g = z(g = \pi \rho_0) - z(g = 0) = 0,  
\end{equation}
is established, where $z(g=0) = 0$ follows from \citeeq{eqn:G}. From mean-value theorem, the identity implies the existence of $g_* \in (0,\pi \rho_0)$ such that $f(g_*)/g_* = 0$. Thus, $g^*$ must be the unique zero of $f(g)$. Since $f$ is a continuous function in $(0, 1/{2t_1})$, and $g^*$ is the unique zero, it must have constant sign in the interval $(g^*,1/2t_1]$. \\
Combining this with the fact that $\lim_{g \rightarrow 1/2t_1} f(g) = - \infty$, we conclude
\begin{equation}
    - f(g) > 0 ~~~~ \forall g \in (\pi \rho_0, 1/2t_1)
\end{equation}
for the $\sigma_1 = -1$ branch, as claimed.

\section{Subsystems}
\label{SM:subsystems}

\subsection{Strong, weak and intermediate chains}
Suppose $t_\alpha$ is the largest among the $p$ hoppings of the model defined on $B(p)$. Then an arbitrary chain subsystem $[ \beta_1, \beta_2, \ldots \beta_L]$ is said to be strong if $n_\alpha > \sum_{\beta \neq \alpha} n_\beta$, is said to be weak if $n_\alpha = 0$, and is said to be intermediate otherwise. We only consider chains of odd length (or equivalently, even number of sites).

\subsection{SSH chains}
We only consider SSH chains with an even number of sites, and denote the SSH chains with alternating $t_\alpha$ and $t_\beta$ bonds as $[\alpha \beta \ldots \beta \alpha]$. Here, the two extreme sites of the SSH chain are connected to the rest of sites through a link of type $\alpha$. A few examples have been demonstrated in \figref{Fig:SSH_subsystems}.

\begin{figure}
    \includegraphics[width = \columnwidth]{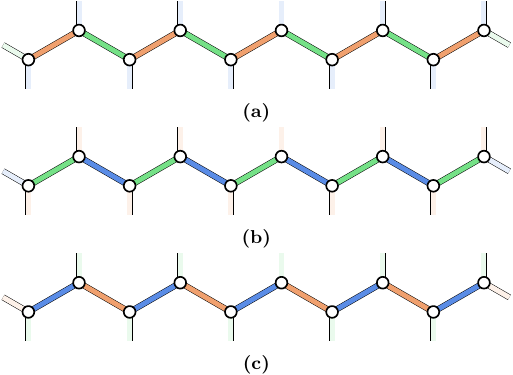}

    \caption{$[12 \cdots 21]$, $[23 \cdots 32]$ and $[31 \cdots 13]$ SSH subsystems of length $9$. }
    \label{Fig:SSH_subsystems}
\end{figure}

\subsection{$B_\alpha(p)$ subsystem}
$B_\alpha(p)$ is defined to be the infinite subsystem of $B(p)$ obtained by removing the the sub-trees connected to
the root node through links of colors other than $\alpha$. As an example, \figref{fig:semi_infinite} depicts the subsystem $B_2(p)$

\begin{figure}
    \centering
    \includegraphics[width = \columnwidth]{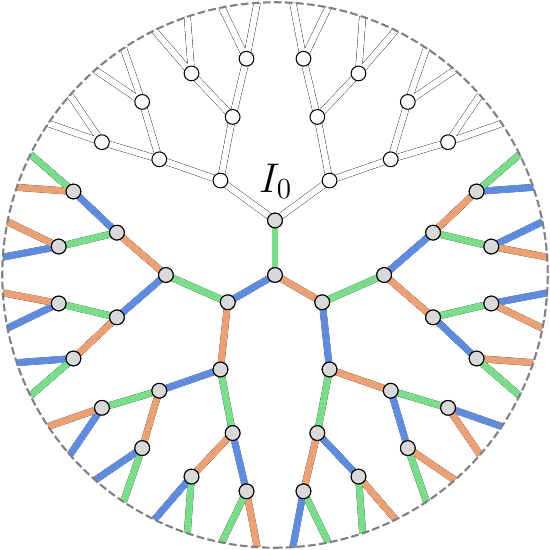}
    \caption{Semi-infinite subsystem $B_2(p)$.}
    \label{fig:semi_infinite}
\end{figure}

\begin{figure}
    \includegraphics[width = \columnwidth]{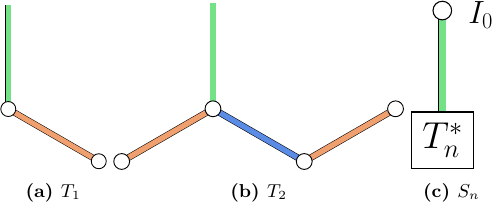}
    \label{fig:SM_T_base}
    \caption{\textbf{(a)} System $T_1$ with two sites. \textbf{(b)} System $T_2$ with four sites. \textbf{(c)} Construction of $S_n$ system by removing site $I_{n+1}$ from $T_n$ (referred to as the $T_n^*$ subsystem) and subsequent inclusion of the root site $I_0$.}

\end{figure}

\begin{figure}

    \includegraphics[width = \columnwidth]{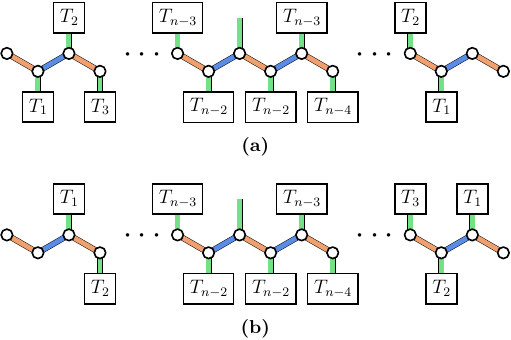}
    \label{fig:T_n_recursion}

    \caption{Iterative procedure to construct the $T_n$ system when \textbf{(a)} $n$ is odd, and \textbf{(b)} $n$ is even.}
    
\end{figure}

\subsection{$S_n$ subsystem}
We provide an algorithm to construct the subsystem $S_n$ (where $n$ is odd), which goes as follows:
\begin{enumerate}
    \item Define $T_1$ and $T_2$ to be subsystems with $2$ and $4$ sites respectively as shown in \figref{fig:SM_T_base} (a-b). An open green link has been indicated to aid in the next step.
    \item Construct $T_k$ recursively using procedure outlined in \figref{fig:T_n_recursion} for $k = 3, 4, \ldots n$.
    \item Attach site $I_0$ to the open green link in $T_n$. In the resulting finite tree, consider the site $I_{n+1}$ connected to $I_0$ through chain $[21 \ldots 121]$ of length $n+1$. Remove this site from the tree, and call the remaining finite subsystem $S_n$. See \figref{fig:SM_T_base} (c)
\end{enumerate}

It is useful to define $I_n$ as the site in $S_n$ connected to $I_0$ through chain $[21 \ldots 12]$ of length $n$. The construction of the $S_n$ system is ideal because

\begin{itemize}
    \item when $t_2 = 0$ and $t_1, t_3 >0$, ~ $I_0$ and $I_n$ are the only isolated sites in the system, ie, fermions at $I_0$ or $I_n$ cannot hop to any other site. All other sites in $S_n$ are part of SSH chains of the form $[13 \ldots 31]$.
    \item $I_0$ and $I_n$ are separated by an SSH chain $[21 \ldots 12]$ which is $n$ links long.
    \item The site furthest from $I_0$ is at a separation of $n + 1$ links. All sites belonging to  generation $g \in \{1, 2, \ldots n - 1 \}$ have exactly $3$ nearest neighbours.
    \item For $t_1, t_2, t_3 > 0$, the system has $0$ nullity, that is, there are no eigenstates with exactly zero energy for finite $n$. This is ensured by keeping number of sites on each sublattice equal, along with more general considerations discussed in  \cite{IvanGutman2011}.
\end{itemize}

\subsection{$S_n^*$ subsystem}
We define the $S_n^*$ subsystem to be the $S_n$ subsystem with the sites $I_0$ and $I_n$ removed. To infer the presence of zero-energy edge states in the subsystem $S_n$ hosted by $I_0$ and $I_n$, it is useful to compare the energy spectra of $S_n$ and $S^*_n$, as has been done in \ref{SM:filling_anomaly}.

\section{Band properties, dos and phase boundaries}
\label{SM:band_properties}

\subsection{Band edges}
The function $z(G)$ \citeeq{eqn:G} is analytic in $G$ at all points except the branch points of the form $\pm \ci /2t_\alpha$ and possible poles at $G =0$ and $G = \infty$ (depending on choice of $\sigma_\alpha$). To reconcile this with the fact that $G(\omega^+)$ displays a non-analyticity at the band edges in the spectrum (which correspond to $G(\pm \varepsilon + \ci \eta) \in \mathbb{R}$), we require ${dz}/{dG} = 0$ at these points. In the case where $p \geq 3$, this leads to an implicit equation in the variable $G(\varepsilon^+)$ 

\begin{equation}
(p-2) = \sum_\alpha \sigma_\alpha (1 + 4t_\alpha^2 G^2)^{-1/2}. \label{eqn:BandEdges}
\end{equation}
Numerical investigation of \citeeq{eqn:BandEdges} reveals that in all the gapless phases, the density of states form a single band, characterised by the band bottom and band top ($\omega = \pm \varepsilon_b$). These correspond to the unique pair of solutions of \eqref{eqn:BandEdges} with $\sigma_\alpha = + 1 ~~ \forall \alpha$. \\
In the gapped phase $I_\alpha$, the bottom of the conduction band and the top of the valence band ($\omega = \pm \varepsilon_g/2$) correspond to the solutions of \eqref{eqn:BandEdges} with $\sigma_\alpha = -1$, and $\sigma_\beta = + 1 ~ \forall \beta \neq \alpha$. \\

The exponent characterising the band edges for $p \geq 3$ can be obtained through a Taylor expansion of $z(G)$ about the point $G_\varepsilon$ where $dz/dG = 0$,

\begin{equation}
	z - \varepsilon = \frac{1}{2} \frac{d^2 z}{dG^2} |_{G_\epsilon} (G - G_\varepsilon)^2 + \cdots . 
\end{equation}
yielding $(G - G_\varepsilon) \sim (z - \varepsilon)^{1/2}$. This argument rests on the fact that at  ${d^2 z}/{dG^2} |_{G_\epsilon} \neq 0$, which can be proven for systems with $p \geq 3$. When $p = 2$, the exponent characterizing the band edges is $-1/2$ instead, as is expected for one-dimensional models such as the SSH chains. 

\subsection{Distinction between gapped phases}
Since our model belongs to the BDI symmetry class, it is endowed with particle-hole (and sublattice) symmetry which manifests itself in the language of Green's functions as $G(-z) = -G(z)$. Along with the analyticity property $G^*(z) = G(z^*)$ of resolvent Green's functions, this implies that for arbitrary points in the phase diagram, we have
\begin{align}
	\Re G(-\omega^+) &= - \Re G(\omega^+), \label{eqn:RealParticleHole} \\
	\Im G(-\omega^+) &= + \Im G(\omega^+).
\end{align}
A gapped spectrum requires the the density of states to vanish at the chemical potential ($\Im G(0^+) = 0$), which combined with the particle-hole constraint \eqref{eqn:RealParticleHole} gives rise to $G(0^+) = 0$. Substituting this in \eqref{eqn:G}, we obtain a constraint on the sign functions $\sigma_\alpha$
\begin{equation}
	\sum_\alpha \sigma_\alpha (0^+) = p - 2.
\end{equation}
The above equation has exactly $p$ solutions, with each corresponding to a topologically distinct insulating phase. The phases can be characterised by the tuple $(\sigma_1(0^+), \sigma_2(0^+), \cdots, \sigma_p(0^+))$, among which exactly one will equal $-1$ while the other $p-1$ will equal $+1$. This distinction manifests itself in multiple ways, which are subject of sections \ref{SM:edge_mode}, \ref{SM:winding_characteristic} and \ref{SM:correlation_lengths}.

\subsection{Phase boundaries}
For a point in either a gapped phase \textbf{I}$_\alpha$ or a semi-metallic phase $\mathcal{C}_\alpha$, we can expand \eqref{eqn:G} in powers of $G$ about $z = G = 0$ to obtain the low frequency behaviour
\begin{equation}
	z = \left( - t_\alpha^2 + \sum_{\beta \neq \alpha} t_\beta^2 \right) G - \left( - t_\alpha^4 + \sum_{\beta \neq \alpha} t_\beta^4 \right) G^3 + \mathcal{O}(G^5). \label{eqn:z_G_expansion}
\end{equation}
For physical consistency, we require $ - \frac{1}{\pi} \Im G > 0$ for all $z = \omega^+$, which places a constraint on \citeeq{eqn:z_G_expansion}
\begin{equation}
	t_\alpha^2 \leq \sum_{\beta \neq \alpha} t_\beta^2.
\end{equation}
Since the only assumption made was regarding the vanishing of density of states at $\omega = 0$, we conclude that the phase boundaries are described as

\begin{align}
t^2_\alpha < 1/2 ~~~& (\textrm{ for \textbf{I}}_\alpha ~ ), \\
t^2_\alpha = 1/2 ~~~& (\textrm{ for }\mathcal{C}_\alpha ~ ), 
\end{align}
(where the convention $\sum_\beta t^2_\beta = 1$ has been used). As an example, the phase boundaries for $p = 4$ are shown in \figref{fig:SM_p4} (a).

\begin{figure}
    \centering
    \includegraphics{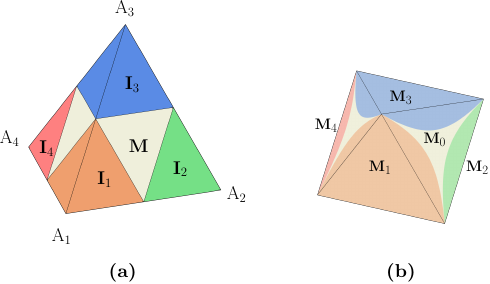}
    \caption{(a) Regions \Insulator$_\alpha$ and \Metal~ in the $p = 4$ phase diagram. (b) Blow up of the \Metal~ region, with $4 + 1$ distinct regimes corresponding to \Metal$_\alpha$ and \Metal$_0$  respectively. }
    \label{fig:SM_p4}
\end{figure}

\subsection{Density of states at chemical potential}
In a metallic phase, the density of states at the chemical potential (denoted $\rho_0$) is finite and may be obtained by substituting $z = 0^+$ and $G = -\ci \pi \rho_0$ in \eqref{eqn:G},

\begin{equation}
 p - 2 = \sum_\alpha \sigma_\alpha(0^+) \sqrt{1 - 4 \pi ^2 t_\alpha^2 \rho_0^2} \label{eqn: dos_0}.
\end{equation}
The existence of solutions requires $\rho_0 \leq {1}/{2 \pi t_\alpha}$, which here emerges as a natural bound on $\rho_0$. Interestingly, even in the gapless regime, the tuple $(\sigma_1(0^+), \sigma_2(0^+), \cdots, \sigma_p(0^+))$ can take on $p+1$ distinct values, which correspond to the $p+1$ distinct metallic regions referred to in the main text.\\
The boundary between the $M_\alpha$ and $M_0$ regions corresponds to a change in the sign of $\sigma_\alpha$, which can only happen in the circumstance when the bound on $\rho_0$ is saturated ($\rho_0 = 1/2 \pi t_\alpha$). Thus, the boundary is described by the equation

\begin{equation}
 p - 2 = \sum_{\beta \neq \alpha} \sqrt{1 - (t_\beta / t_\alpha)^2 },
\end{equation}
which the distinct regimes of \Metal ~ portrayed for the case $p = 4$ in \figref{fig:SM_p4}(b)

\section{Analytic solution of edge state}
\label{SM:edge_mode}

The subsystem $B_\alpha(p)$ as defined in \ref{SM:subsystems} allows us to analyse the edge-modes associated with various phases in an analytic fashion. As discussed in the main text \citeeq{eqn:Gambeta}, one can define an on-site greens function for the edge site as
\begin{equation}
	\Gamma_\alpha (z) = \langle I_0 | \mathbf{G}(z)_{B_\alpha(p)} | I_0 \rangle.
\end{equation}
Analogous to $G_\alpha$, the relevant Dyson equation reads
\begin{equation}
	\Gamma^{-1}_\alpha (z) = z - t_\alpha^2 G_\alpha (z). \label{eqn: GammaAlpha}
\end{equation}
It is useful to study the equation near $z = 0$ to infer about the zero energy modes, and is accomplished through analysis methods developed in \ref{SM:GreenFn_analytic}. We discuss features of $\Gamma_\alpha$, phase by phase.

\subsection{Metallic phase}
For metallic phases, we know that $G(z \rightarrow 0^+) = -\ci \pi \rho_0 \neq 0$, which implies, through \citeeq{eqn:SM_Galpha} and \citeeq{eqn: GammaAlpha}, that $\Gamma_\alpha (z \rightarrow 0^+)$ if non-zero and suggests a flat dispersion of $- \Im \Gamma_\alpha (\omega^+)$ about $\omega =  0$. This can be verified through numerics, and also by expanding $\Gamma_\alpha(\omega^+)$ upto leading order in $\omega$.

\subsection{Gapped phase}
In \ref{SM:band_properties}, we showed that gapped (and also semimetallic) phases are characterized by $G(z \rightarrow 0) = 0$. This suggests expanding $z(G)$ \citeeq{eqn:G} about $G = 0$. For the phase \textbf{I}$_\alpha$, this is accomplished by choosing $\sigma_\alpha = -1$ and $\sigma_\beta = +1 ~ \forall ~ \beta \neq \alpha$ to yield the expansion

\begin{equation}
	z = (1 - 2 t_\alpha^2) G(z) + \mathcal{O}(G^3). \label{eqn:zGlinear}
\end{equation}
Eliminating $G$ using \eqref{eqn:SM_Galpha}, and using $\beta$ to denote indices not equal to $\alpha$, we obtain
\begin{align}
	G_\alpha(z) &= - \frac{1}{t^2_\alpha G(z)} + \mathcal{O}(G) \\  &= \frac{1}{z} \left(2 - \frac{1}{t_\alpha^2} \right) + \mathcal{O}(z), \\
	 G_\beta(z) &= G(z) + \mathcal{O}(G^3)  \\ &= \frac{z}{(1 - 2 t_\alpha^2)} + \mathcal{O}(z^3)
\end{align}
which implies
\begin{align}
\Gamma_\alpha(\omega^+) &= \frac{\omega}{1 - 2 t^2_\alpha} + \mathcal{O}(\omega^3), \\
\Gamma_\beta(\omega^+) &=  \left(\frac{1}{\omega} -\ci \pi \delta(\omega)\right) \left( \frac{2 t_\alpha^2 - 1}{2 t_\alpha^2 + t_\beta^2 - 1} \right) + \mathcal{O}(\omega). 
\end{align}
This suggests the existence of edge modes in the subsystems $B_\beta(p)$ in the phase $I_\alpha$. This is a characteristic of the $I_\alpha$ phase and reflects its nature as an obstructed atomic insulator.

\subsection{Semimetals}
For points on $\mathcal{C}_\alpha$, the leading order term in \eqref{eqn:zGlinear} vanishes and gives rise a leading order dependence $z \sim G^3$. Following steps analogous to the previous section, we find
\begin{align}
	\Gamma_\beta(z) &\sim \omega^{\frac{1}{3}} & (\beta = \alpha), \\
	&\sim \omega^{-\frac{1}{3}} & (\beta \neq \alpha).
\end{align}
Thus, the pole singularity in the gapped phase and the featureless gapless phase is bridged through a powerlaw behaviour in the critical phase. This is discussed further in a later section \ref{SM:critical_properties}.

\subsection{Wavefunction of the edge mode}
One can analytically determine the exact wave-function of the edge mode in the gapped phase $I_\alpha$ through an explicit solution of the Schrodinger equation. Inspired from the pole singularity found exactly at $\omega = 0$ in the previous sections, we posit the existence of a zero energy mode described by wave-function $\psi_I$ ($I \in B_\beta(p)$) and write the relevant Schrodinger equation

\begin{equation}
 - \sum_{\alpha'} t_{\alpha'} \psi_{I + {\alpha'}} = E \psi_I = 0 ~~ \forall ~ I \in B_\beta(p) - \{g_0\} 
\end{equation}

Naively, the above system of linear equations has infinitely many solutions, but we shall see demanding normalizability of the wave function leads to a unique answer. \\
To exploit the generation structure of the Bethe lattice, it is useful to consider the edge site of $B_\beta(p)$ as $I_0$ and let $g_k = \{I ~ | ~ g(I) = k\}$ denote the collection of sites in generation $k$. Since $g_0 = \{ I_0 \} $, we shall use the two interchangeably. \\
The bipartite nature of $B_\beta(p)$ allows us to denote the sites in generations $g_0, g_2, g_4, \cdots, g_{2n}, \cdots$ as belonging to sublattice A, while the sites in the odd generations $g_1, g_3, g_5, \cdots, g_{2n +1}, \cdots$ belong to the sublattice B. From the sub-lattice symmetry of the problem, the zero energy modes can always be chosen to either have support on sublattice A, or on sublattice B. To discuss the edge mode associated with site $g_0$, we restrict ourself to wave-functions having support on sublattice A, and hence assert that $\psi_I = 0$ for all sites $I$ in the B sublattice. \\

Let the wave-function amplitude at the root site be $\psi_{g_0} = K$, which will be determined from normalization. Site $I$ in the generation $g_n$ is labelled by the sequence of links from $g_0$ to $I$: $[\beta \beta_1 \beta_2 \cdots \beta_{n-1}]$. We claim that for the site $[\beta \beta_1] $ in generation $g_2$, the wave-function amplitude $\psi_I \neq 0$ if and only if $\beta = \alpha$. 

\subsubsection{Proof of claim}
To show this, assume the contrary, i.e. $\psi_{[\beta \beta_1]} = K' \neq 0$ and $\beta_1 \neq \alpha$. Schrodinger equation for the site $I = [\beta \beta_1 \alpha]$ implies

\begin{align}
&\sum_{\beta_3 \neq \alpha} t_{\beta_3} \psi_{I + \beta_3} = - t_\alpha K' \\
\implies &\sum_{\beta_3 \neq \alpha} t_{\beta_3} |\psi_{I + \beta_3}| > t_\alpha |K'| \\
\implies &\left( \sum_{\beta_3 \neq \alpha}  t^2_{\beta_3} \right) \left( \sum_{\beta_3 \neq \alpha} |\psi_{I + \beta_3}|^2 \right) > t^2_\alpha |K'|^2 \\
\implies & \sum_{\beta_3 \neq \alpha} |\psi_{[\beta \beta_1 \alpha \beta_3]}|^2 > \frac{t^2_\alpha}{1 - t^2_\alpha} |K'|^2 \label{eqn:SM_ew1}
\end{align}
where $\sum_{\alpha'} t^2_{\alpha'} = 1$. Using arguments similar to the above, we can show that for sites in generation $g_6$,

\begin{align}
 & \sum_{\beta_5 \neq \alpha} |\psi_{[\beta \beta_1 \alpha \beta_3 \alpha \beta_5]}|^2 > \frac{t^2_\alpha}{1 - t^2_\alpha} |\psi_{[\beta \beta_1 \alpha \beta_3]}|^2 \\
\Rightarrow & \sum_{\beta_3 \neq \alpha} \sum_{\beta_5 \neq \alpha} |\psi_{[\beta \beta_1 \alpha \beta_3 \alpha \beta_5]}|^2 > \left( \frac{t^2_\alpha}{1 - t^2_\alpha} \right)^2 |K'|^2
\end{align}
Inductively, a bound for the sum of $|\psi_I|^2$ for $I$ in the generation $g_{2n}$ and having form $[\beta \beta_1 \alpha \beta_3 \cdots \alpha \beta_{2n - 1}]$ is obtained.

\begin{equation}
\begin{split}
\sum_{\beta_3 \neq \alpha} \sum_{\beta_5 \neq \alpha} \cdots \sum_{\beta_{2n-1} \neq \alpha}  |\psi_{[\beta \beta_1 \alpha \beta_3 \cdots \alpha \beta_{2n - 1}]}|^2 & \\ > \left( \frac{t^2_\alpha}{1 - t^2_\alpha} \right)^{n - 1} & |K'|^2
\end{split}
\end{equation}
From normalizability constraint, the LHS must be bounded above by $1$, which leads to
\begin{equation}
	|K'|^2 < \left( \frac{1 - t_\alpha^2}{t_\alpha^2} \right)^{n-1}. \label{eqn:SM_ew2}
\end{equation}
Since the gapped phase \textbf{I}$_\alpha$ is characterized by $t_\alpha^2 > 1/2$ and the above holds for all $n$, we are forced to conclude $K' = 0$, proving the assertion claimed.
\subsubsection{Generalizing the statement}
The claim $\psi_{[\beta \beta_1]} \neq 0 \implies  \beta_1 = \alpha$ can be generalized to $\psi_{[\beta \beta_1 \cdots \beta_{2n - 1}]} \neq 0 \implies \beta_1 = \alpha_1$. Suppose it were false and $\beta_1 \neq \alpha$, then it would imply the existence of zero-modes in the bulk, as the edge $g_0$ is disconnected from the entire sub-tree due to $\psi_{[\beta]} = \psi_{[\beta \beta_1]} = 0$. However the existence of zero-modes in the bulk contradicts the gap in the bulk spectrum in the insulating \textbf{I}$_\alpha$ phase, lending credence to the assertion. \\
Similarly, one can generalize even further to $\psi_{[\beta \beta_1 \beta_2 \dots \beta_{2n-1}]} \neq 0 \implies \alpha = \beta_1 = \beta_3 =  \cdots = \beta_{2n-1}$. This follows from induction on $n$, and using the key strategy of showing $\sum_I |\psi_I|^2$ over sites $I$ of the form $[\beta \alpha \beta_2 \cdots \alpha \beta_{2n-1} \alpha \beta_{2n+1} \alpha \cdots \beta_{2n + 2m - 1}]$ is an unbounded function of $m$ . \\

Thus, the edge-state wave-function has non-zero amplitude possible only on sites of the form $[\beta \alpha \beta_2 \alpha \cdots \beta_{2n-2} \alpha \beta_{2n}]$ and is forced to be $0$ everywhere else.

\subsubsection{Determining the wave-function}
From the Schrodinger equation for the zero mode $\psi$ about the site $I = [\beta \alpha \cdots \beta_{2n-2} \alpha \beta_{2n}]$, along with the constrained support of the wavefunction, we get

\begin{align}
& t_{\beta_{2n}} \psi_{I + \beta_{2n}} + t_{\alpha} \psi_{I + \alpha} + \sum_{\beta_{2n+1} \neq \alpha, \beta_{2n}} t_{\beta_{2n + 1}} \psi_{I + \beta_{2n+1}}  = 0 \\
\implies & \psi_{[\beta \alpha \cdots \beta_{2n-2} \alpha \beta_{2n} \alpha]} = - (t_{\beta_{2n}}/{t_\alpha}) ~ \psi_{[\beta \alpha \cdots \beta_{2n-2} \alpha]}
\end{align}
From induction, the wave-function of the edge mode with $\psi_{g_0} = K$ is 
\begin{equation}
	\psi_{[\beta \alpha \cdots \beta_{2n-2} \alpha]} = K \prod_{j = 0}^{n - 1} (-t_{\beta_{2j}}/t_\alpha)
\end{equation}

where $\beta_0 \equiv \beta$ and $\psi_I= 0$ everywhere else. The normalization constant $K$ is obtained

\begin{align}
	\sum_{I \in B_\beta(p)} |\psi_I|^2 &= |K|^2 \left[ 1 + (t_{\beta}/t_\alpha)^2 \sum_{n = 1}^{\infty} \prod_{j = 1}^{ n -1} \sum_{\beta_{2j} \neq \alpha} (t_{\beta_{2j}}/t_{\alpha})^2 \right], \\
\implies  1 &= |K|^2 \left[ 1 + (t_\beta/t_\alpha)^2 \sum_{n = 1}^{\infty} \left( \frac{1 - t^2_\alpha}{t^2_\alpha}  \right)^{n-1} \right], \\
\implies |K|^2 &= \left( \frac{2 t^2_\alpha - 1}{2 t^2_\alpha + t^2_\beta - 1} \right). \label{eqn:SM_ew3}
\end{align}
This coincides with the weight of the pole singularity in the spectral function for root site $I_0$, which was discussed in a previous section.

\subsubsection{Stability to symmetry preserving disorder}
Remarkably, the edge mode wavefunction survives even in the presence of disorder in the hopping amplitudes $t^{(I)}_{\alpha}$, which now explicitly depend on the choice of link $I - I+\alpha$ (breaking translation symmetry). We are in interested in weak disorder that maintains the BDI symmetries of the model, without introducing higher order hoppings (which would effectively introduce loops in the model) and without closing the gap. An example of such a disorder is one in which the hopping amplitude for each link is sampled from an independent (uniform) distribution $t^{I}_\beta \sim \textrm{Unif}[t_\beta (1-\delta), t_\beta (1 + \delta)]$, where $\delta$ controls the disorder strength and the tuple $\{t_\beta\}$ corresponds to a point in \Insulator$_\alpha$. \\
If the disorder strength $\delta$ is sufficiently weak such that
\begin{equation}
    \kappa = \frac{t_\alpha^2}{\sum_{\beta \neq \alpha} t_\beta^2} \left(\frac{1 - \delta}{1 + \delta} \right)^2 > 1,    
\end{equation}
then for each disorder realization $\{ t^{(I)}_\alpha \}$, the relation
\begin{equation}
    |t^{(I)}_\alpha|^2 > \kappa \sum_{\beta \neq \alpha} |t^{(I)}_\beta|^2 \label{eqn:SM_ew4}
\end{equation}
holds for all sites $I \neq I_0$. With this bound, we find that the arguments presented in previous sections readily generalize. For example \citeeq{eqn:SM_ew1} now reads
\begin{equation}
\begin{split}
    \sum_{\beta_3 \neq \alpha} |\psi_{[\beta \beta_1 \alpha \beta_3]}|^2 &> \frac{|t^{(I)}_\alpha|^2}{\sum_{\beta \neq \alpha} |t^{(I)}_\beta|^2} |K'|^2 \\
    &> \kappa ~ |K'|^2
\end{split}
\end{equation}
and \citeeq{eqn:SM_ew2} generalizes to 
\begin{equation}
    |K'|^2 < \kappa^{-(n-1)}.
\end{equation}
Following a very similar chain of arguments, we obtain a lower bound for the normalization constant 
\begin{equation}
   |K|^2 > \left[ 1 + \frac{\kappa}{\kappa - 1} \frac{t^{(I_0)}_\beta}{t^{(I_0 + \beta)}_\alpha} \right]^{-1}  
\end{equation}
suggesting a finite weight at $I_0$ of the edge-mode wavefunction, even after disorder averaging! Furthermore, there is a unique edge mode for each disorder realization, reaffirming the fact that the edge mode is topological in character. \\

Similar results can be obtained for other distributions of disorder realizations, with the key requirement being \citeeq{eqn:SM_ew4} holds for each disorder realization for some $\kappa > 1$.

\section{Details of filling anomaly calculations}
\label{SM:filling_anomaly}

\begin{figure*}
    \includegraphics[width = \textwidth]{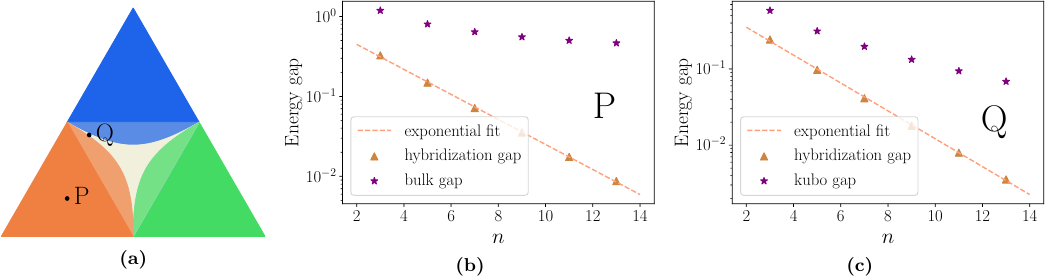}
    \caption{\textbf{(a)} Location of points P: (0.816, 0.408, 0.408) and Q: (0.667, 0.333, 0.667) on the phase diagram. \textbf{(b)} Demonstration of filling anomaly in the gapped phase at P. \textbf{(c)} Demonstration of filling anomaly in the gapless phase at Q.}

    \label{fig:SM_finsize}
\end{figure*}

Subsystems $S_n$ and $S_n^*$ defined in \ref{SM:subsystems} are used to demonstrate filling anomaly in the system. 

\subsection{Gapped phases}
For point P of the \Insulator$_1$ phase (see \figref{fig:SM_finsize}\textbf{(a)}), $S_n$ is expected to harbor zero-energy edge modes at the sites $I_0$ and $I_n$ as $n \rightarrow \infty$. In contrast, $S_n^*$ is expected to be fully gapped, even in the thermodynamic limit. For finite values of $n$, the two edge-modes of $S_n$ are expected to hybridize, and showcase a hybridization gap which goes to $0$ as $n \rightarrow \infty$. We compare this hybridization gap with the excitation gap of $S_n^*$, which serves as a proxy for the bulk gap. \figref{fig:SM_finsize}\textbf{(b)} confirms our expectations, and suggests an exponential decay of the hybridization gap with $n$. \\

This exponential decay can be understood on analytic grounds making use of the exact form of the edge mode wavefunction developed in \ref{SM:edge_mode}. The hybridization gap, to the leading order, is proportional to the overlap between the edge-mode wave-functions of $I_0$ and $I_n$, which goes as $ \sim (t_2/t_1)^{n/2} = 0.71^{n}$. This matches well with the numerically determined behaviour of the hybridization gap $\sim 0.70^{n}$.

\subsection{Gapless phases}
The point Q of the \Metal$_0$ phase (see \figref{fig:SM_finsize}\textbf{(a)}) is gapless, and thus, the excitation gaps of both $S_n$ and $S_n^*$ are expected to vanish in the limit $n \rightarrow \infty$. Despite this, we expect edge modes hosted by $I_0$ and $I_n$ to manifest in the spectrum owing to the decoupled limit where $t_2 \rightarrow 0$. To make this notion precise, we expect the hybridization gap of these edge modes (owing to finite separation between $I_0$ and $I_n$) to be much smaller than the Kubo gap of the bulk, ie, the spacing of energy states in the bulk in a finite system. Using the excitation gaps of $S_n$ and $S_n^*$ to represent the hybridization and Kubo gaps respectively, filling anomaly is equivalent to  

\begin{equation}
    \frac{\textrm{Excitation gap of } S_n}{\textrm{Excitation gap of } S_n^*} \rightarrow 0 ~~~~ \textrm{as } n \rightarrow \infty.
\end{equation}
\figref{fig:SM_finsize}\textbf{(c)} demonstrates this feature. \\

Although hybridization gap displays an exponential decay with $n$, it goes as $\sim 0.66^{n}$ instead of $\sim (t_2/t_1)^{n/2} = 0.71^{n}$ as a consequence of gaplessness of the bulk itself.

\subsection{Caveats}
The feature of filling anomaly in the $S_n$ system in the gapless phases is a consequence of its design. 
Observation of filling anomaly in gapless phases rests on the approximate `radial symmetry' of the $S_n$ system, as the length of the SSH chain separating $I_0$ and $I_n$ is of the same order of magnitude as the length of the longest chain in the system. \\
This can be motivated from the expected scaling of Kubo and hybridization gaps with system size. While the hybridization gap is expected to go down exponentially with the separation between $I_0$ and $I_n$ sites, the Kubo gap is instead expected to go as (length of longest metallic chain)$^{-1}$. For the point Q, the longest metallic chain refers to the longest chain of the form $[13 \ldots 31]$ or $[31 \ldots 13]$. From these scaling relationships, we can see that in the limit of large system size, the hybridization gap is expected to go to $0$ faster than the Kubo gap when $(\textrm{separation between $I_0$ and $I_n$}) \sim (\textrm{length of longest metallic chain})$, producing a filling anomaly. \\
For arbitrary subsystems there are several subtleties in taking the limit of large system sizes, which are avoided by the $S_n$ system.

\section{Winding characteristic}
\label{SM:winding_characteristic}

The sign functions $\sigma_\alpha(z)$ were introduced in \ref{SM:GreenFn_analytic} to account for the multi-valued nature of the complex $\sqrt{...}$ function in \citeeq{eqn:SM_Galpha}, which possesses a pair of branch points ($G = \pm i/2 t_\alpha$). In \ref{SM:band_properties}, the tuple $\left( \sigma_\alpha(0^+) \right)$ was found to characterize different gapped phases, as well as the different metallic regions $M_\alpha$. In section \ref{SM:correlation_functions}, we discussed how $G(\omega^+)$ winds about the branch point $ G = - \ci / 2t_\alpha $ when $\sigma_\alpha(0^+) = -1$, suggesting a winding characteristic. \\
However, in the $G$ plane, the branch point $- \ci / 2 t_\alpha$ is a quantity that is informed by the hamiltonian, and thus winding about this point cannot be computed from the many body ground state alone, and sensitively depends on the model parameters. To remedy this, we instead consider the parameter-independent dimensionless quantity $W_\alpha(z) = G/G_\alpha$. From \eqref{eqn:SM_Galpha}, this quantity can be written in the form
\begin{equation}
W_\alpha(z) = \frac{1 + \sigma_\alpha(z) \sqrt{1 + 4 t^2_\alpha G(z)^2}}{2} .\label{eqn:SM_Walpha}
\end{equation}
The branch point thus corresponds to $W_\alpha = 1/2$, with the curve $W_\alpha(\omega^+)$ exhibiting different winding characteristics about this point depending on the tuple $\sigma_\alpha(0^+)$, as shown in \figref{fig:M1}(c) of the main-text. 

Discontinuities in $\sigma_\alpha(\omega^+)$ are associated with crossing of the branch cut associated with the radical $\sqrt{1 + 4 t^2_\alpha G(z)^2}$, and manifests as a winding in the complex $W_\alpha$ plane about $1/2$. In \textbf{M}$_0$, $\sigma(\omega^+) = +1 ~~~ \forall \omega$, and thus none of the $W_\alpha(\omega^+)$ wind about the point $1/2$. In contrast, in \textbf{M}$_\alpha$ and \textbf{I}$_\alpha$, $W_\alpha(\omega^+)$ winds a single time about the point $1/2$, while all the other functions $W_\beta(\omega^+)$ for $\beta \neq \alpha$ do not wind about $1/2$  in the complex plane.

Using \eqref{eqn:GalphaG}, one can relate $W_\alpha$ to the two point function between nearest neighbours separated by an $\alpha$ link $G_{I, I + \alpha} (z) = -t_\alpha G_\alpha(z) G(z), $

\begin{equation}
G_{I, I+\alpha}(z) = \frac{1 - W_\alpha}{t_\alpha}.
\end{equation}
In the gapped phase \textbf{I}$_\alpha$, $G(0) = 0$, which leads to $W_\alpha = 0$ and $W_\beta = 1 ~~ \forall \beta \neq \alpha$ at $z = 0^+$. This manifests itself itself in the static response function, as $W_{\beta} = 1 \implies G_{I, I+\beta}(0^+) = 0 $ (No long-time correlation across weak links), and $W_{\beta} = 0 \implies G_{I, I+\alpha}(0^+) = 1/2 t_\alpha $ (large long-time correlation across strong links). \\

In the gapless phase, the zero-frequency response becomes non-zero  $ \forall \beta$, but the winding characteristic instead manifests as quantitative bounds on the response functions
\begin{align}
\sigma_\alpha(0^+) = -1 & \implies W_\alpha (0^+) <	 1/2 
 \\ &\implies  G_{I, I+\alpha}(0^+) > 1/2 t_\alpha, \\
\sigma_\alpha(0^+) = +1 & \implies W_\alpha (0^+) >	 1/2 \\ &\implies  G_{I, I+\alpha}(0^+) < 1/2 t_\alpha.
\end{align}

Another way in which the winding characteristic plays a role in the metallic phases is through the long-distance correlation functions and their respective functional forms, which we discuss in section
\ref{SM:correlation_lengths}.

\section{properties of correlation functions}
\label{SM:correlation_lengths}

\subsection{Sum rule constraint}
Consider an arbitrary non-interacting fermionic system, with single-orbitalled sites $I \in \{1, 2, \cdots , N_s \}$, and corresponding creation and annihilation operators $c_I, c_I^\dagger$ satisfying the algebra $\{c_I, c_J \} = 0, \{c^\dagger_I, c_J \} = \delta_{IJ}$. The corresponding number operators are denoted $n_I = c^\dagger_I c_I$ with the total particle number denoted as $\hat{N} = \sum_I n_I$. \\

For a system with fixed number of particles $\hat{N} = N$, we have the identity $\langle \hat{N} n_J \rangle = \langle \hat{N} \rangle \langle n_J \rangle $. Expanding $\hat{N}$, we obtain

\begin{equation}
	\sum_I \langle n_I n_J \rangle = \sum_I  \langle n_I \rangle \langle n_J \rangle.
\end{equation}
For a non-interacting hamiltonian, employing Wick's theorem for the ground state yields the sum rule
\begin{equation}
	\sum_I | \langle c^\dagger_I c_J \rangle |^2 = \langle n_J \rangle. \label{eqn: gen_sumrule}
\end{equation}
The only assumptions employed in obtaining the above were: (i) correlation functions are evaluated over ground state of non-interacting hamiltonian (ii) finite system (iii) fixed number of particles. In order to extend the applicability of the sum rule to infinite systems, one needs to work with a definite filling $f$ instead of a definite number $N$, and \eqref{eqn: gen_sumrule} holds as-is.\\

The symmetries of the problem imply that the correlation functions $\langle c^\dagger_I c_{I_0} \rangle$ depend only on the generation in which $I$ belongs to, i.e.,
\begin{equation}
	\langle c^\dagger_I c_{I_0} \rangle \equiv C_{p}( g(I) ).
\end{equation}
We apply the sum-rule for $B(p)$ with  $t_\alpha = 1/\sqrt{p}$, letting $J$ be the root site $I_0$ and $\langle n_{J} \rangle = 1/2$ (half-filling) to yield
\begin{equation}
	\sum_{g = 1}^{\infty} p (p-1)^{g - 1}  |C_p(g)|^2 = 1/4.
\end{equation}
Requiring the LHS to converge, we get
\begin{equation}
	\lim_{g \rightarrow \infty} \left[ \frac{C_p(g+1)}{C_p(g)} \right]^2 \leq \frac{1}{p - 1}.
\end{equation}

Thus, even for the gapless, highly symmetric parameter choice for $B(p)$ ($p \geq 3$), the correlation function is constrained to decay as (or faster than) $(p-1)^{-g/2}$. Let us posit $C(g)$ to be well described by an exponential for large $g$ ~ $C(g) \sim e^{-g/\xi}$, then the ratio test in fact implies a bound $\xi \leq 2/ \log(p-1)$ on the correlation length. This is a departure from conventional models on a lattice, where gapless phases are generically associated with long range correlations which decay as power laws.
This suggests that generic points in the parameter space of $B(p)$, gapless or not, are expected to possess exponentially decaying correlation functions. We find this expectation to be true, with the key exception being the critical points C$_{IJ}$.
\subsection{Formulation for asymptotic behaviour}
The integral form of the correlation function \citeeq{eqn:corrfn_intform} reads

\begin{equation}
C(\{ n_\alpha \}) = \Im ~ \ci^{\Sigma_\alpha n_\alpha} \int dg f(g) ~~ \prod_\alpha g_\alpha^{n_\alpha} .
\end{equation}
Where the value of $\sigma_\alpha$ and the domain of integration depend on the region of parameter space as discussed in \ref{SM:correlation_functions}. The integrand $\prod_\alpha g_\alpha^{n_\alpha}$ is real and strictly positive, which suggests that in the limit $\sum_\alpha n_\alpha \gg 1$, the largest contribution to the integral comes from where the integrand is maximized. Corrections due to the width of the peak contribute a factor of either $N^{-1}$ or $N^{-1/2}$ depending on whether the first derivative vanishes at the peak or not, but the exponential behaviour is captured by
\begin{equation}
C(\{ n_\alpha \} ) \sim  \left(\Im ~ \ci^{\Sigma n_\alpha} \right) ~~ \max_g  \prod_\alpha g_\alpha(g)^{n_\alpha}, \label{eqn: AsymptoticCorrelation}
\end{equation}
where the maximum is taken over $g$ belonging to respective domain of integration. It is useful to define a correlation length $\xi$ for infinite chains as

\begin{align}
	C(\{ n_\alpha \}) &\sim \exp \left(- L / \xi(\{ n_\alpha \}) \right) ~ (L \rightarrow  \infty), \\
	\implies \xi(\{ n_\alpha \}) &= - \lim_{L \rightarrow \infty} \frac{L}{\log |C(\{ n_\alpha \})|}, 
\end{align}
where $L = \sum_\alpha n_\alpha$ is the length of the chain across which correlation is computed, and the limit of $L \rightarrow \infty$ is taken in a manner such that the limit $\lim_{L \rightarrow \infty} n_\alpha/L$ converges to the desired finite value.

\subsection{\textbf{M}$_0$ region}
\textbf{M}$_0$ is characterized by $\sigma_\alpha(0^+) = +1 ~ \forall \alpha$, and the domain of integration over $g$ is $[0, \pi \rho_0]$. In this domain, $g_\alpha(g) \neq 1$ is a strictly increasing function in $g$ and thus the maximum value of the integrand must occur at the end-point of the domain (i.e. at $\rho_0$). Therefore, we have
\begin{align}
C(\{ n_\alpha \} ) &\sim \prod_\alpha [ g_\alpha( 2 \pi \rho_0) ]^{n_\alpha} \\
				&\sim \prod_\alpha \left( \frac{2 \pi t_\alpha \rho_0}{1 + \sqrt{1 - (2 \pi t_\alpha \rho_0)^2}} \right)^{n_\alpha}.
\end{align}
This is an explicit demonstration of the exponential nature of correlation functions in the $M_0$ region, and allows us to define emergent length-scales $\xi_\alpha$ such that
\begin{align}
C(\{ n_\alpha \} ) &\sim \exp\left( - \Sigma_\alpha n_\alpha / \xi_\alpha \right), \\
\xi_\alpha &= \left[ \log \frac{1 + \sqrt{1 - (2 \pi t_\alpha \rho_0)^2}}{2 \pi t_\alpha \rho_0} \right]^{-1}.
\end{align} 
The $p$ values of $\xi_\alpha$ completely determine the long-range correlation function in this phase. It is interesting to note that the functional form of $C(\{ n_\alpha \} )$ does not distinguish between different link color indices $\alpha$, and has the same functional form for each. This will fail in the $I_\alpha$ and $M_\alpha$ regions, where the form of the correlation functions will explicitly distinguish between the strong link $t_\alpha$ and the weak links $t_\beta, ~\beta \neq \alpha$. \\ 
The lack of distinction between $t_\alpha$'s also manifests as an emergent long-range property of the system
\begin{equation}
	C(\{ n_\alpha + n_\alpha' \} ) \sim C(\{ n_\alpha \} ) ~ C(\{ n_\alpha' \} ) 
\end{equation}
for $n_\alpha, n_\alpha' \gg 1$. This suggests that all the points in the $M_0$ region correspond to a single phase, characterized by a metallic density of states and long range correlation functions of the described analytic form. 

\subsection{\textbf{I}$_\alpha$ region}
Before discussing the \textbf{M}$_\alpha$ region of the phase diagram, it is helpful to understand the \textbf{I}$_\alpha$ region. Here, an explicit computation of the form of the correlation function is presented specifically for weak chains and strong chains, as it becomes possible to determine the maximum in \eqref{eqn: AsymptoticCorrelation} analytically, and intermediate chains are discussed briefly. The domain of integration in this phase is $[0,1/2t_\alpha]_{\sigma_\alpha = +1} \cup [0,1/2t_\alpha]_{\sigma_\alpha = -1}$, with $\sigma_\beta = +1 ~~ \forall \beta \neq \alpha$ for both the segments in the domain. \\

Weak chains (i.e. chains of length $L$ of the form $[\beta_1 \beta_2 \cdots \beta_l \cdots \beta_L]$ with $\beta_l \neq \alpha ~ \forall l$) are characterized by $n_\alpha = 0$. Since $g_{\beta}(g)$ is a strictly increasing function of $g$ in the domain $[0, 1/2t_\alpha]$ regardless of the value taken by $\sigma_\alpha$, the integrand is guaranteed to be maximized at the point $g = 1/2t_\alpha$, which allows us to write the correlation function along the weak chains as

\begin{equation}
	C_{\textrm{weak}}(\{ n_\beta \}) \sim \prod_{\beta \neq \alpha} \left( \frac{t_\beta/t_\alpha}{1 + \sqrt{1 - (t_\beta/t_\alpha)^2}} \right)^{n_\beta}.
\end{equation}
This suggests defining emergent length scales $\xi^{\textrm{weak}}_\beta$ associated with weak chains such that $C_{\textrm{weak}}(\{ n_\beta \}) \sim \exp \left( - \sum_{\beta \neq \alpha} n_\beta/\xi^{\textrm{weak}}_\beta \right)$, which are given by
\begin{equation}
\xi^{\textrm{weak}}_\beta = \left[ \log \frac{1 + \sqrt{1 - (t_\beta / t_\alpha)^2}}{t_\beta/ t_\alpha} \right]^{-1}.
\end{equation}

Strong chains (i.e. chains of length $2L+1$ of the form $[\alpha \beta_2 \alpha \cdots \beta_{2L} \alpha]$) are instead characterized by $n_\alpha = 1 + \sum_{\beta \neq \alpha} n_\beta$. $ g_{\alpha}(g) g_{\beta}(g), ~~ \beta \neq \alpha$ can be demonstrated to be strictly increasing on the segment $[0,1/2 t_\alpha]$ with $\sigma_\alpha = +1$, and strictly decreasing on the segment $ [0,1/2 t_\alpha]$ with $\sigma_\alpha = -1$, implying that the maximum of the integrand is achieved at $I_0$ with $\sigma_\alpha = -1$ and thus
\begin{equation}
	C_{\textrm{strong}}(\{ n_\beta \}) \sim \prod_{\beta \neq \alpha} \left( - \frac{t_\beta}{t_\alpha} \right)^{n_\beta}.
\end{equation}
One can define emergent length scales $\xi^{\textrm{strong}}_\beta$ associated with the strong chains analogously using $C_{\textrm{strong}}(\{ n_\beta \}) \sim \exp \left( - \sum_{\beta \neq \alpha} 2 n_\beta / \xi^{\textrm{strong}}_\beta \right)$ (factor of $2$ has been introduced to account for the fact that length of a strong chain is $2 \sum_{\beta \neq \alpha} n_\beta  + 1$),
\begin{equation}
	\xi^{\textrm{strong}}_\beta = 2 ~ \left[ \log \frac{t_\alpha}{t_\beta} \right]^{-1}.
\end{equation}
It is interesting to note that within the $I_\alpha$ phase, all the weak links enter in the same form into the correlation function. That is, a distinction between $t_\beta$ for $\beta \neq \alpha$ isn't made in the functional forms explicitly, and manifests as

\begin{align}
C_{\textrm{weak}}(\{ n_\beta + n_\beta' \} ) & \sim C_{\textrm{weak}}(\{ n_\beta \} ) ~ C_{\textrm{weak}}(\{ n_\beta' \} ), \\
C_{\textrm{strong}}(\{ n_\beta + n_\beta' \} ) & \sim C_{\textrm{strong}}(\{ n_\beta \} ) ~ C_{\textrm{strong}}(\{ n_\beta' \} ).
\end{align} 
For generic chains (strong, weak or intermediate), a closed form is unavailable, but the procedure for obtaining the max value of integrand is similar. In each case, $g_\beta(g)$ is a strictly increasing function of $g$ for all $\beta$ on the segment $[0, 1/2t_\alpha]$ with $\sigma_\alpha = +1$, implying that the maximum must be achieved on the segment where $\sigma_\alpha = -1$. It can be shown that on the segment with $\sigma_\alpha = -1$, the integrand $\prod_{\beta} g_\beta^{n_\beta}$ achieves a unique global maximum at some value of $g$, which we denote as $g_*(\{n_\beta \}, \{ t_\beta \})$. As we have seen, for all choices of weak chains we have $g_* = 1/2 t_\alpha$, while for all choices of strong chains $g_* = 0$. For intermediate chains, expanding the integrand about the points $g = 0$ and $g = 1 / 2 t_\alpha$ leads to the conclusion that $g_* \in (0,1/2t_\alpha)$ and can thus be obtained using
\begin{equation}
	0 = \frac{d}{dg} \prod_\beta g_\beta(g)^{n_\beta} |_{g = g_*, \sigma_\alpha = -1}. \label{eqn:gstar}
\end{equation}
Once $g^*$ is obtained, the correlation function for an arbitrary chain is given by
\begin{widetext}
\begin{equation}
C(\{ n_\beta \}) \sim \left(  \frac{1 + \sqrt{1 - (2 g_*  t_\alpha)^2}}{2 g_* t_\alpha} \right)^{n_\alpha} \prod_{\beta \neq \alpha} \left( \frac{2 g_* t_\beta}{1 + \sqrt{1 - (2 g_* t_\beta)^2}} \right)^{n_\beta} ,
\end{equation}
\end{widetext}
where the dependence of $g_*$ on $\{n_\beta
 \}$ and $\{t_\beta \}$ has been suppressed. $g_*$ does not have a closed form expression in terms of the parameters, but can be computed numerically.
 
\subsection{\textbf{M}$_\alpha$ region}
In the \textbf{M}$_\alpha$ region, the functional form of the correlation function for weak chains turns out to be identical to that in the \textbf{I}$_\alpha$ phase, while the functional form of the correlation function of strong chains is analogous to that in \textbf{M}$_0$. This allows us to interpret \textbf{M}$_\alpha$ as an extended regime separating the \textbf{I}$_\alpha$ and \textbf{M}$_0$ phases. Correlation functions along intermediate chains reveal non-analyticities in the interior of the \textbf{M}$_\alpha$ region, signalling a cascade of subsystem transitions in going from the \textbf{I}$_\alpha$ to the \textbf{M}$_0$ phase. \\

The domain of integration in \textbf{M}$_\alpha$ consists of two segments, $[0,1/2t_\alpha]_{\sigma_\alpha = +1}$ and $[\pi \rho_0,1/2t_\alpha]_{\sigma_\alpha = -1}$, with $\sigma_\beta = +1 ~~ \forall \beta \neq \alpha$ for both the segments. $g_\beta$ is a strictly increasing function in the $\sigma_\alpha = +1$ segment for all values of $\beta$. The integrand is thus maximized in the $\sigma_\alpha = -1$ segment, and one can use monotonicity arguments to obtain analytic forms for weak and strong chains. \\

For $\beta \neq \alpha$, $g_\beta(g)$ is a strictly increasing function of $g$ on the segment $[\pi \rho_0,1/2t_\alpha]_{\sigma_\alpha = -1}$, which implies a maxima in the integrand for $C_\textrm{weak}$ is achieved at $g = 1/2t_\alpha$. Similarly, $g_\alpha(g) g_\beta(g)$ is a strictly decreasing function of $g$ on the segment $[\pi \rho_0, 1/2t_\alpha]_{\sigma_\alpha = -1}$, which implies a maxima in the integrand for $C_\textrm{strong}$ is achieved at $g = \pi \rho_0$.

\begin{align}
	C_{\textrm{weak}}(\{ n_\beta \}) &\sim \prod_{\beta \neq \alpha}  \left( \frac{t_\beta/t_\alpha}{1 + \sqrt{1 - (t_\beta/t_\alpha)^2}} \right)^{n_\beta}, \\
	C_{\textrm{strong}}(\{ n_\beta \}) &\sim \prod_{\beta \neq \alpha} \left( - \frac{t_\beta}{t_\alpha} ~ \frac{1 + \sqrt{1 - (2 \pi \rho_0 t_\alpha)^2}}{1 + \sqrt{1 - (2 \pi \rho_0 t_\beta)^2}} \right)^{n_\beta} .
\end{align}
The emergent length scales associated with these can be defined analogously

\begin{align}
	\xi^{\textrm{weak}}_\beta &= \left[ \log \frac{1 + \sqrt{1 - (t_\beta/t_\alpha)^2}}{t_\beta/t_\alpha} \right]^{-1}, \\
	\xi^{\textrm{strong}}_\beta &= 2 \left[ \log \left( \frac{t_\alpha}{t_\beta} ~ \frac{1 + \sqrt{1 - (2 \pi \rho_0 t_\beta)^2}}{1 + \sqrt{1 - (2 \pi \rho_0 t_\alpha)^2}} \right) \right]^{-1}.
\end{align}
For large separation, results analogous to the $\mathbf{I}_\alpha$ region hold
\begin{align}
    C_{\textrm{weak}}(\{ n_\beta + n_\beta' \} )  &\sim C_{\textrm{weak}}(\{ n_\beta \} ) ~ C_{\textrm{weak}}(\{ n_\beta' \} ), \\
    C_{\textrm{strong}}(\{ n_\beta + n_\beta' \} )  &\sim C_{\textrm{strong}}(\{ n_\beta \} ) ~ C_{\textrm{strong}}(\{ n_\beta' \}).
\end{align}

For the intermediate chains, it is useful to consider the earlier introduced notion of $g_*(\{ n_\beta \}, \{ t_\beta \})$, which is the unique solution to \eqnref{eqn:gstar} in the domain $g \in [0,1/2t_\alpha]$. From uniqueness of $g_*$, we can assert that the integrand $\prod_\beta g_\beta(g)^{n_\beta}$ with $\sigma_\alpha = -1$ must be strictly increasing in the interval $[0,g_*]$ and strictly decreasing in $[g_*, 1/2 t_\alpha]$. Thus, for the case of \textbf{M}$_\alpha$ region the maximum value of the integrand depends on the precise relationship between $g_*$ and $\pi \rho_0$, with the distinction arising as
\begin{widetext}
\begin{align*}
C(\{ n_\beta \}) &\sim \left(  \frac{1 + \sqrt{1 - (2 g_*  t_\alpha)^2}}{2 g_* t_\alpha} \right)^{n_\alpha} \prod_{\beta \neq \alpha} \left( \frac{2 g_* t_\beta}{1 + \sqrt{1 - (2 g_* t_\beta)^2}} \right)^{n_\beta} & \mathrm{if} ~~ g_* \geq \pi \rho_0, \\
& \sim \left(  \frac{1 + \sqrt{1 - (2 \pi \rho_0  t_\alpha)^2}}{2 \pi \rho_0 t_\alpha} \right)^{n_\alpha} \prod_{\beta \neq \alpha} \left( \frac{2 \pi \rho_0 t_\beta}{1 + \sqrt{1 - (2 \pi \rho_0 t_\beta)^2}} \right)^{n_\beta} & \mathrm{if}~~ g_* \leq \pi \rho_0.
\end{align*}
\end{widetext}

For any given chain $\{ n_\beta \}$, the quantities $g_*(\{ n_\beta \})$ and $\rho_0$ are expected to depend continuously on the parameters $t_\beta$. Since $\rho_0 = 0 \leq g_*(\{ n_\beta \})$ for points in parameter space on the critical curve $\mathcal{C}_\alpha$, and $\rho_0 = 1/2 t_\alpha \geq g_*(\{ n_\beta \})$ for points on the \textbf{M}$_\alpha$ - \textbf{M}$_0$ boundary, continuity implies the existence of a locus of points separating the $I_\alpha$ and $M_0$ phases where we have $ g_*\{ n_\beta \} =  \pi \rho_0$. Each point on this locus is associated with a non-analyticity in $\xi_{n_\alpha}$, indicated by the change in the functional form of $ C(\{ n_\beta \}) $. We refer to this non-analyticity as the subsystem transition associated with the chain $\{ n_\beta \}$, which suggests that the $M_\alpha$ region consists of multiple gapless phases distinguished by their correlation functions. \\
Since the above argument holds for any choice of chain $\{ n_\beta \}$, we conclude that in going from any point in \textbf{I}$_\alpha$ to any point in \textbf{M}$_0$, one encounters a cascade of subsystem transitions in the \textbf{M}$_\alpha$ region, where the correlation functions undergo a change in their functional form. This is a generic feature for arbitrary $p \geq 3$, and is expected to hold in a broader class of arboreal systems. Numerical investigation of these subsystem transitions further suggests that each point in \textbf{M}$_\alpha$ lies on the subsystem-transition boundary for some chain, indicating that each point $M_\alpha$ displays a non-analyticity in some long-range two point function.

\section{Critical properties}
\label{SM:critical_properties}

In this section we analyse various aspects of the model on and near the $\mathcal{C}_\alpha$ surface, and explore a collection of critical exponents characterizing the transition. Since the expression for phase boundary is given by
\begin{equation}
	t_\alpha^2 = \sum_{\beta \neq \alpha} t_\beta^2,
\end{equation}
we find it useful to define the distance to the transition as $D^{(2)}_\alpha = - t_\alpha^2 + \sum_{\beta \neq \alpha} t_\beta^2$. To distinguish points on $\mathcal{C}_\alpha$ from points C$_{\alpha \beta}$, we introduce an additional notion of distance $D^{(4)}_\alpha = - t_\alpha^4 + \sum_{\beta \neq \alpha} t_\beta^4$. 

\subsection{Density of states}
In our new notation, \eqref{eqn:z_G_expansion} for \textbf{I}$_\alpha, \mathcal{C}_\alpha$ can be written as
\begin{equation}
z = D^{(2)}_\alpha G - D^{(4)}_\alpha G^3 + \mathcal{O}(G^5). \label{eqn: new_z_G_expansion}
\end{equation}
Points on $\mathcal{C}_\alpha$ have $D^{(2)}_\alpha = 0$, which implies that as $z \rightarrow 0$,
\begin{equation}
	G(z) = [ - z / D^{(4)}_\alpha ]^{1/3} + \cdots \label{eqn: G_critical}
\end{equation}
Taking the appropriate branch cut for the complex cube root function under the consistency condition $\Im G(z) \leq 0$, we obtain the leading order behaviour of the density of states,
\begin{equation}
	\rho(\omega) = \frac{\sqrt{3}}{2 ~ |D^{(4)}_\alpha|^{1/3}} ~~ |\omega|^{1/3} + \cdots ,
\end{equation}
where the exponent $1/3$ is a characteristic of all points on the critical surface $\mathcal{C}_\alpha$.

\subsection{Gap closing}
For a point in \Insulator$_\alpha$ close to the critical surface, \eqref{eqn: new_z_G_expansion} holds, and one can obtain the gap  $\varepsilon_g$ as a function of parameters, by first finding $G_g = G( \varepsilon_g / 2)$. From the subsection on band-edges in \ref{SM:band_properties}, we have
\begin{equation}
\frac{dz}{dG} \Big|_{G = G_g} = 0 \implies 0 = D^{(2)}_\alpha - 3 D^{(4)}_\alpha G_g^2  + \cdots .
\end{equation}
Substituting the leading order term of $G_g$ back into the expression for $z(G)$ \citeeq{eqn:G}, we obtain
\begin{equation}
\varepsilon_g = \frac{2}{\sqrt{3 |D^{(4)}_\alpha}|} ~~ | D^{(2)}_\alpha |^{3/2} + \cdots .
\end{equation}
Thus, the band gap scales as the distance to transition raised to the power $3/2$, and defines another characteristic exponent.

\subsection{Density of states at chemical potential}
As we approach the critical curve $\mathcal{C_\alpha}$ from the metallic side, we expect the density of states at the chemical potential to vanish with a certain critical exponent. To determine this, we expand \eqref{eqn: dos_0} in powers of $\rho_0$ for metallic points close to $\mathcal{C_\alpha}$ and obtain
\begin{equation}
	\rho_0 = \frac{1}{\pi} \left[ - \frac{D^{(2)}_\alpha}{D^{(4)}_\alpha} \right]^{1/2} + \cdots . \label{eqn: rho_0_critical}
\end{equation}
Thus, the density of states at the chemical potential approaches $0$ as the square-root of the distance to transition.

\subsection{Fate of the edge mode}
Substituting \citeeq{eqn: G_critical} into \citeeq{eqn:SM_Galpha} for a point on the critical surface $\mathcal{C}_\alpha$, we get an expansion 
\begin{align}
	G_\beta (z) & = [ - z / D^{(4)}_\alpha ]^{1/3} + \cdots & (\beta \neq \alpha), \\
	& = t_\alpha^{-2} [ - z / D^{(4)}_\alpha ]^{-1/3} + \cdots & (\beta = \alpha).
\end{align}
Using \eqref{eqn: GammaAlpha}, we conclude that on the curve $\mathcal{C}_\alpha$, the melting of the edge mode manifests itself as
\begin{align}
- \frac{\Im ~ \Gamma_\beta}{\pi} (\omega^+) &= \frac{\sqrt{3}}{2 \pi} t^{-2}_\beta \frac{|D^{(4)}_\alpha|^{1/3}}{|\omega|^{1/3}} & (\beta \neq \alpha), \\
&= \frac{\sqrt{3}}{2 \pi} \frac{|\omega|^{1/3}}{|D^{(4)}_\alpha|^{1/3}} & (\beta = \alpha).
\end{align}
Thus the pole singularity of the gapped phase has melted into the $z^{-1/3}$ singularity on the critical surface.

\subsection{Correlation function at finite length}
In section \ref{SM:correlation_lengths} we determined the non-analyticities in the long range correlation functions encountered in the \textbf{M}$_\alpha$ region, which was made precise using the notion of a correlation length. In this section, we instead look at the correlation function for given points $I \neq J$, and the non-analyticities that can be shown by this finite-length correlation function. \\

Let the shortest path from $I$ to $J$ be of length $L = 2l + 1$ and be described by the sequence of links $[\beta_1 \beta_2 \cdots \beta_{2l + 1}]$ Then the correlation function can be expressed in an integral form \citeeq{eqn:corrfn_intform},

\begin{equation}
	C_{IJ} = (-1)^{l} ~ \int dg f(g) ~~ \prod_{\beta} g_\beta(g)^{n_\beta}.
\end{equation}
The integration measure $f(g)$ along with $g_\beta(g)$ are smooth functions of the parameters $\{ t_\alpha \}$, and hence cannot give rise to a singularity in the correlation function. Instead, the domain of integration carries a dependence on the parameters, as $\rho_0$ explicitly appears as a limit of integration in \Metal. From \eqref{eqn: dos_0}, one expects $\rho_0$ to be a smooth function of the parameters within the gapped and metallic phases, but features a notable singularity at the critical surface $\mathcal{C}_\alpha$ separating them \citeeq{eqn: rho_0_critical}, where $\rho_0 \sim \sqrt{ D_\alpha^{(2)} }$. \\

We focus on the points in the \Metal$_\alpha$ region close to the boundary $\mathcal{C}_\alpha$, and separate the correlation function into two parts
\begin{equation}
	C_{IJ} = (-1)^l C_{IJ}^{\textrm{singular}} + (-1)^l C_{IJ}^{\textrm{smooth}},
\end{equation}
where
\begin{widetext}
\begin{align}
C_{IJ}^{\textrm{singular}} &= - \int dg f(g) ~~ \prod_\beta g_\beta^{n_\beta} & \textrm{over domain: } g \in [0, \pi \rho_0]_{\sigma_\alpha = -1} \\
C_{IJ}^{\textrm{smooth}} &= \int dg f(g) ~~ \prod_\beta g_\beta^{n_\beta} & \textrm{over domain: } g \in [0,1/2 t_\alpha]_{\sigma_\alpha = +1} \cup [0,1/2 t_\alpha]_{\sigma_\alpha = -1}
\end{align}
\end{widetext}

As $D^{(2)}_\alpha \rightarrow 0$, one approaches the critical surface $\mathcal{C}_\alpha$, and $\rho_0 \sim [D^{(2)}_\alpha]^{1/2} \rightarrow 0$. This allows us to expand $C_{IJ}^{\textrm{singular}}$ in powers of $D^{(2)}_\alpha$. $f(g)$ and $g_\beta(g)$ are expanded about $g = 0$, and each term is integrated from $0$ to $\pi \rho_0$ to obtain an expansion in $\rho_0$. Next, \eqref{eqn: rho_0_critical} is used to obtain an expansion in powers of $D^{(2)}_\alpha$. \\
For the chain in question, we find that if the integer $m \equiv 1 - n_\alpha + \sum_{\beta \neq \alpha} n_\beta$ satisfies $m > 0$, then
\begin{widetext}
\begin{equation}
C_{\textrm{singular}} = \theta({D^{(2)}_\alpha}) \left[ \frac{1}{\pi} \left( \frac{1}{m + 1} - \frac{3}{m + 3} \right) \left( t_\alpha^{-n_\alpha} \prod_{\beta \neq \alpha} t_\beta^{n_\beta} \right) ~ |D^{(4)}_\alpha|^{\frac{-1 - m}{2}} \right] |D^{(2)}_\alpha|^{\frac{1 + m}{2}} + \cdots .
\end{equation}
\end{widetext}
In the case where $m = 0$ (correlation length along a strong chain), a cancellation occurs and we obtain the expression
\begin{widetext}
\begin{equation}
C_{\textrm{singular}} = \theta({D^{(2)}_\alpha}) \left[ \frac{4}{ 15 \pi} \left( \sum_\alpha n_\alpha t_\alpha^2 \right)  \left( t_\alpha^{-n_\alpha} \prod_{\beta \neq \alpha} t_\beta^{n_\beta} \right) ~ |D^{(4)}_\alpha|^{- \frac{3}{2}} \right] |D^{(2)}_\alpha|^{\frac{5}{2}} + \cdots .
\end{equation}
\end{widetext}
This non-analyticity in the correlation function can be seen to diminish as the length of the chain $L \rightarrow \infty$. For large $L$, the ideas discussed in \ref{SM:correlation_lengths} regarding subsystem transitions are applicable.

\subsection{Entanglement spectrum}
Since the entanglement spectrum is determined by $\tilde{h}(k)$, which is essentially the Fourier transform of appropriate correlation functions, we expect that the features of the spectrum ( such as entanglement gap $\Delta_g$ ) will inherit non-analytic behaviour at the critical curves $\mathcal{C}_\alpha$ from the correlation functions $C_{IJ}$. For points close to the critical curve $\mathcal{C}_\alpha$, the correlation functions can be expanded in powers of $D^{(2)}_\alpha$ to obtain the form for the entanglement gap:

\begin{equation}
\Delta_g ( \{ t_\beta \} ) = \Delta^{\textrm{analytic}}_g ( \{ t_\beta \} ) + \theta(D^{(2)}_\alpha) ~ ( \cdots ) ~ |D^{(2)}_\alpha |^{\frac{5}{2}} + \cdots
\end{equation}

This is verified numerically, and can be seen in \figref{fig:critical} (e) of the main text

\section{Entanglement in manifold systems}
\label{SM:manifold_entanglement}
We consider a non-interacting, tight binding hamiltonian on a square lattice at half filling with parameters $t_1$ and $t_2$. The unit cell consists of four sites with nearest neighbour hoppings as shown in \figref{fig:SM_SSH2D}, which we number from $ a= 1$ (bottom left site of unit cell) to $4$ (proceeding counterclockwise). All hoppings are taken to be real, implying the hamiltonian lies in the BDI class. Labelling each site with the lattice vector $\vec{r} = (x, y)$, the hamiltonian is
\begin{figure}
    \centering
    \includegraphics{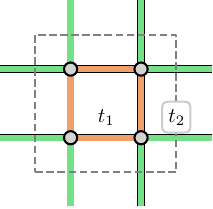}
    \caption{Unit cell for the square lattice model defined in \ref{SM:manifold_entanglement}.}
    \label{fig:SM_SSH2D}
\end{figure}
\begin{equation}
    \mathcal{H} = \sum_{\vec{\delta} \in  \{\pm \hat{x}, \pm \hat{y} \} } \sum_{a,b \in \{1,2,3,4\} } H_{ab}(\vec{\delta}) c^\dagger_{\vec{r} + \vec{\delta}, a} c_{\vec{r}, b}
\end{equation}
where the hopping matrix $H_{ab}(\vec{\delta})$ is as shown in \figref{fig:SM_SSH2D}. Fourier transforming to $\vec{k} = (k_x, k_y)$, we obtain
\begin{equation}
    \mathcal{H} = \sum_{\vec{k}} \sum_{a, b \in \{1, 2, 3, 4\} } H_{ab}(\vec{k}) c^\dagger_{\vec{k},a} c_{\vec{k},b}
\end{equation}
where
\begin{widetext}
\begin{equation}
    H(\vec{k}) = - \begin{pmatrix}
        0 & t_1 + t_2 e^{\ci k_x} & 0 & t_1 + t_2 e^{\ci k_y} \\
        t_1 + t_2 e^{-\ci k_x} & 0 & t_1 + t_2 e^{\ci k_y} & 0 \\
        0 & t_1 + t_2 e^{-\ci k_y} & 0 & t_1 + t_2 e^{-\ci k_x} \\
        t_1 + t_2 e^{-\ci k_y} & 0 & t_1 + t_2 e^{\ci k_x} & 0
    \end{pmatrix}.
\end{equation}
\end{widetext}
Subsequently, the entanglement hamiltonian for a SSH chain subsystem, which is specified by restricting to $a = 1$ and $a = 2$ sites of the unit cell with height variable $y = 0$, can be written in terms of its fermionic operators $d$, $d^\dagger$ as
\begin{equation}
    \tilde{\mathcal{H}} = \sum_{k_x} \sum_{a,b \in \{1, 2\} } \tilde{H}_{ab}(k_x)  d^\dagger_{(k_x, y = 0),a} d_{(k_x, y = 0),b} 
\end{equation}
where
\begin{equation}
    \tilde{H}_{ab}(k_x) = \langle c^\dagger_{(k_x, y = 0),a} c_{(k_x, y = 0),b} \rangle.  
\end{equation}

We claim that $\tilde{H}_{12}(k_x = \pi) = 0$. To show this, eigenvalues and eigenvectors of 
\begin{widetext}
\begin{equation}
    H(\vec{k} = (\pi, k_y)) = - \begin{pmatrix} 0 & t_1 - t_2 & 0 & t_1 + t_2 e^{\ci k_y} \\
    t_1 - t_2 & 0 & t_1 + t_2 e^{\ci k_y} & 0 \\
    0 & t_1 + t_2 e^{ - \ci k_y} & 0 & t_1 - t_2 \\
    t_1 + t_2 e^{ - \ci k_y} & 0 & t_1 - t_2 & 0
    \end{pmatrix} 
\end{equation}
\end{widetext}
 are computed in order to determine $\langle c^\dagger_{(\pi, k_y),1} c_{(\pi, k_y),2} \rangle$. This is subsequently Fourier-transformed in the $k_y$ coordinate to obtain
\begin{equation}
    \langle c^\dagger_{(k_x, y = 0),1} c_{(k_x, y = 0),2} \rangle = \int_{-\pi}^{\pi} dk_y \langle c^\dagger_{\vec{k},1} c_{\vec{k},2} \rangle.
\end{equation}

Since $\tilde{H}_{12}(k_x = \pi) = 0$ regardless of the values of $t_1$ and $t_2$, the entanglement spectrum of the SSH chain subsystem is always gapless, indicating that these subsystems are not topological in the metallic phase.

% \begin{figure}
%     \centering
%     \includegraphics[width = \columnwidth]{NewFigs/Contour_(1,0.5,0.5).png}
%     \caption{\textbf{$G(\omega^{+})$ curve in the $I_{\alpha}$ phase:} Notice how the curve winds about the branch point $1/2t_{\alpha}$ while crossing the respective branch cut \ttd{Change to $G/G_\alpha$ curve}}
%     \label{fig:enter-label}
% \end{figure}

\fi

\end{document}